\documentclass[pdftex,twocolumn,epjc3]{svjour3}          

\usepackage{amsmath}
\usepackage{amssymb}
\usepackage{multirow}
\usepackage{hyperref}
\usepackage{widetext}
\usepackage{color}
\usepackage{epsfig}
\usepackage{longtable,lscape}
\usepackage{amssymb}
\usepackage{indentfirst}
\usepackage{feynmf}
\usepackage{epstopdf}
\usepackage{slashed}
\usepackage{cases}
\usepackage{enumerate}
\usepackage{float}
\usepackage{longtable}
\usepackage{widetext}
\usepackage{subfigure}

\allowdisplaybreaks[3]

\newcommand{\mev}{\textrm{ MeV}}

\newcommand{\gev}{\textrm{ GeV}}

\usepackage[T1]{fontenc}

\smartqed  

\journalname{Eur. Phys. J. A}

\begin{document}
\begin{sloppypar}

\title{Molecular pentaquark states with open charm and bottom flavors}

\author{Jia-Xin Lin\thanksref{addr1} \and Hua-Xing Chen\thanksref{addr1} \and Wei-Hong Liang\thanksref{addr2,addr3} \and Wen-Ying Liu\thanksref{addr1} \and Dan Zhou\thanksref{addr4}}

\institute{School of Physics, Southeast University, Nanjing 210094, China\label{addr1}
\and
Department of Physics, Guangxi Normal University, Guilin 541004, China\label{addr2}
\and
Guangxi Key Laboratory of Nuclear Physics and Technology, Guangxi Normal University, Guilin 541004, China\label{addr3}
\and
Department of Physics and Hebei Key Laboratory of Photophysics Research and Application, Hebei Normal University, Shijiazhuang 050024, China\label{addr4}
}

\date{Received: date / Accepted: date}

\maketitle

\begin{abstract}
We study the possibly-existing molecular pentaquark states with open charm and bottom flavors, {\it i.e.}, the states with the quark contents $c\bar{b}qqq$ and $b\bar{c}qqq$ ($q=u,d,s$). We investigate the meson-baryon interactions through the coupled-channel unitary approach within the local hidden-gauge formalism, and extract the poles by solving the Bethe-Salpeter equation in coupled channels. These poles qualify as molecular pentaquark states, which are dynamically generated from the meson-baryon interactions through the exchange of vector mesons. Our results suggest the existence of the $\Sigma_c^{(*)} B^{(*)}$ and $\Sigma_b^{(*)} \bar{D}^{(*)}$ molecular states with isospin $I=1/2$, the $\Xi_c^{(\prime,*)} B^{(*)}$ and $\Xi_b^{(\prime,*)} \bar{D}^{(*)}$ molecular states with isospin $I=0$ and $I=1$, as well as the $\Omega_c^{(*)} B^{(*)}$ and $\Omega_b^{(*)} \bar D^{(*)}$ molecular states with isospin $I=1/2$.
\keywords{Local hidden-gauge formalism \and Meson-baryon interaction \and Molecular pentaquark state}
\end{abstract}

%
\noindent
{\bf Program Summary and Specifications}\\
\begin{small}
\noindent
{Program title:}\\
{Licensing provisions:}\\
{Programming language:}\\
{Repository and DOI:}\\
{Description of problem:} The $P_c(4312)^+$, $P_c(4440)^+$, $P_c(4457)^+$ with the quark content $c\bar{c}uud$ are just below the $\bar{D}^{(*)} \Sigma_c$ thresholds, and the $P_{cs}(4459)^0$ with the quark content $c\bar{c}sud$ is just below the $\bar{D}^* \Xi_c$ threshold. The observation of these hidden-charm pentaquark states motivate us to investigate the hadronic molecular states with the quark content $c\bar{b}qqq$ and $b\bar{c}qqq$ ($q=u, d, s$).\\
{Method of solution:} We investigate the meson-baryon interactions through the coupled-channel unitary approach within the local hidden-gauge formalism, and extract the poles by solving the Bethe-Salpeter equation in coupled channels.\\
{Additional comments:}\\
\end{small}
%

%
\section{Introduction}
\label{sec:Intro}
%

The exotic hadrons, such as the compact multiquark states and hadronic molecular states, can not be explained in the conventional quark model as the $q \bar q$ mesons and $qqq$ baryons. Studies on these states have become a crucial subject in hadron physics and received much attention in the past twenty years~\cite{Chen:2016qju,Liu:2019zoy,Chen:2022asf,Lebed:2016hpi,Esposito:2016noz,Hosaka:2016pey,Guo:2017jvc,Ali:2017jda,Olsen:2017bmm,Karliner:2017qhf,Bass:2018xmz,Brambilla:2019esw,Guo:2019twa,Ketzer:2019wmd,Yang:2020atz,Fang:2021wes,Jin:2021vct,JPAC:2021rxu,Meng:2022ozq,Brambilla:2022ura}. Especially, a number of hidden-charm pentaquark states were reported, including the $P_c(4312)^+$, $P_c(4337)$, $P_c(4380)^+$, $P_c(4440)^+$, $P_c(4457)^+$, $P_{cs}(4338)^0$, and $P_{cs}(4459)^0$~\cite{LHCb:2015yax,LHCb:2019kea,LHCb:2020jpq,LHCb:2021chn,LHCb:2022jad}. These structures contain the quark contents $\bar c c u u d$ or $\bar c c u d s$. They are perfect candidates for the hidden-charm pentaquark states (with strangeness), and various theoretical interpretations were proposed to explain them, such as the tightly-bound compact pentaquark states~\cite{Maiani:2015vwa,Lebed:2015tna,Stancu:2019qga,Giron:2019bcs,Ali:2019npk,Weng:2019ynv,Eides:2019tgv,Cheng:2019obk,Ali:2019clg,Wang:2019got,Wang:2020eep}, loosely-bound hadronic molecular states~\cite{Chen:2019asm,Chen:2015moa,Chen:2016otp,Chen:2019bip,Chen:2020pac,Chen:2020opr,Chen:2021erj,Wang:2015pcn,Liu:2019tjn,Huang:2019jlf,Guo:2019kdc,Fernandez-Ramirez:2019koa,Meng:2019ilv,Wu:2019adv,Yamaguchi:2019seo,Valderrama:2019nbk,Liu:2019zvb,Wang:2019ato,Gutsche:2019mkg,Du:2019pij,Ozdem:2018qeh,Wang:2019hyc,Zhang:2019xtu,Azizi:2016dhy}, and kinematical effects~\cite{Liu:2015fea,Kuang:2020bnk}. A suggestion of the peak at $4450 \mev$ to be due to a triangle singularity was made in Ref.~\cite{Guo:2015umn}, but it was dismissed in Ref.~\cite{Bayar:2016ftu} since it involved a $p$-wave amplitude at threshold in the triangle loop. See Ref.~\cite{Nakamura:2021qvy} for more relevant studies. Since the $P_c(4312)^+$, $P_c(4440)^+$, and $P_c(4457)^+$ are just below the $\bar D^{(*)} \Sigma_c$ thresholds, it is natural to explain them as the $\bar D^{(*)} \Sigma_c$ hadronic molecular states, whose existence has been predicted in Refs.~\cite{Wu:2010jy,Wang:2011rga,Yang:2011wz,Karliner:2015ina,Wu:2012md}. Similarly, the $P_{cs}(4459)^0$ is just below the $\bar D^{*} \Xi_c$ threshold, so it is natural to explain it as the $\bar D^{*} \Xi_c$ molecular state~\cite{Chen:2015sxa,Chen:2020uif,Peng:2020hql}.

Among various theoretical methods, the coupled-channel unitary approach within the local hidden-gauge formalism has been extensively applied to study the hidden-charm pentaquark states (with strangeness). In Refs.~\cite{Wu:2010jy,Wu:2010vk} the authors studied the interaction between the charmed mesons and charmed baryons, based on which they predicted several pentaquark states as the $\bar{D}^{(*)}\Sigma_c$ hadronic molecular states. This work was updated in Ref.~\cite{Xiao:2019aya}, where the authors reproduced the $P_c(4312)^+$, $P_c(4440)^+$, and $P_c(4457)^+$ from the interaction inside the $\bar{D}^{(*)} \Sigma_c^{(*)}$, $\bar{D}^* \Lambda_c$, $\eta_c N$, and $J/\psi N$ channels through the heavy quark spin symmetry. More coupled-channel studies can be found in Refs.~\cite{He:2019ify,Xiao:2020frg,Burns:2019iih,Wang:2022oof,He:2019rva,Yamaguchi:2016ote}. Besides, this method was applied in Refs.~\cite{Debastiani:2017ewu,Liang:2017ejq,Yu:2018yxl,Yu:2019yfr,Dias:2018qhp,Dias:2019klk,Wang:2022aga,Montana:2017kjw} to study the excited heavy baryons observed in recent years~\cite{LHCb:2017uwr,LHCb:2020tqd,BaBar:2007xtc,Belle:2017jrt,Belle:2018yob,LHCb:2018vuc,LHCb:2017iph}.

In this paper we shall apply the coupled-channel unitary approach within the local hidden-gauge formalism to study the possibly-existing molecular pentaquark states with the quark contents $c\bar{b}qqq$ and $b\bar{c}qqq$ ($q=u,d,s$), especially,
\begin{itemize}

\item[$\bullet$] For the $c\bar{b}uud$ system with isospin $I=1/2$, we consider the interactions inside the $N B_c^{(*)}$, $\Lambda_c B^{(*)}$, and $\Sigma_c^{(*)} B^{(*)}$ channels. Seven poles are extracted in this system.

\item[$\bullet$] For the $b\bar{c}uud$ system with isospin $I=1/2$, we consider the interactions inside the $N \bar{B}_c^{(*)}$, $\Lambda_b \bar{D}^{(*)}$, and $\Sigma_b^{(*)} \bar{D}^{(*)}$ channels. Seven poles are extracted in this system.

\item[$\bullet$] For the $c\bar{b}sud$ system with isospin $I=0$, we consider the interactions inside the $\Lambda B_c^{(*)}$, $\Lambda_c B_s^{(*)}$, $\Xi_c^\prime B^{(*)}$, and $\Xi_c^{(*)} B^{(*)}$ channels. Ten poles are extracted in this system.

\item[$\bullet$] For the $b\bar{c}sud$ system with isospin $I=0$, we consider the interactions inside the $\Lambda \bar{B}_c^{(*)}$, $\Lambda_b \bar{D}_s^{(*)}$, $\Xi_b^\prime \bar{D}^{(*)}$, and $\Xi_b^{(*)} \bar{D}^{(*)}$ channels. Ten poles are extracted in this system.

\item[$\bullet$] For the $c\bar{b}sud$ system with isospin $I=1$, we consider the interactions inside the $\Sigma^{(*)} B_c^{(*)}$, $\Xi_c^\prime B^{(*)}$, $\Sigma_c^{(*)} B_s^{(*)}$, and $\Xi_c^{(*)} B^{(*)}$ channels. Seven poles are extracted in this system.

\item[$\bullet$] For the $b\bar{c}sud$ system with isospin $I=1$, we consider the interactions inside the $\Sigma^{(*)} \bar B_c^{(*)}$, $\Xi_b^\prime \bar D^{(*)}$, $\Sigma_b^{(*)} \bar D_s^{(*)}$, and $\Xi_b^{(*)} \bar D^{(*)}$ channels. Seven poles are extracted in this system.

\item[$\bullet$] For the $c\bar{b}ssu$ system with isospin $I=1/2$, we consider the interactions inside the $\Xi^{(*)} B_c^{(*)}$, $\Xi_c^\prime B_s^{(*)}$, $\Xi_c^{(*)} B_s^{(*)}$, and $\Omega_c^{(*)} B^{(*)}$ channels. Seven poles are extracted in this system.

\item[$\bullet$] For the $b\bar{c}ssu$ system with isospin $I=1/2$, we consider the interactions inside the $\Xi^{(*)} \bar B_c^{(*)}$, $\Xi_b^\prime \bar D_s^{(*)}$, $\Xi_b^{(*)} \bar D_s^{(*)}$, and $\Omega_b^{(*)} \bar D^{(*)}$ channels. Seven poles are extracted in this system.

\end{itemize}
The above poles can be possibly qualified as the molecular pentaquark states, dynamically generated from the meson-baryon interactions through the exchange of vector mesons. We shall calculate their masses and widths as well as their couplings to various coupled channels. Especially, our results suggest the existence of the $\Sigma_c^{(*)} B^{(*)}$ and $\Sigma_b^{(*)} \bar{D}^{(*)}$ molecular states with $I=1/2$, the $\Xi_c^{(\prime,*)} B^{(*)}$ and $\Xi_b^{(\prime,*)} \bar{D}^{(*)}$ molecular states with $I=0$ and $I=1$, as well as the $\Omega_c^{(*)} B^{(*)}$ and $\Omega_b^{(*)} \bar D^{(*)}$ molecular states with $I=1/2$.

This paper is organized as follows. In Sec.~\ref{sec:form} we derive the transition potentials from the meson-baryon interactions through the coupled-channel unitary approach within the local hidden-gauge formalism. These transition potentials are used in Sec.~\ref{sec:results} to study the $c\bar{b}uud$ and $b\bar{c}uud$ systems, and extract the poles by solving the Bethe-Salpeter equation in coupled channels. We apply the same method to study the other $c\bar{b}qqq$ and $b\bar{c}qqq$ systems ($q=u,d,s$) in Sec.~\ref{sec:others}. A brief conclusion is given in Sec.~\ref{sec:Con}.

%
\section{Formalism}
\label{sec:form}
%

The $c\bar{c}uud$ system has been systematically studied in Ref.~\cite{Wu:2010jy} through the coupled-channel unitary approach within the local hidden-gauge formalism, where the authors considered the interactions of the $N \eta_c$, $N J/\psi$, $\Lambda_c \bar{D}^{(*)}$, and $\Sigma_c \bar{D}^{(*)}$ channels. Similarly, the $b\bar{b}uud$ system has been studied in Ref.~\cite{Wu:2010rv}, where the authors considered the interactions of the $N \eta_b$, $N \Upsilon$, $\Sigma_b B^{(*)}$, and $\Lambda_b B^{(*)}$ channels; the $ccsq\bar{q}$, $bbsq\bar{q}$, and $bcsq\bar{q}$ systems $(q=u, d)$ have been studied in Ref.~\cite{Wang:2022aga}, where the authors considered the interactions of the $\Xi_{cc}^{(*)} \bar{K}^{(*)}$, $\Omega_{cc}^{(*)} \eta$, $\Omega_{cc}^{(*)} \omega$, and $\Xi_c^{(\prime,*)} D^{(*)}$ channels for the $ccsq\bar{q}$ system, the interactions of the $\Xi_{bb}^{(*)} \bar{K}^{(*)}$, $\Omega_{bb}^{(*)} \eta$, $\Omega_{bb}^{(*)} \omega$, and $\Xi_b^{(\prime,*)} \bar{B}^{(*)}$ channels for the $bbsq\bar{q}$ system, and the interactions of the $\Xi_{bc}^{(\prime,*)} \bar{K}^{(*)}$, $\Xi_c^{(\prime,*)} \bar{B}^{(*)}$, $\Xi_b^{(\prime,*)} D^{(*)}$, $\Omega_{bc}^{(\prime,*)} \eta$, and $\Omega_{bc}^{(\prime,*)} \omega$ channels for the $bcsq\bar{q}$ system; the $c\bar{b}uds$ system has been studied in Ref.~\cite{Shen:2022rpn}, where the authors considered the interactions of the $B^{(*)} \Xi_c$, $B^{(*)} \Xi_c^\prime$, $B_s^{(*)} \Lambda_c$, and $B_c^{(*)} \Lambda$ channels.

In this paper we follow the same approach to study the $c\bar{b}qqq$ and $b\bar{c}qqq$ systems ($q=u,d,s$) to search for possibly-existing molecular pentaquark states with open charm and bottom flavors. In this section we study the $c\bar{b}uud$ and $b\bar{c}uud$ systems, and the results for the other systems will be given in Sec.~\ref{sec:others}.

For the former $c\bar{b}uud$ system we consider
\begin{enumerate}[a)]
  \item $PB_{1/2}$ channels: $N B_c$, $\Lambda_c B$, $\Sigma_c B$;
  \item $VB_{1/2}$ channels: $N B_c^*$, $\Lambda_c B^*$, $\Sigma_c B^*$;
  \item $PB_{3/2}$ channel: $\Sigma_c^* B$;
  \item $VB_{3/2}$ channel: $\Sigma_c^* B^*$.
\end{enumerate}
In the above expressions, $P$ and $V$ stand for the pseudoscalar and vector mesons, respectively; $B_{1/2}$ and $B_{3/2}$ stand for the baryons with the spin-parity $J^P=1/2^+$ and $J^P=3/2^+$, respectively. We also use the symbols $\mathcal{M}$ and $\mathcal{B}$ to generally denote the meson and baryon, respectively.

For the latter $b\bar{c}uud$ system we consider
\begin{enumerate}[a)]
  \item $PB_{1/2}$ channels: $N \bar{B}_c$, $\Lambda_b \bar{D}$, $\Sigma_b \bar{D}$;
  \item $VB_{1/2}$ channels: $N \bar{B}_c^*$, $\Lambda_b \bar{D}^*$, $\Sigma_b \bar{D}^*$;
  \item $PB_{3/2}$ channel: $\Sigma_b^* \bar{D}$;
  \item $VB_{3/2}$ channel: $\Sigma_b^* \bar{D}^*$.
\end{enumerate}

\begin{table}[t]
  \renewcommand\arraystretch{1.5}
  \centering
  \caption{Threshold masses of the 16 channels with $I=1/2$ and $I_3=1/2$ for the $c\bar{b}uud$ and $b\bar{c}uud$ systems, in units of MeV.}
  \label{tab:mass}
  \setlength{\tabcolsep}{6mm}{
  \begin{tabular}{l|ccc}
      \hline\hline
      \multirow{2}{*}{$PB_{1/2}$}  & $N B_c$& $\Lambda_c B$ & $\Sigma_c B$ \\
      ~ & $7213$ & $7566$ & $7733$\\
      \hline
      \multirow{2}{*}{$VB_{1/2}$}  & $N B_c^*$ & $\Lambda_c B^*$ & $\Sigma_c B^*$ \\
      ~ & $7270$ & $7611$ & $7778$\\
      \hline
      \multirow{2}{*}{$PB_{3/2}$}  & $\Sigma_c^* B$ \\
      ~ & $7798$\\
      \hline
      \multirow{2}{*}{$VB_{3/2}$}  & $\Sigma_c^* B^*$ \\
      ~ & $7843$\\
      \hline\hline
      \multirow{2}{*}{$PB_{1/2}$}  & $N \bar{B_c}$ & $\Lambda_b \bar{D}$ & $\Sigma_b \bar{D}$ \\
      ~  & $7213$ & $7487$ & $7680$\\
      \hline
      \multirow{2}{*}{$VB_{1/2}$}  & $N \bar{B_c}^*$ & $\Lambda_b \bar{D}^*$ & $\Sigma_b \bar{D}^*$ \\
      ~  & $7270$& $7628$ & $7822$\\
      \hline
      \multirow{2}{*}{$PB_{3/2}$} & $\Sigma_b^* \bar{D}$ \\
      ~  & $7702$\\
      \hline
      \multirow{2}{*}{$VB_{3/2}$}  & $\Sigma_b^* \bar{D}^*$ \\
      ~  & $7843$\\
      \hline\hline
  \end{tabular} }
\end{table}

There are two isospin possibilities: $I=1/2$ and $I=3/2$. Similar to the $c\bar{c}uud$ and $b\bar{b}uud$ systems~\cite{Wu:2010jy,Wu:2010rv}, we find the interactions of the $c\bar{b}uud$ and $b\bar{c}uud$ systems in the $I=3/2$ sector to be also repulsive (see Sec.~\ref{sec:uuu}), so we only investigate the $I=1/2$ sector in this section. The isospin states for the $c\bar{b}uud$ and $b\bar{c}uud$ systems with $I=1/2$ and $I_3=1/2$ are
\begin{eqnarray}
\Big| N B_c^{(*)} \Big\rangle &=& \Big| p B_c^{(*)+} \Big\rangle, \\[2mm]
\Big| \Lambda_c B^{(*)} \Big\rangle &=& \Big| \Lambda_c^+ B^{(*)+}\Big\rangle, \\[2mm]
\Big| \Sigma_c^{(*)} B^{(*)} \Big\rangle &=& \sqrt{\frac{2}{3}} \Big|\Sigma_c^{(*)++} B^{(*)0} \Big\rangle - \sqrt{\frac{1}{3}}\Big| \Sigma_c^{(*)+} B^{(*)+} \Big\rangle, \nonumber\\[2mm]
\\
\Big| N \bar{B_c}^{(*)} \Big\rangle &=& \Big| p B_c^{(*)-} \Big\rangle, \\[2mm]
\Big| \Lambda_b \bar{D}^{(*)} \Big\rangle &=& \Big| \Lambda_b^0 \bar{D}^{(*)0} \Big\rangle, \\[2mm]
\Big| \Sigma_b^{(*)} \bar{D}^{(*)} \Big\rangle &=& \sqrt{\frac{2}{3}} \Big| \Sigma_b^{(*)+} D^{(*)-} \Big\rangle - \sqrt{\frac{1}{3}}\Big| \Sigma_b^{(*)0} \bar{D}^{(*)0} \Big\rangle .\nonumber\\
\label{eq:isospin}
\end{eqnarray}

The threshold masses of these channels are summarized in Table~\ref{tab:mass}, with the masses of heavy mesons and baryons taken from PDG2022~\cite{pdg}. However, the $B_c^*$ meson has not been observed in experiments yet, and we take its mass from Ref.~\cite{Mathur:2018epb}, where the authors estimated it using the lattice QCD method.

\begin{figure}[htb]
  \centering
  \includegraphics[width=0.48\textwidth]{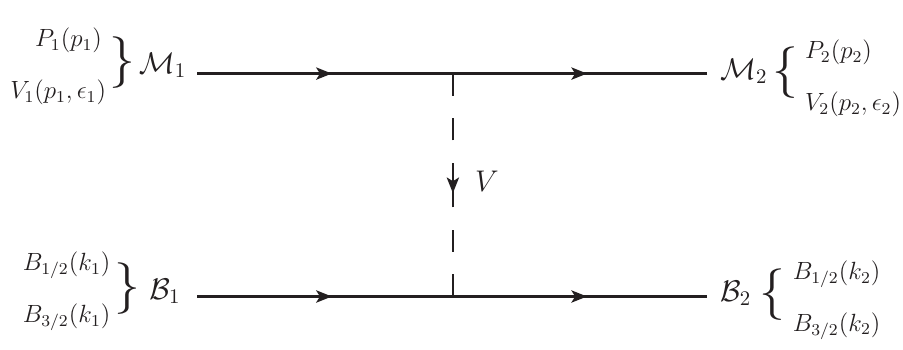}
  \caption{Feynman diagram for the interaction $\mathcal{M}_1 \mathcal{B}_1 \to \mathcal{M}_2 \mathcal{B}_2$ via the exchange of the vector meson $V$. $\mathcal{M}_1 (\mathcal{M}_2)$ and $\mathcal{B}_1 (\mathcal{B}_2)$ are the initial (final) meson and baryon, respectively.}
  \label{Fig.ppvvbb}
\end{figure}

To study the interactions of the $c\bar{b}uud$ and $b\bar{c}uud$ systems, we consider the exchange of vector mesons in the $t$-channel, as illustrated in Fig.~\ref{Fig.ppvvbb}. We shall evaluate its induced transition potentials through the local hidden-gauge approach (LHG)~\cite{Bando:1984ej,Bando:1987br,Meissner:1987ge}. There are two vertices, the upper $V\mathcal{M}_1\mathcal{M}_2$ and the lower $V\mathcal{B}_1\mathcal{B}_2$, which will be separately discussed as follows.

\subsection{The vertex $V\mathcal{M}_1\mathcal{M}_2$}
%

We use two Lagrangians to describe the upper vertex $V\mathcal{M}_1\mathcal{M}_2$ of Fig.~\ref{Fig.ppvvbb}~\cite{Oset:2010tof}:
\begin{eqnarray}
  \mathcal{L}_{VPP} &=& -ig \big\langle \left[ P, \partial_\mu P \right] V^\mu \big\rangle,
  \label{eq:vpp}
  \\[2mm]
  \mathcal{L}_{VVV} &=& ig \big\langle \left( V^\mu \partial_\nu V_\mu - \partial_\nu V_\mu V^\mu \right) V^\nu \big\rangle,
  \label{eq:vvv}
\end{eqnarray}
where the symbol $\langle\cdots\rangle$ denotes the trace of a matrix, and the coupling constant $g$ is defined as $g={M_V}/({2f_{\pi}})$ with $f_{\pi}=93\,\rm MeV$ and $M_V=800 \, \rm MeV$. $P$ and $V$ are two matrices for the pseudoscalar and vector mesons, respectively:
\begin{widetext}
\setlength{\arraycolsep}{1.5pt}
\renewcommand{\arraystretch}{1.5}
\begin{eqnarray}
P &=&
\left(\begin{array}{ccccc}
\frac{\eta}{\sqrt{3}}+\frac{\eta'}{\sqrt{6}}+\frac{\pi^0}{\sqrt{2}} & \pi^+                                                               & K^+                                            & ~~~~~~\bar{D}^0~~~~~~ & B^+   \\
\pi^-                                                               & \frac{\eta}{\sqrt{3}}+\frac{\eta'}{\sqrt{6}}-\frac{\pi^0}{\sqrt{2}} & K^{0}                                          & D^-                   & B^0   \\
K^{-}                                                               & \bar{K}^{0}                                                         & -\frac{\eta}{\sqrt{3}}+\sqrt{\frac{2}{3}}\eta' & D_s^-                 & B_s^0 \\
D^0 & D^+       & D_s^+       & \eta_c & B_c^+ \\
B^- & \bar{B}^0 & \bar{B}^0_s & B_c^-  & \eta_b
\end{array} \right)\, ,
\label{eq-pfields}
\\ V &=&
\left( \begin{array}{ccccc}
\frac{\omega+\rho^0}{\sqrt{2}} & \rho^+                         & K^{*+}         & \bar{D}^{*0} & B^{*+} \\
\rho^-                         & \frac{\omega-\rho^0}{\sqrt{2}} & K^{*0}         & D^{*-}       & B^{*0} \\
K^{*-}                         & \bar{K}^{*0}                   & \phi           & D_s^{*-}     & B_s^{*0} \\
~~~D^{*0}~~~                   & ~~~D^{*+}~~~                   & ~~~D_s^{*+}~~~ & ~~~J/\psi~~~ & ~~~B_c^{*+}~~~ \\
B^{*-}                         & \bar{B}^{*0}                   & \bar{B}^{*0}_s & B_c^{*-}     & \Upsilon
\end{array} \right) \, .
\label{eq-vfields}
\end{eqnarray}
\end{widetext}
Although formally we work within the flavor $SU(5)$ symmetry, the vertices of Eqs.~\eqref{eq:vpp} and \eqref{eq:vvv} only use the overlap of $q \bar q$ in the external mesons and the exchanged vector mesons \cite{Sakai:2017avl}. In addition, we apply wave functions for baryon states as shown in Sec.~\ref{subsec:vbb}, where the heavy quarks are singled out with the flavor-spin symmetry imposed in the remaining light quarks. Hence, the $SU(5)$ symmetry is not used in baryon sector. The dominant contributions come from the exchange of light vectors, since the propagators are bigger, and the heavy quarks are spectators in the process. Then the interaction follows the flavor $SU(3)$ symmetry of light quarks. Hence one is projecting over $SU(3)$ the original $SU(5)$ amplitudes, and also, since the interaction does not depend on the spectator heavy quarks, then the heavy quark spin symmetry (HQSS) is automatically fulfilled for the dominant terms


As discussed in Ref.~\cite{Debastiani:2017ewu}, the exchange of light vector mesons can be well described by the flavor $SU(3)$ symmetry. However, some non-diagonal transitions can exchange heavy vector mesons, {\it e.g.}, the $N B_c$ and $\Lambda_c B$ channels interact through the exchange of the $D^*$ meson, and the $N \bar{B}_c$ and $\Lambda_b \bar{D}$ channels interact through the exchange of the $\bar{B}^*$ meson. Here the heavy quarks are no longer spectators, and these terms do not comply with the heavy quark spin symmetry, but the propagators of $D^*$ and $\bar{B}^*$ go as $\mathcal{O}(1/m_Q^2)$ that is subleading in the $\mathcal{O}(1/m_Q)$ counting ($m_Q$ stands for the heavy quark mass). To describe these interactions we need to explicitly work within the flavor $SU(5)$ symmetry, which is violated to some extent when using the physical masses for the interacting mesons and baryons. More relevant discussions can be found in Sec.~\ref{subsec:resultsuud}.

\subsection{The vertex $V\mathcal{B}_1\mathcal{B}_2$}
\label{subsec:vbb}
%

The lower vertex $V\mathcal{B}_1\mathcal{B}_2$ of Fig.~\ref{Fig.ppvvbb} has been systematically studied in Refs.~\cite{Klingl:1997kf,Oset:2010tof} for the flavor $SU(3)$ symmetry through the Lagrangian
\begin{equation}
  \mathcal{L}_{VBB}=g \left[ \big\langle \bar{B} \gamma_\mu V^\mu B - \bar{B} \gamma_\mu B V^\mu \big\rangle + \big\langle \bar{B} \gamma_\mu B \big\rangle \big\langle V^\mu \big\rangle\right],
\end{equation}
where $B$ and $V$ are two $3\times3$ matrices for the flavor $SU(3)$ baryons and vector mesons, respectively. However, this Lagrangian can not be directly extended to the flavor $SU(5)$ symmetry, and it is more straightforward and intuitive to use the baryon wave functions~\cite{Lutz:2003jw,Hofmann:2006qx,Roberts:2007ni}. This has been done in Ref.~\cite{Debastiani:2017ewu}, and in this paper we follow the same approach to calculate the lower vertex $V\mathcal{B}_1\mathcal{B}_2$.

We need the following wave functions for the $J^P=1/2^+$ baryons:
\begin{eqnarray}
    \big| p \big\rangle &=& \frac{1}{\sqrt{2}} \left(\big| \phi_{MS} \big\rangle \big| \chi_{MS} \big\rangle + \big| \phi_{MA} \big\rangle \big| \chi_{MA} \big\rangle \right), \\[2mm]
     \big| \Lambda_c^+ \big\rangle &=& \Big | \frac{1}{\sqrt{2}}c\left(ud-du\right)\Big\rangle \big| \chi_{MA} \big\rangle, \\[2mm]
    \big| \Sigma_c^{++} \big\rangle &=& \big| cuu \big\rangle \big| \chi_{MS} \big\rangle, \\[2mm]
    \big| \Sigma_c^+ \big\rangle &=& \Big| \frac{1}{\sqrt{2}}c\left(ud+du\right) \Big\rangle \big| \chi_{MS} \big\rangle, \\[2mm]
    \big| \Lambda_b^0 \big\rangle &=& \Big| \frac{1}{\sqrt{2}} b \left(ud-du\right) \Big\rangle \big| \chi_{MA} \big\rangle, \\[2mm]
    \big| \Sigma_b^{+} \big\rangle &=& \big| buu \big\rangle \big| \chi_{MS} \big\rangle, \\[2mm]
    \big| \Sigma_b^0 \big\rangle &=&\Big| \frac{1}{\sqrt{2}} b \left(ud+du\right) \Big\rangle \big| \chi_{MS} \big\rangle,
\end{eqnarray}
where
\begin{eqnarray}
   &\,& \big| \phi_{MS} \big\rangle=\frac{1}{\sqrt{6}}\left(\big| udu+uud-2duu \big\rangle\right), \\[2mm]
    &\,& \big| \phi_{MA} \big\rangle=\frac{1}{\sqrt{2}}\left(\big| uud-udu \big\rangle\right),
\end{eqnarray}
and $\chi_{MS}$ and $\chi_{MA}$ stand for the mixed-symmetric and mixed-antisymmetric spin wave functions, respectively. These two functions are orthogonal to each other:
\begin{eqnarray}
 \big\langle  \chi_{MS}\big| \chi_{MS} \big\rangle & =&1,\\ [2mm]
 \big\langle  \chi_{MA}\big| \chi_{MA} \big\rangle &=&1,\\  [2mm]
 \big\langle  \chi_{MS}\big| \chi_{MA} \big\rangle &=&0.
\end{eqnarray}

Besides, we also need the following wave functions for the $J^P=3/2^+$ baryons:
\begin{eqnarray}
  \big| \Sigma_c^{*++} \big\rangle &=& \big| cuu \big\rangle \big| \chi_{S} \big\rangle, \\[2mm]
  \big| \Sigma_c^{*+} \big\rangle &=& \Big | \frac{1}{\sqrt{2}}c\left(ud+du\right)\Big\rangle \big| \chi_{S} \big\rangle, \\[2mm]
  \big| \Sigma_b^{*+} \big\rangle &=& \big| buu \big\rangle \big| \chi_{S} \big\rangle, \\[2mm]
  \big|\Sigma_b^{*0} \big\rangle &=& \Big | \frac{1}{\sqrt{2}} b \left(ud+du\right)\Big\rangle \big| \chi_{S} \big\rangle,
\end{eqnarray}
where $\chi_{S}$ stands for the fully-symmetric spin wave function. All the other relevant baryon wave functions are summarized in \ref{Sec:baryon}.

Based on the above baryon wave functions, we can write the lower vertex $V\mathcal{B}_1\mathcal{B}_2$ for the exchanged light vector mesons through the quark operators as
\begin{equation}
  \mathcal{L}_{V\mathcal{B}\mathcal{B}} \to
  \left\{\begin{array}{l}
    g~u\bar{d}, \text{for $\rho^+$}\\[2mm]
    \frac{g}{\sqrt{2}} \left( u\bar{u}-d\bar{d} \right), \text{for $\rho^0$}\\[2mm]
    g~d\bar{u}, \text{for $\rho^-$}\\[2mm]
    \frac{g}{\sqrt{2}} \left( u\bar{u}+d\bar{d} \right), \text{for $\omega$}
  \end{array}\right.,
  \label{eq:VBB1}
\end{equation}
and we can write it for the exchanged heavy vector mesons through the quark operators as
\begin{equation}
  \mathcal{L}_{V\mathcal{B}\mathcal{B}} \to
  \left\{\begin{array}{l}
    g~c\bar{u}, \text{for $D^{*0}$}\\[2mm]
    g~c\bar{d}, \text{for $D^{*+}$}\\[2mm]
    g~b\bar{d}, \text{for $\bar{B}^{*0}$}\\[2mm]
    g~b\bar{u}, \text{for $B^{*-}$}
  \end{array}\right..
  \label{eq:VBB2}
\end{equation}
Note that we have taken $\gamma^\mu \to \gamma^0$ in the above formulae, which is valid when the transferred momentum is small. This approximation makes the lower vertex $V\mathcal{B}_1\mathcal{B}_2$ spin-independent, {\it i.e.},
\begin{equation}
  \mathcal{L}_{V\mathcal{B}\mathcal{B}} = \mathcal{L}_{VB_{1/2}B_{1/2}} = \mathcal{L}_{VB_{3/2}B_{3/2}} \, .
\end{equation}

\subsection{Scattering matrices}

Based on the Lagrangians given in Eqs.~\eqref{eq:vpp}, \eqref{eq:vvv}, \eqref{eq:VBB1}, and \eqref{eq:VBB2}, we calculate the upper and lower vertices of Fig.~\ref{Fig.ppvvbb} to derive the transition amplitudes, which can be generally written as
\begin{eqnarray}
  V^{P\mathcal{B}}_{ij}(s)&=& \frac{C^{P\mathcal{B}}_{ij}}{4f_{\pi}^2} \left(p^0_i + p^0_j\right) \, ,
  \label{eq:Vij1}
  \\[2mm]
  V^{V\mathcal{B}}_{ij}(s)&=& \frac{C^{V\mathcal{B}}_{ij}}{4f_{\pi}^2} \left(p^0_i + p^0_j\right) \times \vec{\epsilon}_i \cdot \vec{\epsilon}_j \, ,
  \label{eq:Vij2}
\end{eqnarray}
where $i$ and $j$ denote the initial and final channels, respectively; $p^0_i$ and $p^0_j$ are the energies of the initial and final mesons, respectively; $\vec{\epsilon}_i$ and $\vec{\epsilon}_j$ are the polarization vectors of the initial and final vector mesons, respectively.

The matrices $C^{\mathcal{M}\mathcal{B}}_{ij}$ for the $c\bar{b}uud$ system are
\begin{eqnarray}
  \label{eq:cbCijPB12}
  C^{PB_{1/2}}_{ij}&=&\left(
  \renewcommand{\arraystretch}{2}
  \renewcommand{\arraycolsep}{1.3pt}
  \begin{array}{c|ccc}
  J={1\over2}            & N B_c                          & \Lambda_c B                     & \Sigma_c B
  \\ \hline
  N B_c                & 0                              & \frac{1}{\sqrt{2}} \lambda_1    & -\frac{1}{\sqrt{2}} \lambda_1 \\
  \Lambda_c B          & \frac{1}{\sqrt{2}} \lambda_1   & 1                               & 0 \\
  \Sigma_c B           & -\frac{1}{\sqrt{2}} \lambda_1  & 0                               & -1
  \end{array}
  \right),
\\
  \label{eq:cbCijVB12}
  C^{VB_{1/2}}_{ij}&=&\left(
  \renewcommand{\arraystretch}{2}
  \renewcommand{\arraycolsep}{1.3pt}
  \begin{array}{c|ccc}
  J={1\over2}/{3\over2}       & N B_c^*                         & \Lambda_c B^*                     & \Sigma_c B^*
  \\ \hline
  N B_c^*                & 0                              & \frac{1}{\sqrt{2}} \lambda_1    & -\frac{1}{\sqrt{2}} \lambda_1 \\
  \Lambda_c B^*          & \frac{1}{\sqrt{2}} \lambda_1   & 1                               & 0 \\
  \Sigma_c B^*           & -\frac{1}{\sqrt{2}} \lambda_1  & 0                               & -1
  \end{array}
  \right),
\\
  \label{eq:cbCijPB32}
  C^{PB_{3/2}}_{ij}&=&\left(
  \renewcommand{\arraystretch}{2}
  \renewcommand{\arraycolsep}{1.3pt}
  \begin{array}{c|c}
  J={3\over2}       & \Sigma_c^* B
  \\ \hline
  \Sigma_c^* B    & -1
  \end{array}
  \right),
\\
  \label{eq:cbCijVB32}
  C^{VB_{3/2}}_{ij}&=&\left(
  \renewcommand{\arraystretch}{2}
  \begin{array}{c|c}
  J={1\over2}/{3\over2}/{5\over2}       & \Sigma_c^* B^*
  \\ \hline
  \Sigma_c^* B^*              & -1
  \end{array}
  \right),
\end{eqnarray}
and the matrices $C^{\prime\mathcal{M}\mathcal{B}}_{ij}$ for the $b\bar{c}uud$ system are
\begin{eqnarray}
  \label{eq:bcCijPB12}
  C^{\prime PB_{1/2}}_{ij}&=&\left(
  \renewcommand{\arraystretch}{2}
  \renewcommand{\arraycolsep}{1.3pt}
  \begin{array}{c|ccc}
  J={1\over2}            & N \bar{B}_c                          & \Lambda_b \bar{D}                     & \Sigma_b \bar{D}
  \\ \hline
  N \bar{B}_c                & 0                              & \frac{1}{\sqrt{2}} \lambda_2    & -\frac{1}{\sqrt{2}} \lambda_2 \\
  \Lambda_b \bar{D}          & \frac{1}{\sqrt{2}} \lambda_2   & 1                               & 0 \\
  \Sigma_b \bar{D}           & -\frac{1}{\sqrt{2}} \lambda_2  & 0                               & -1
  \end{array}
  \right),
\\
  \label{eq:bcCijVB12}
  C^{\prime VB_{1/2}}_{ij}&=&\left(
  \renewcommand{\arraystretch}{2}
  \renewcommand{\arraycolsep}{1.3pt}
  \begin{array}{c|ccc}
  J={1\over2}/{3\over2}       & N \bar{B}_c^*         & \Lambda_b \bar{D}^*                     & \Sigma_b \bar{D}^*
  \\ \hline
  N \bar{B}_c^*                & 0                              & \frac{1}{\sqrt{2}} \lambda_2    & -\frac{1}{\sqrt{2}} \lambda_2 \\
  \Lambda_b \bar{D}^*          & \frac{1}{\sqrt{2}} \lambda_2   & 1                               & 0 \\
  \Sigma_b \bar{D}^*           & -\frac{1}{\sqrt{2}} \lambda_2  & 0                               & -1
  \end{array}
  \right),
\\
  \label{eq:bcCijPB32}
  C^{\prime PB_{3/2}}_{ij}&=&\left(
  \renewcommand{\arraystretch}{2}
  \renewcommand{\arraycolsep}{1.3pt}
  \begin{array}{c|c}
  J={3\over2}       & \Sigma_b^* \bar{D}
  \\ \hline
  \Sigma_b^* \bar{D}    & -1
  \end{array}
  \right),
\\
  \label{eq:bcCijVB32}
  C^{\prime VB_{3/2}}_{ij}&=&\left(
  \renewcommand{\arraystretch}{2}
  \renewcommand{\arraycolsep}{1.3pt}
  \begin{array}{c|c}
  J={1\over2}/{3\over2}/{5\over2}       & \Sigma_b^* \bar{D}^*
  \\ \hline
  \Sigma_b^* \bar{D}^*              & -1
  \end{array}
  \right).
\end{eqnarray}

The above $C^{\mathcal{M}\mathcal{B}}_{ij}$ and $C^{\prime \mathcal{M}\mathcal{B}}_{ij}$ matrices contain two parameters $\lambda_1$ and $\lambda_2$, which are the reduction factors due to the exchanged heavy vector mesons, {\it e.g.}, the $N B_c$ and $\Lambda_c B$ channels interact through the exchange of the $D^*$ meson, and the $N \bar{B}_c$ and $\Lambda_b \bar{D}$ channels interact through the exchange of the $\bar{B}^*$ meson. As discussed above, we need to explicitly work within the flavor $SU(5)$ symmetry to describe these interactions; however, this symmetry is badly broken, so we need to add these reduction factors. The authors of Refs.~\cite{Debastiani:2017ewu,Liang:2017ejq} estimated this factor to be $\lambda \simeq 0.25$ for the exchange of the $D^*_s$ meson, and to be $\lambda \simeq 0.1$ for the exchange of the $B^*_s$ meson. We follow the same approach to obtain
\begin{equation}
  \lambda_1= \frac{-M_V^2}{\left(m_{B}-m_{B_c}\right)^2-m_{D^*}^2} \simeq 0.21,
\end{equation}
for the exchange of the $D^*$ meson, and
\begin{equation}
  \lambda_2=\frac{-M_V^2}{\left(m_{\bar{D}}-m_{B_c}\right)^2-m_{\bar{B}^*}^2} \simeq 0.07.
\end{equation}
for the exchange of the $\bar{B}^*$ meson.

%
\section{Numerical Results}
\label{sec:results}
%

\begin{figure*}[hbt]
  \centering
  \includegraphics[width=1.0\textwidth]{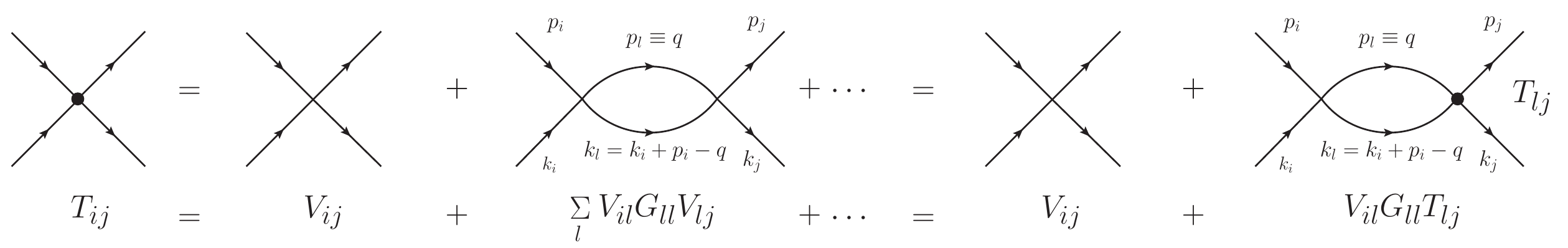}
  \caption{Scattering matrix calculated by solving the Bethe-Salpeter equation in coupled channels.}
  \label{Fig.bs}
\end{figure*}

In this section we perform numerical analyses to search for the possibly-existing molecular pentaquark states in the $c\bar{b}uud$ and $b\bar{c}uud$ systems. As depicted in Fig.~\ref{Fig.bs}, we can use the transition amplitudes of Eq.~\eqref{eq:Vij1} and Eq.~\eqref{eq:Vij2} to evaluate the scattering matrices $T^{\mathcal{MB}}$ by solving the Bethe-Salpeter equation in coupled channels~\cite{Oller:1998zr,Oller:2000fj}:
\begin{equation}
  \label{eq:T}
  T^{\mathcal{M}\mathcal{B}}(s) = {V^{\mathcal{M}\mathcal{B}}(s) \over {\bf 1} - V^{\mathcal{M}\mathcal{B}}(s) G^{\mathcal{M}\mathcal{B}}(s) },
\end{equation}
where $G^{\mathcal{M}\mathcal{B}}(s)$ is the meson-baryon loop function. This loop function is divergent, and we can renormalize it through either the dimensional regularization or the cut-off regularization. In the present study we use the dimensional regularization to regularize it as
\begin{eqnarray}
  G_{ll}^{\mathcal{M}\mathcal{B}}(s) &=& i \int \frac{\mathrm{d}^4 q}{(2\pi)^4}\frac{2M_l}{(P-q)^2 - M_l^2 + i\varepsilon} \frac{1}{q^2 - m_l^2+i\varepsilon} \nonumber\\[2mm]
  &=& {2M_l \over 16\pi^2}\Big\{a(\mu) + \ln{M_l^2 \over \mu^2} + {m_l^2-M_l^2+s \over 2s}\ln{m_l^2 \over M_l^2} \nonumber\\[2mm]
  & &+{k\over\sqrt{s}} \big[\ln\big(s-(M_l^2 - m_l^2) + 2k\sqrt{s}\big) \nonumber\\[2mm]
  & & + \ln\big(s+(M_l^2 - m_l^2) + 2k\sqrt{s}\big) \nonumber\\[2mm]
  & & - \ln\big(-s-(M_l^2 - m_l^2) + 2k\sqrt{s}\big) \nonumber\\[2mm]
  & & -\ln\big(-s+(M_l^2 - m_l^2) + 2k\sqrt{s}\big)\big]\Big\}.
  \label{eq:G}
\end{eqnarray}
In the above expression, $l$ denotes the intermediate channel; $m_l$ and $M_l$ are the masses of the intermediate meson and baryon, respectively; $s=P^2$ with $P=p_i+k_i$ the total four-momentum of the baryon and meson; $p_l \equiv q$ and $k_l = p_i + k_i - q$ are the momenta of the intermediate meson and baryon, respectively; $k = \sqrt{ (s-(m_l+M_l)^2)(s-(m_l-M_l)^2) } / (2\sqrt s)$ is the three-momentum of the particle in the center of mass frame; $\mu$ and $a(\mu)$ are the regularization scale and the subtraction constant, respectively. Here we take $\mu=1 \, \rm GeV$ so that $a(\mu)$ is the only free parameter. For examples, $a(\mu=1\gev)=-2.3$ was used in Refs.~\cite{Wu:2010jy,Wu:2010vk} to investigate the hidden-charm pentaquark states in the $c\bar{c}uud$ system, and $a(\mu=1\gev)=-3.71$ was used in Ref.~\cite{Wu:2010rv} to investigate the hidden-bottom pentaquark states in the $b\bar{b}uud$ system. Since the $c\bar{b}uud$ and $b\bar{c}uud$ systems are just between the $c\bar{c}uud$ and $b\bar{b}uud$ systems, in this work we take the value of $a(\mu)$ to be $a(\mu=1 \, \rm GeV)=-3.2$. More discussions on this parameter can be found at the end of this section.

In the present study we have also investigated the cut-off regularization. Some of the results are given in \ref{Sec:discussion}, where we discuss the difference between the results obtained using the dimensional regularization and those obtained using the cut-off regularization. Especially, for the $PB_{1/2}$ sector of the $b\bar{c}uud$ system with $I={1/2}$, we choose the subtraction constant for the dimensional regularization to be $a(\mu=1 \gev)=-3.2$, and we choose the three-momentum cutoff for the cut-off regularization to be $q_{\max}=450 \mev$. Using these two parameters, the loop functions $G_{\Sigma_b \bar{D}}^{\mathcal{MB}}(s)$ and $G^{\prime \mathcal{MB}}_{\Sigma_b \bar{D}}(s)$ have the same value at the $\Sigma_b \bar{D}$ threshold. Besides, for the $PB_{1/2}$ sector of the $c \bar b uud$ system with $I={1/2}$, the subtraction constant $a(\mu=1 \gev)=-3.2$ corresponds to the cutoff $q_{\max}=685 \mev$, where the loop functions $G_{\Sigma_c B}^{\mathcal{MB}}(s)$ and $G_{\Sigma_c B}^{\prime \mathcal{MB}}(s)$ have the same value at the $\Sigma_c B$ threshold.

Eq.~(\ref{eq:G}) is valid on the first Riemann sheet, {\it i.e.}, the physical sheet. Sometimes we search for poles in the second Riemann sheet for Re$\sqrt{s} > M_l+m_l$, and in this case we need to use
\begin{equation}
G_{ll}^{\mathcal{M}\mathcal{B},II}(s) = G_{ll}^{\mathcal{M}\mathcal{B}}(s) + i {2M_l k \over 4 \pi \sqrt s} \, ,
\end{equation}
for Im$(k)>0$.

Within the coupled-channel framework, a pole can couple to several different channels. We introduce the parameter $g_i$ to describe the coupling strength and define it in the vicinity of the pole as
\begin{equation}
T_{ij}(s) = \frac{g_ig_j}{\sqrt{s}-\sqrt{s_p}} \, ,
\label{eq:gi1}
\end{equation}
where $\sqrt{s_p}$ is the position of pole on the $\sqrt{s}$ complex plane, and $g_i$ is the coupling between the pole and the channel $i$. We can also write them in the residue form as
\begin{equation}
g_i^2 = \lim_{\sqrt{s} \to \sqrt{s_p}} (\sqrt{s}-\sqrt{s_p})~T_{ii}(s) \, .
\label{eq:gi2}
\end{equation}
Once the couplings are calculated, we can evaluate the probabilities of different channels for a given state as~\cite{Montana:2017kjw}:
\begin{equation}
  P_i = -g_i^2 \, \frac{\partial G_i}{\partial \sqrt{s}} \bigg |_{\sqrt{s}=\sqrt{s_p}}.
\end{equation}

\subsection{Poles in the $c\bar{b}uud$ system}
\label{subsec:resultsuud}
\begin{table*}[htb]
  \centering
  \renewcommand{\arraystretch}{1.4}
  \caption{The poles extracted from the $c\bar{b}uud$ and $b\bar{c}uud$ systems with $I=1/2$, obtained using $a(\mu=1 \, \rm GeV)=-3.2$. Pole positions, binding energies ($E_B$), widths, and threshold masses of various coupled channels are in units of MeV. The couplings $g_i$ and $P_i$ have no dimension.}
  \setlength{\tabcolsep}{1mm}{
  \begin{tabular}{c|c|c|c|c|c|c|c|c|c}
  \hline\hline
  \multicolumn{2}{c|}{Content:$c\bar{b}uud$} & $(I)J^P$ & Pole &~~~$E_B$~~~& \,Width\, & \,Channel \, & Threshold &~~~$|g_i|$~~~ & $P_i$
  \\ \hline\hline
  \multirow{3}{*}{$PB_{1/2}$} & \multirow{3}{*}{$|\Sigma_c B\rangle$} & \multirow{3}{*}{$(\frac{1}{2})\frac{1}{2}^-$} & \multirow{3}{*}{7701.0 + i2.2} & \multirow{3}{*}{32} & \multirow{3}{*}{4.4} & $N B_c$  & 7213 & 0.48 & $-$ 0.00 $-$ i0.00
  \\ \cline{7-10}
                                                                                                                                               &&&&   &  & $\Lambda_c B$        & 7566 & 0.04 & $-$ 0.00 + i0.00
  \\ \cline{7-10}
                                                                                                                                               &&&&   &  & $\Sigma_c B$        & 7733 & 2.93 & 1.00 + i0.00
  \\ \hline
  \multirow{3}{*}{$VB_{1/2}$} & \multirow{3}{*}{$|\Sigma_c B^*\rangle$} & \multirow{3}{*}{$(\frac{1}{2})\frac{1}{2}^-, \frac{3}{2}^-$} & \multirow{3}{*}{7749.5 + i2.0}  & \multirow{3}{*}{28} & \multirow{3}{*}{4.0} & $N B_c^*$     & 7270   & 0.46 & $-$ 0.00 $-$ i0.00
  \\ \cline{7-10}
                                                                                                                                               &&&&   &  & $\Lambda_c B^*$        & 7611 & 0.04 & $-$ 0.00 + i0.00	
  \\ \cline{7-10}
                                                                                                                                               &&&&   &  & $\Sigma_c B^*$        & 7778 & 2.85 & 1.00 + i0.00
  \\\hline
  $PB_{3/2}$ & $|\Sigma_c^* B\rangle$ & $(\frac{1}{2})\frac{3}{2}^-$ & 7766.9 + i0.0 & 31 & 0 & $\Sigma_c^* B$    & 7798   & 2.86 & 1.00 + i0.00
  \\\hline
  $VB_{3/2}$ & $|\Sigma_c^* B^*\rangle$ & $(\frac{1}{2})\frac{1}{2}^-, \frac{3}{2}^-, \frac{5}{2}^-$ & 7815.3 + i0.0 & 28 & 0 & $\Sigma_c^* B^*$    & 7843    &2.78 & 1.00 + i0.00
  \\ \hline\hline
  \multicolumn{2}{c|}{Content:$b\bar{c}uud$} & $(I)J^P$ &Pole  &~$E_B$~& \,Width\, & \,Channel \,  & Threshold & ~$|g_i|$~ & $P_i$
  \\ \hline\hline
  \multirow{3}{*}{$PB_{1/2}$} & \multirow{3}{*}{$|\Sigma_b \bar{D}\rangle$} & \multirow{3}{*}{$(\frac{1}{2})\frac{1}{2}^-$} & \multirow{3}{*}{7674.2 + i0.2}  & \multirow{3}{*}{6} & \multirow{3}{*}{0.4} & $N \bar{B}_c$    & 7213    & 0.13 & $-$ 0.00 $-$ i0.00
  \\ \cline{7-10}
                                                                                                                                               &&&&   &  & $\Lambda_b \bar{D}$        & 7487 & 0 & 0.00 + i0.00 	
  \\ \cline{7-10}
                                                                                                                                               &&&&   &  & $\Sigma_b \bar{D}$        & 7680 & 1.23 & 0.94 $-$ i0.01
  \\ \hline
  \multirow{3}{*}{$VB_{1/2}$} & \multirow{3}{*}{$|\Sigma_b \bar{D}^*\rangle$} & \multirow{3}{*}{$(\frac{1}{2})\frac{1}{2}^-, \frac{3}{2}^-$} & \multirow{3}{*}{7815.5 + i0.2}  & \multirow{3}{*}{7} & \multirow{3}{*}{0.4} & $N \bar{B}_c^*$   & 7270     & 0.13 & $-$ 0.00 $-$ i0.00
  \\ \cline{7-10}
                                                                                                                                               &&&&   &  & $\Lambda_b \bar{D}^*$        & 7628 & 0 & 0.00 + i0.00 	
  \\ \cline{7-10}
                                                                                                                                               &&&&   &  & $\Sigma_b \bar{D}^*$        & 7822 & 1.24 & 0.94 $-$ i0.01
  \\\hline
  $PB_{3/2}$ & $|\Sigma_b^* \bar{D}\rangle$ & $(\frac{1}{2})\frac{3}{2}^-$ & 7696.1 + i0.0 & 6 & 0 & $\Sigma_b^* \bar{D}$     & 7702    &1.23 & 0.99 + i0.00
  \\\hline
  $VB_{3/2}$ & $|\Sigma_b^* \bar{D}^*\rangle$ & $(\frac{1}{2})\frac{1}{2}^-, \frac{3}{2}^-, \frac{5}{2}^-$ & 7837.4 + i0.0  & 6 & 0 & $\Sigma_b^* \bar{D}^*$    & 7843    &1.23 & 1.00 + i0.00
  \\ \hline\hline
  \end{tabular}}
  \label{tab:resultuud}
\end{table*}
In this subsection we perform numerical analyses to investigate the $c\bar{b}uud$ system. We use $a(\mu)=-3.2$ and find altogether seven bound states with the binding energies about $28\sim32 \mev$:
\begin{itemize}

\item[$\bullet$] We find one bound state of $(I)J^P = ({1\over2}){1\over2}^-$ below the $\Sigma_c B$ threshold. It locates at around $(7701.0 + i2.2) \mev$, so its binding energy and width are about $32 \mev$ and $4.4 \mev$, respectively. This state is inside the $PB_{1/2}$ channel consisting of the $N B_c$, $\Lambda_c B$, and $\Sigma_c B$ coupled channels. It strongly couples to the $\Sigma_c B$ channel, indicating it to be the $\Sigma_c B$ molecular state of $(I)J^P = ({1\over2}){1\over2}^-$, {\it i.e.}, $| \Sigma_c B; (I)J^P = ({1\over2}){1\over2}^- \rangle$.

\item[$\bullet$] We find two bound states of $(I)J^P = ({1\over2}){1\over2}^-$ and $({1\over2}){3\over2}^-$ below the $\Sigma_c B^*$ threshold. These two states locate at around $(7749.5 + i2.0) \mev$, so their binding energies and widths are both about $28\,\mev$ and $4\,\mev$, respectively. These two states are inside the $VB_{1/2}$ channel consisting of the $N B_c^*$, $\Lambda_c B^*$, and $\Sigma_c B^*$ coupled channels. They both strongly couple to the $\Sigma_c B^*$ channel, indicating them to be the $\Sigma_c B^*$ molecular states of $(I)J^P = ({1\over2}){1\over2}^-$ and $({1\over2}){3\over2}^-$, {\it i.e.}, $| \Sigma_c B^*; (I)J^P = ({1\over2}){1\over2}^- \rangle$ and $| \Sigma_c B^*; (I)J^P = ({1\over2}){3\over2}^- \rangle$.

\item[$\bullet$] We find one bound state of $(I)J^P = ({1\over2}){3\over2}^-$ below the $\Sigma_c^* B$ threshold. It locates at around $7766.9 \mev$, so its binding energy is about $31 \mev$ with a zero width. This state is inside the $PB_{3/2}$ channel consisting of the single $\Sigma_c^* B$ channel. It only couples to the $\Sigma_c^* B$ channel, indicating it to be the $\Sigma_c^* B$ molecular state of $(I)J^P = ({1\over2}){3\over2}^-$, {\it i.e.}, $| \Sigma_c^* B; (I)J^P = ({1\over2}){3\over2}^- \rangle$.

\item[$\bullet$] We find three bound states of $(I)J^P = ({1\over2}){1\over2}^-$, $({1\over2}){3\over2}^-$, and $({1\over2}){5\over2}^-$ below the $\Sigma_c^* B^*$ threshold. These three states locate at around $7815.3 \mev$, so their binding energies are all about $28 \mev$ with a zero width. These three states are inside the $VB_{3/2}$ channel consisting of the single $\Sigma_c^* B^*$ channel. They all only couple to the $\Sigma_c^* B^*$ channel, indicating them to be the $\Sigma_c^* B^*$ molecular states of $(I)J^P = ({1\over2}){1\over2}^-$, $({1\over2}){3\over2}^-$, and $({1\over2}){5\over2}^-$, {\it i.e.}, $| \Sigma_c^* B^*; (I)J^P = ({1\over2}){1\over2}^- \rangle$, $| \Sigma_c^* B^*; (I)J^P = ({1\over2}){3\over2}^- \rangle$, and $| \Sigma_c^* B^*; (I)J^P = ({1\over2}){5\over2}^- \rangle$.
\end{itemize}

\begin{figure}[htb!]
  \centering
  \includegraphics[width=0.5\textwidth]{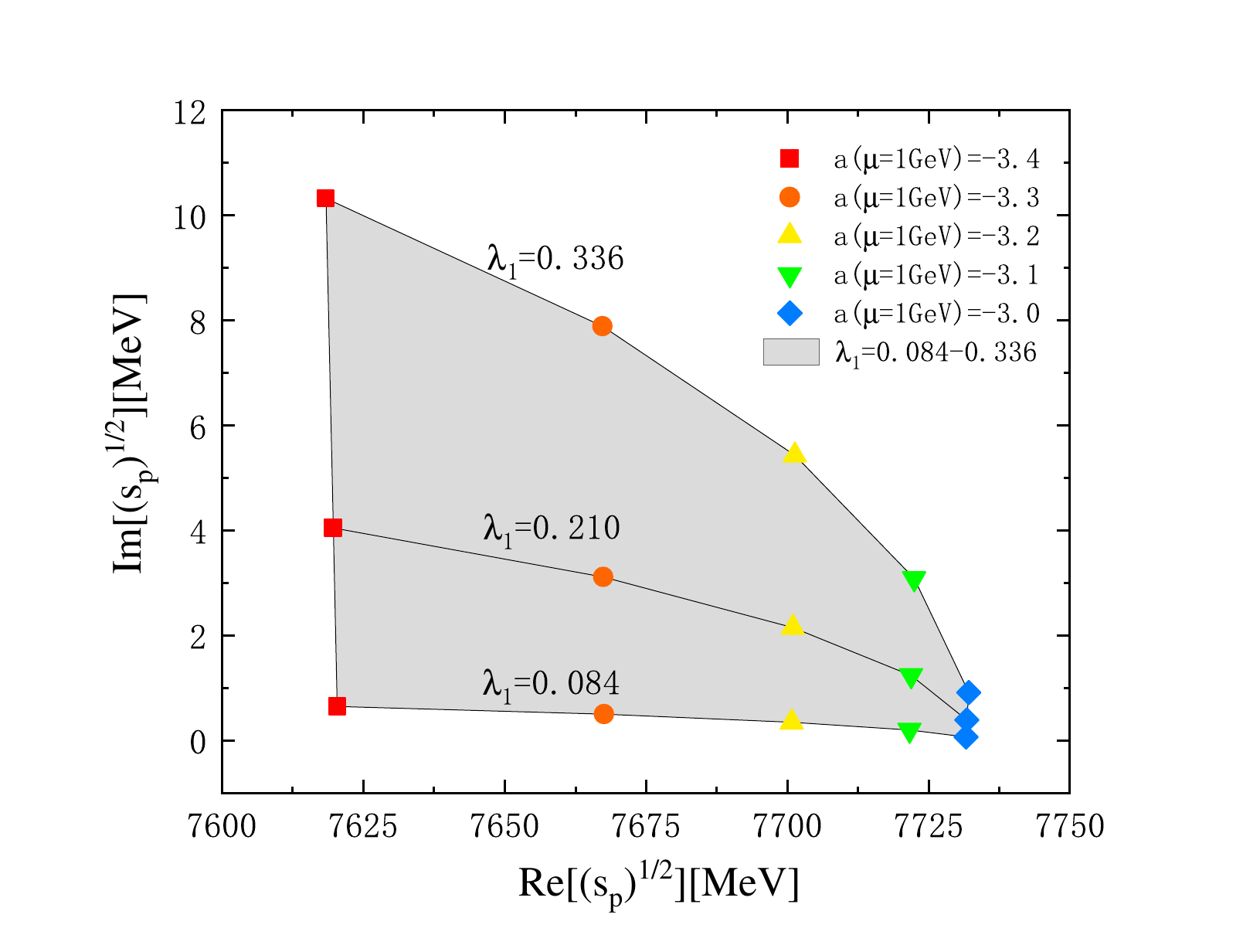}
  \caption{The pole position of the hadronic molecular state $| \Sigma_c B; (I)J^P = ({1\over2}){1\over2}^- \rangle$ with respect to the symmetry violation parameter $\lambda_1 =0.084$, $0.21$, and $0.336$ as well as the subtraction constant $a(\mu)=-3.4$, $-3.3$, $-3.2$, $-3.1$, and $-3.0$.}
  \label{fig:breaking}
\end{figure}

We summarize the former results in Table~\ref{tab:resultuud}. As given in Eq.~(\ref{eq:Vij2}), the transition amplitude $V_{ij}^{V\mathcal{B}}$ contains the polarization vectors $\vec{\epsilon}_i \cdot \vec{\epsilon}_j$, which makes the states extracted from the $V B_{1/2}$ and $V B_{3/2}$ interactions degenerate, {\it e.g.}, the $\Sigma_c B^*$ molecular states of $(I)J^P=({1\over2}){1\over2}^-$ and $({1\over2}){3\over2}^-$ are degenerate, and the $\Sigma_c^* B^*$ molecular states of $(I)J^P=({1\over2}){1\over2}^-$, $({1\over2}){3\over2}^-$, and $({1\over2}){5\over2}^-$ are also degenerate. One may further investigate their mass splittings by coupling these channels to the $P B_{1/2}$ and $P B_{3/2}$ channels via the pion exchanges, but we shall not investigate this in the present study.

It is interesting to study the dependence of our results on the symmetry violation parameters $\lambda_1$ and $\lambda_2$. We use the hadronic molecular state $| \Sigma_c B; (I)J^P = ({1\over2}){1\over2}^- \rangle$ as an example and show its pole position in Fig.~\ref{fig:breaking} with respect to $\lambda_1$ (it does not depend on $\lambda_2$). Besides, we also slightly fine-tune the subtraction constant $a(\mu)$ from $-3.4$ to $-3.0$. We notice that the mass is more sensitive to the subtraction constant $a(\mu)$, while the width is more sensitive to the symmetry violation parameter $\lambda_1$. We also notice that the width increases as the pole moves towards lower energies with decreasing the values of $a(\mu)$, in spite of the fact that the decaying phase space decreases. This is because that the decrease of $a(\mu)$ increases the coupling of this state to the $NB_c$ and $\Lambda_c B$ channels, and so increases the width.

\subsection{Poles in the $b\bar{c}uud$ system}

In this subsection we perform numerical analyses to investigate the $b\bar{c}uud$ system. We use $a(\mu)=-3.2$ and find altogether seven bound states with the binding energies about $6\sim7 \mev$:
\begin{itemize}

\item[$\bullet$] We find one bound state of $(I)J^P = ({1\over2}){1\over2}^-$ below the $\Sigma_b \bar{D}$ threshold. It locates at around $(7674.2 + i0.2)\,\mev$, so its binding energy and width are about $6 \mev$ and $0.4\,\mev$, respectively. This state is inside the $PB_{1/2}$ channel consisting of the $N \bar{B}_c$, $\Lambda_b \bar{D}$, and $\Sigma_b \bar{D}$ coupled channels. It strongly couples to the $\Sigma_b \bar{D}$ channel, indicating it to be the $\Sigma_b \bar{D}$ molecular state of $(I)J^P = ({1\over2}){1\over2}^-$, {\it i.e.}, $| \Sigma_b \bar{D}; (I)J^P = ({1\over2}){1\over2}^- \rangle$.

\item[$\bullet$] We find two bound states of $(I)J^P = ({1\over2}){1\over2}^-$ and $({1\over2}){3\over2}^-$ below the $\Sigma_b \bar{D}^*$ threshold. These two states locate at around $(7815.5 + i0.2)\, \mev$, so their binding energies and widths are both about $7 \mev$ and $0.4\,\mev$, respectively. These two states are inside the $VB_{1/2}$ channel consisting of the $N \bar{B}_c^*$, $\Lambda_b \bar{D}^*$, and $\Sigma_b \bar{D}^*$ coupled channels. They both strongly couple to the $\Sigma_b \bar{D}^*$ channel, indicating them to be the $\Sigma_b \bar{D}^*$ molecular states of $(I)J^P = ({1\over2}){1\over2}^-$ and $({1\over2}){3\over2}^-$, {\it i.e.}, $| \Sigma_b \bar{D}^*; (I)J^P = ({1\over2}){1\over2}^- \rangle$ and $| \Sigma_b \bar{D}^*; (I)J^P = ({1\over2}){3\over2}^- \rangle$.

\item[$\bullet$] We find one bound state of $(I)J^P = ({1\over2}){3\over2}^-$ below the $\Sigma_b^* \bar{D}$ threshold. It locates at around $7696.1 \mev$, so its binding energy is about $6 \mev$ with a zero width. This state is inside the $PB_{3/2}$ channel consisting of the single $\Sigma_b^* \bar{D}$ channel. It only couples to the $\Sigma_b^* \bar{D}$ channel, indicating it to be the $\Sigma_b^* \bar{D}$ molecular state of $(I)J^P = ({1\over2}){3\over2}^-$, {\it i.e.}, $| \Sigma_b^* \bar{D}; (I)J^P = ({1\over2}){3\over2}^- \rangle$.

\item[$\bullet$] We find three bound states of $(I)J^P = ({1\over2}){1\over2}^-$, $({1\over2}){3\over2}^-$, and $({1\over2}){5\over2}^-$ below $\Sigma_b^* \bar{D}^*$ threshold. These three states locate at around $7837.4\, \mev$, so their binding energies are all about $6 \mev$ with a zero width. These three states are inside the $VB_{3/2}$ channel consisting of the single $\Sigma_b^* \bar{D}^*$ channel. They all only couple to the $\Sigma_b^* \bar{D}^*$ channel, indicating them to be the $\Sigma_b^* \bar{D}^*$ molecular states of $(I)J^P = ({1\over2}){1\over2}^-$, $({1\over2}){3\over2}^-$, and $({1\over2}){5\over2}^-$, {\it i.e.}, $| \Sigma_b^* \bar{D}^*; (I)J^P = ({1\over2}){1\over2}^- \rangle$, $| \Sigma_b^* \bar{D}^*; (I)J^P = ({1\over2}){3\over2}^- \rangle$, and $| \Sigma_b^* \bar{D}^*; (I)J^P = ({1\over2}){5\over2}^- \rangle$.
\end{itemize}

\begin{figure}[htb]
  \centering
  \includegraphics[width=0.45\textwidth]{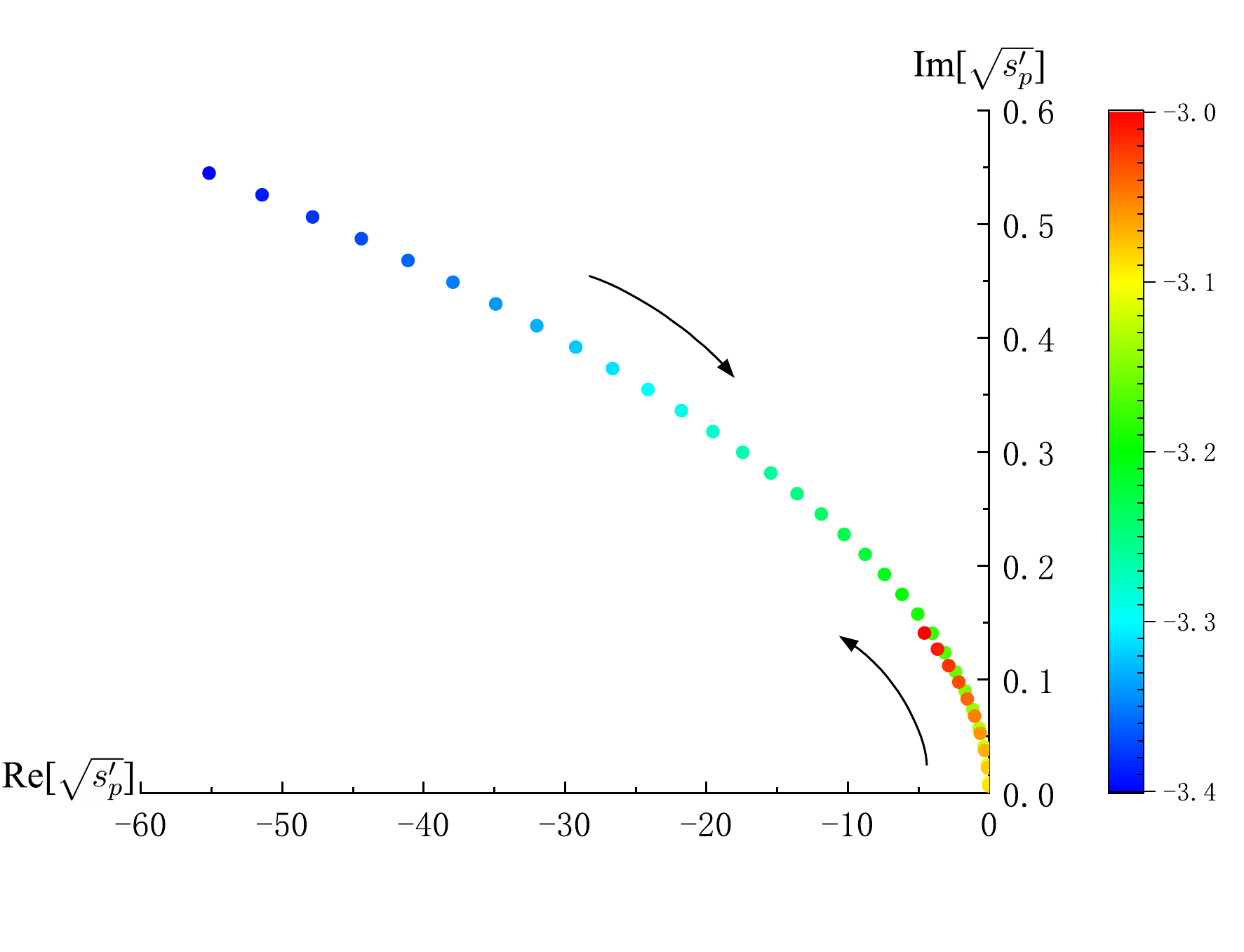}
  \caption{The pole position $\sqrt{s_p^\prime} = \sqrt{s_p}- (M_{\Sigma_b} + M_{\bar{D}})$ of the hadronic molecular state $| \Sigma_b \bar{D}; (I)J^P = ({1\over2}){1\over2}^- \rangle$ with respect to the subtraction constant $a(\mu)$ ranging from $-3.4$ to $-3.0$.}
  \label{fig:polepostion1}
\end{figure}

\begin{figure*}[htb]
  \centering
  \subfigure[$a(\mu)=-3.4$]{\includegraphics[width=0.19\hsize, height=0.19\hsize]{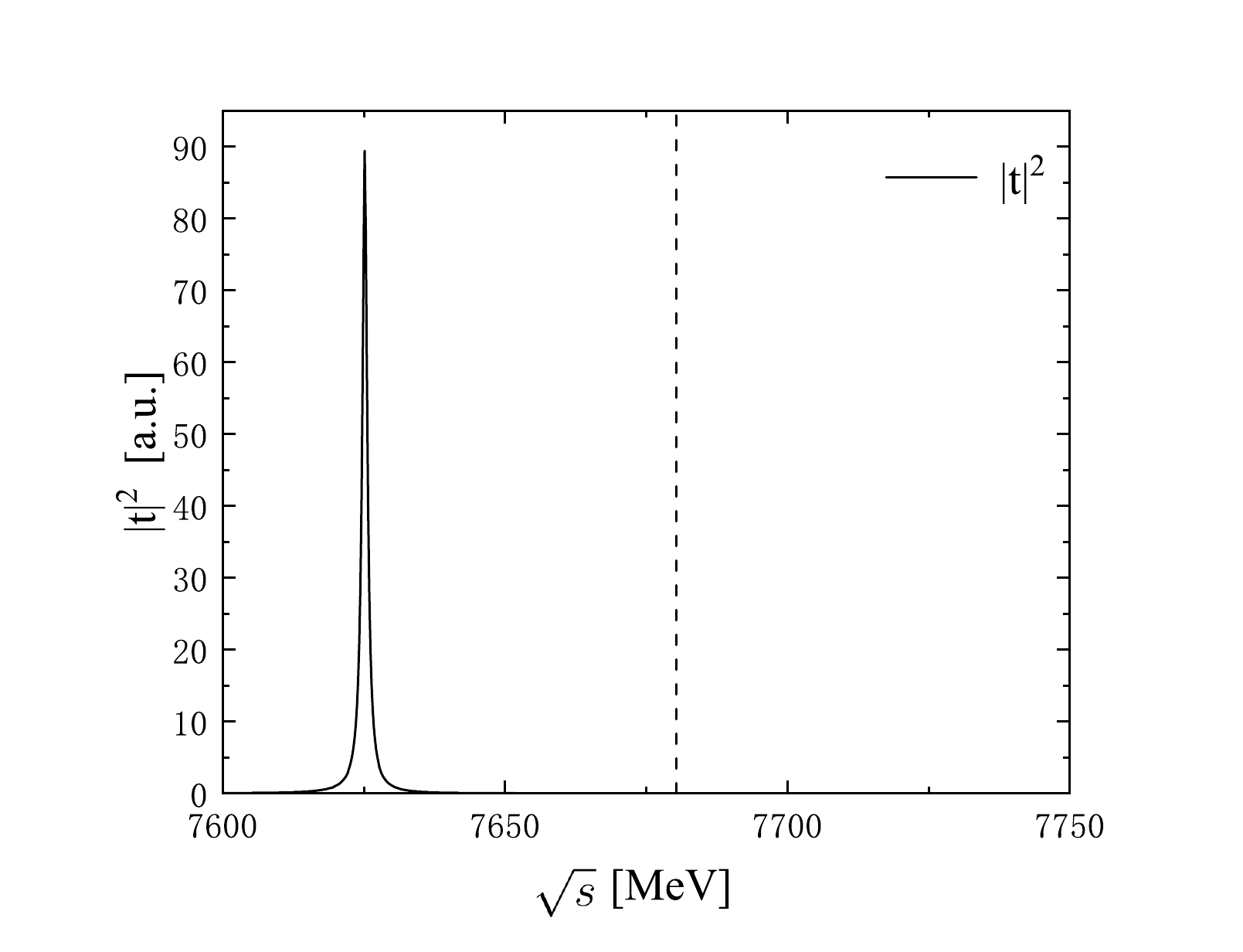}\label{fig:amp1}}
  \subfigure[$a(\mu)=-3.3$]{\includegraphics[width=0.19\hsize, height=0.19\hsize]{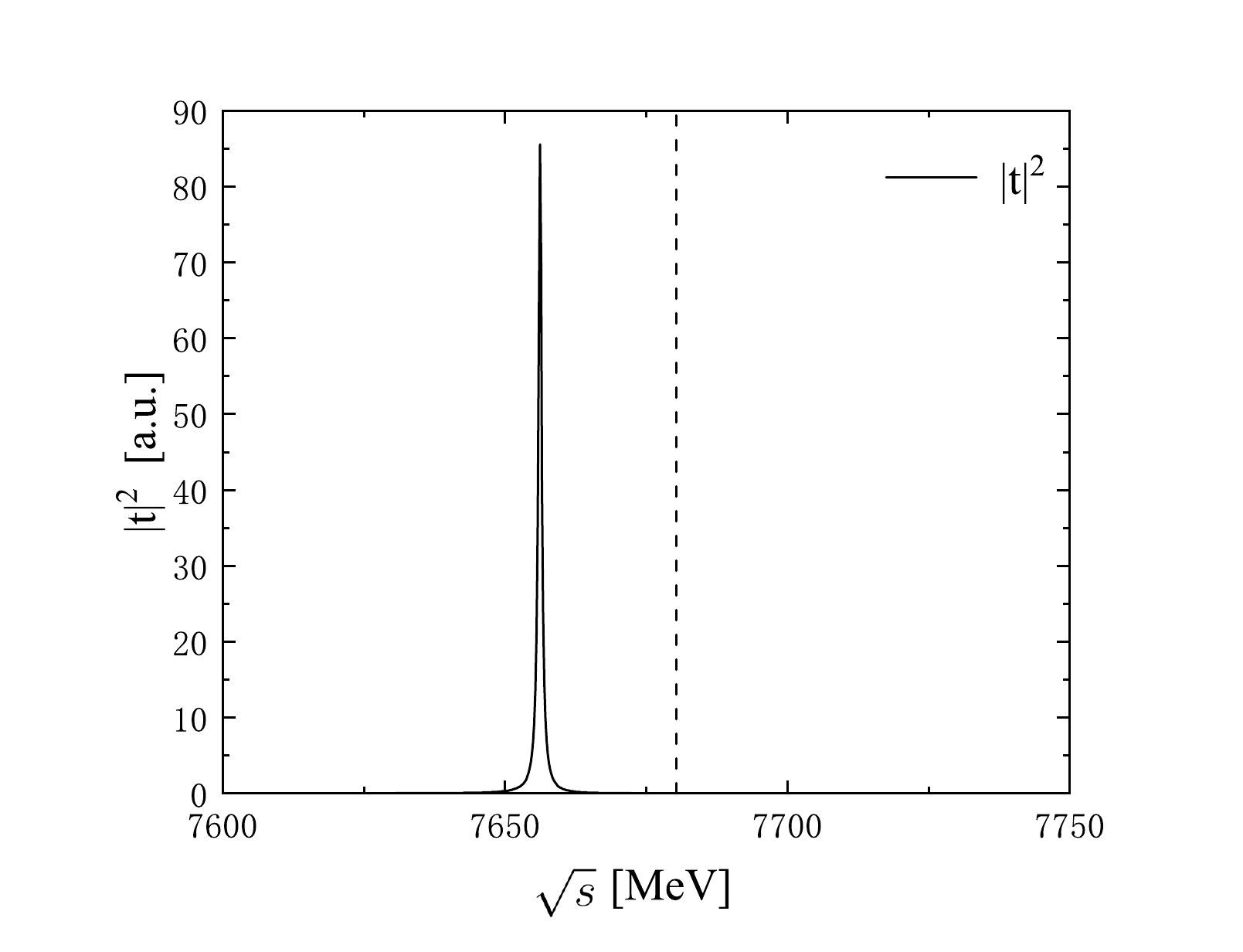}\label{fig:amp2}}
  \subfigure[$a(\mu)=-3.2$]{\includegraphics[width=0.19\hsize, height=0.19\hsize]{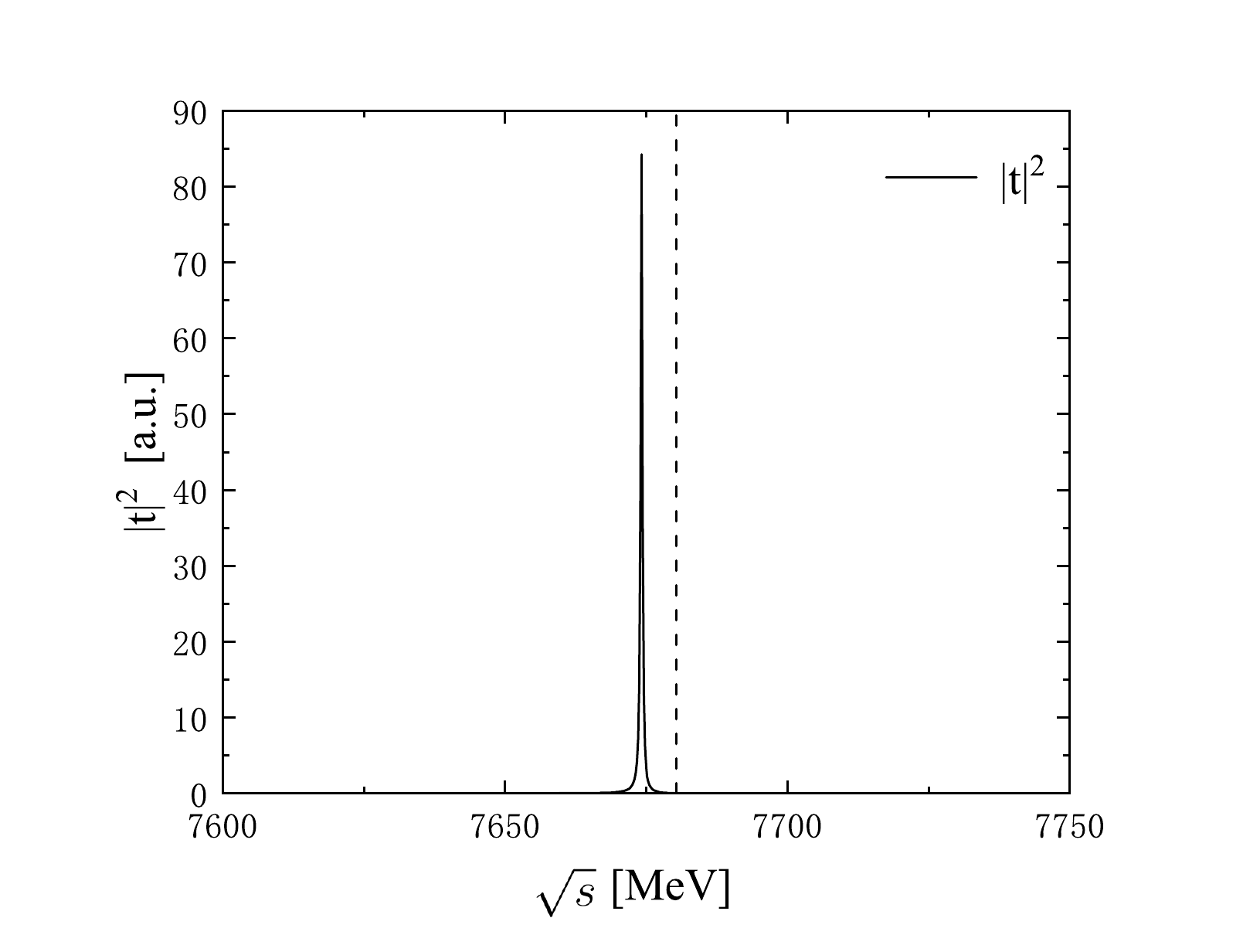}\label{fig:amp3}}
  \subfigure[$a(\mu)=-3.1$]{\includegraphics[width=0.19\hsize, height=0.19\hsize]{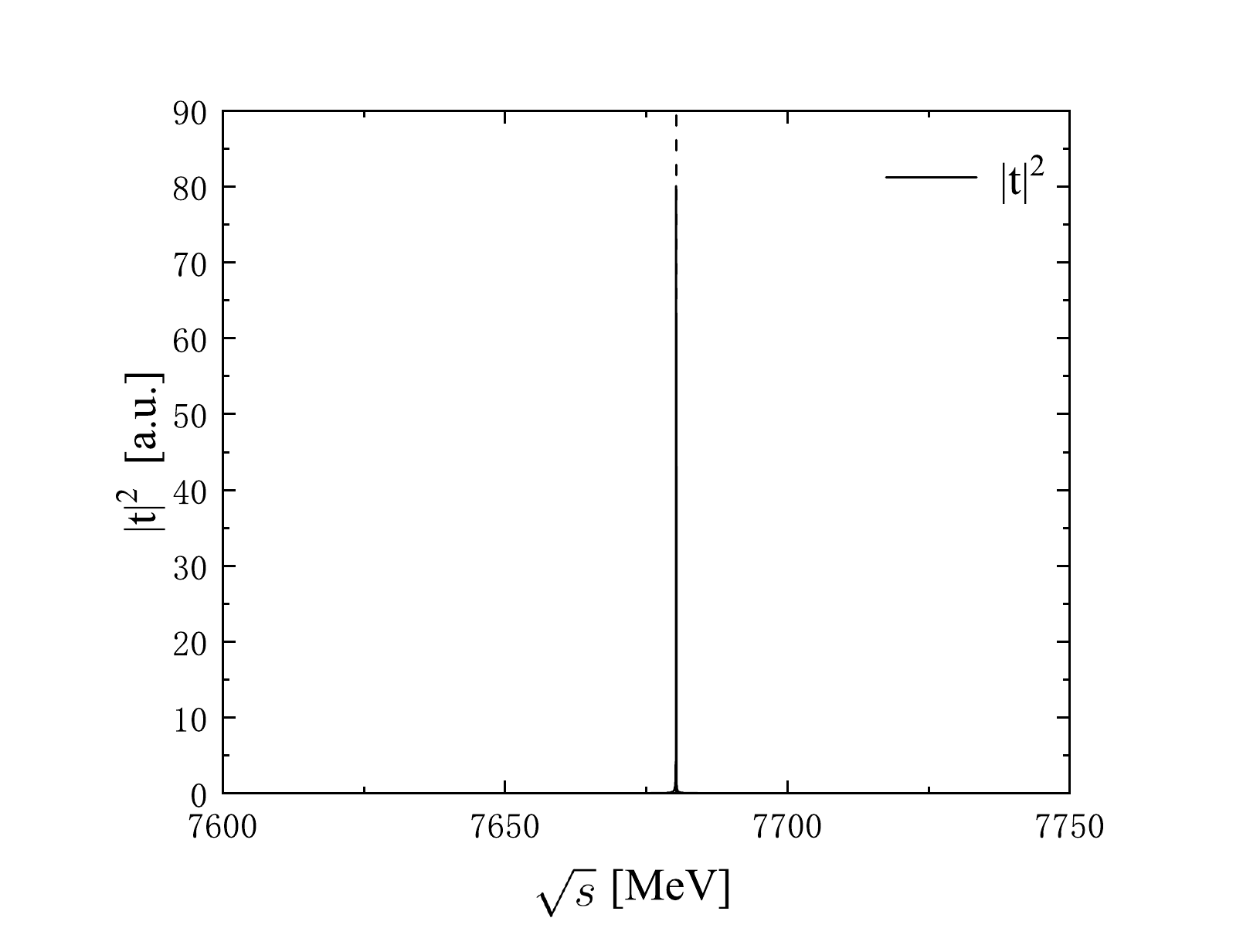}\label{fig:amp4}}
  \subfigure[$a(\mu)=-3.0$]{\includegraphics[width=0.19\hsize, height=0.19\hsize]{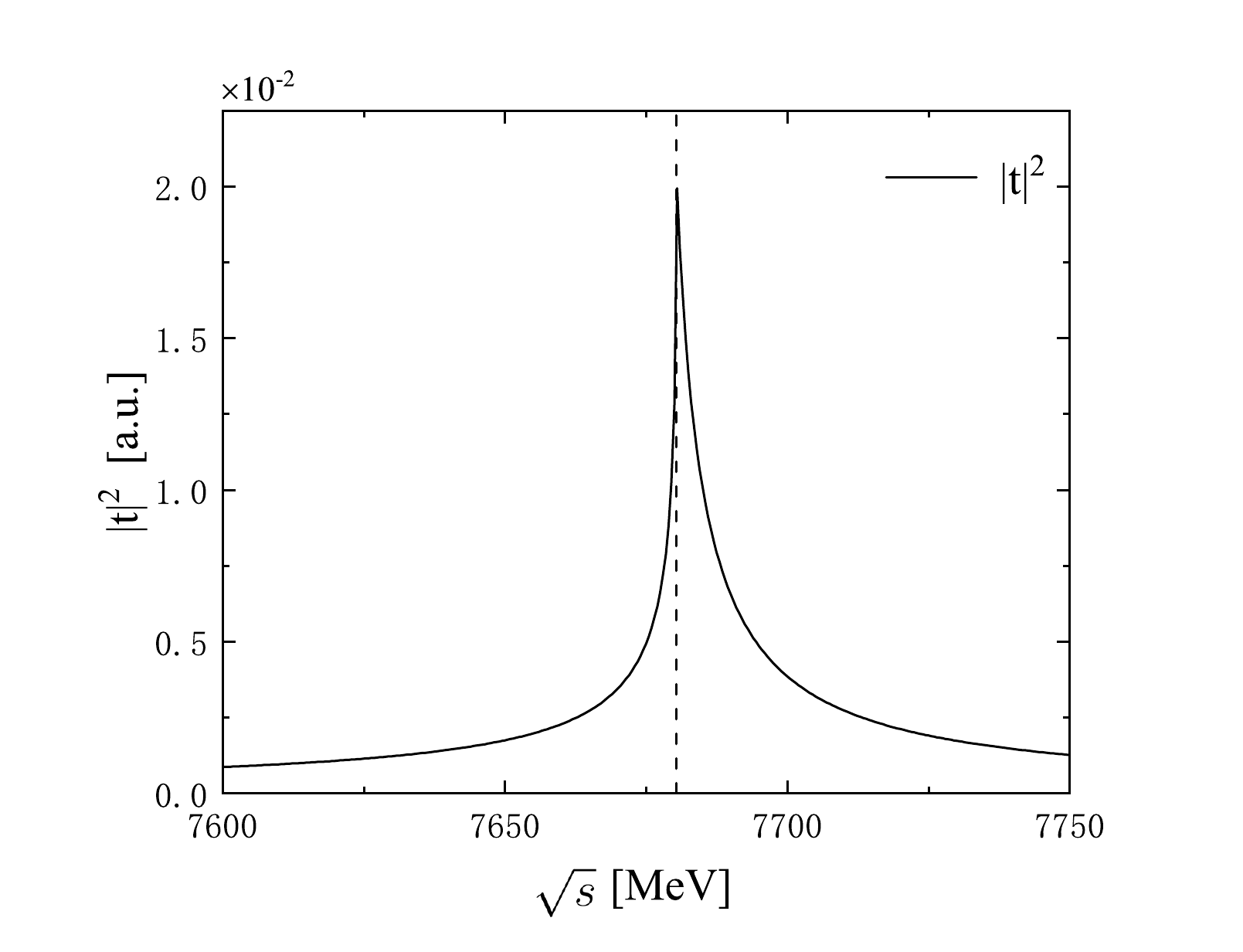}\label{fig:amp5}}
  \caption{The transition amplitude $|t(s)|^2 \equiv |T^{\mathcal{M}\mathcal{B}}_{| \Sigma_b \bar{D}; (I)J^P = ({1\over2}){1\over2}^- \rangle \to | \Sigma_b \bar{D}; (I)J^P = ({1\over2}){1\over2}^- \rangle}(s)|^2$ with the subtraction constant $a(\mu)= -3.4$~(a), $-3.3$~(b), $-3.2$~(c), $-3.1$~(d), and $-3.0$~(e). The $\Sigma_b \bar{D}$ threshold is indicated by a dash line.}
  \label{fig:amp}
\end{figure*}

It is interesting to study the dependence of our results on the subtraction constant $a(\mu)$. We use the hadronic molecular state $| \Sigma_b \bar{D}; (I)J^P = ({1\over2}){1\over2}^- \rangle$ as an example and show its pole position in Fig.~\ref{fig:polepostion1} with respect to this parameter. We find that this pole becomes a bound state when taking $a(\mu) <-3.1$, while it becomes a virtual state when taking $a(\mu) >-3.1$. Actually, when taking $a(\mu) >-3.1$, we do not find any bound state, and instead we find a pole on the sixth Riemann sheet that is also below the $\Sigma_b \bar{D}$ threshold. We also show the transition amplitude
\begin{equation}
t(s) \equiv T^{\mathcal{M}\mathcal{B}}_{| \Sigma_b \bar{D}; (I)J^P = ({1\over2}){1\over2}^- \rangle \to | \Sigma_b \bar{D}; (I)J^P = ({1\over2}){1\over2}^- \rangle}(s) \, ,
\end{equation}
in Fig.~\ref{fig:amp} for the subtraction constant $a(\mu) = -3.4$, $-3.3$, $-3.2$, $-3.1$, and $-3.0$. This pole is identified as a bound state in Fig.~\ref{fig:amp}(a,b,c,d), which appears as the singularity under the threshold. Differently, this pole is identified as a virtual state in Fig.~\ref{fig:amp}(e), which can significantly enhance the near-threshold cusp effect to produce a sharp peak at the threshold. We refer to Refs.~\cite{Dong:2021bvy,Dong:2020hxe,Liu:2023hrz} for more discussions on the near-threshold virtual states.

%
\section{Other systems}
\label{sec:others}

In the previous sections we have applied the coupled-channel unitary approach within the local hidden-gauge formalism to search for possibly-existing molecular pentaquark states in the $c\bar{b}uud$ and $b\bar{c}uud$ systems. In this section we follow the same approach to study the other $c\bar{b}qqq$ and $b\bar{c}qqq$ systems ($q=u,d,s$), separately in the following subsections.

\subsection{Poles in the $c\bar{b}sud$ and $b\bar{c}sud$ system}
\label{sec:sud}

In this subsection we investigate the $c\bar{b}sud$ and $b\bar{c}sud$ systems. There are two isospin possibilities: $I=0$ and $I=1$. In this subsection we investigate the $I=0$ sector, and the results for the $I=1$ sector are the same as those given in Sec.~\ref{sec:suu} for the $c\bar{b}suu$ and $b\bar{c}suu$ systems.

For the former $c\bar{b}sud$ system we consider
\begin{enumerate}[a)]
  \item $PB_{1/2}$ channels: $\Lambda B_c$, $\Lambda_c B_s$, $\Xi_c B$, $\Xi_c^\prime B$;
  \item $VB_{1/2}$ channels: $\Lambda B_c^*$, $\Lambda_c B_s^*$, $\Xi_c B^*$, $\Xi_c^\prime B^*$;
  \item $PB_{3/2}$ channels: $\Xi_c^* B$;
  \item $VB_{3/2}$ channels: $\Xi_c^* B^*$.
\end{enumerate}
Their corresponding $C^{\mathcal{M}\mathcal{B}}_{1}$ matrices are
\begin{eqnarray}
  \nonumber
  \label{eq:sudPB}
  C^{PB_{1/2}}_{1}&=&\left(
  \renewcommand{\arraystretch}{2}
  \renewcommand{\arraycolsep}{1.3pt}
  \begin{array}{c|cccc}
  J={1\over2}      & \Lambda B_c      & \Lambda_c B_s       & \Xi_c B       & \Xi_c^\prime B
  \\ \hline
  \Lambda B_c      & 0             & -{1\over\sqrt{3}}\lambda_3           & -{1\over\sqrt{6}}\lambda_1       & {1\over\sqrt{2}}\lambda_1 \\
  \Lambda_c B_s   & -{1\over\sqrt{3}}\lambda_3      & 0          & \sqrt{2}         & 0 \\
  \Xi_c B          & -{1\over\sqrt{6}}\lambda_1     & \sqrt{2}   & -1     & 0 \\
  \Xi_c^\prime B   & {1\over\sqrt{2}}\lambda_1      & 0          & 0      & -1
  \end{array}
  \right),
\\
\\
  \nonumber
  \label{eq:sudVB}
  C^{VB_{1/2}}_{1}&=&\left(
  \renewcommand{\arraystretch}{2}
  \renewcommand{\arraycolsep}{1.3pt}
  \begin{array}{c|cccc}
  J={1\over2}/{3\over2}      & \Lambda B_c^*      & \Lambda_c B_s^*       & \Xi_c B^*       & \Xi_c^\prime B^*
  \\ \hline
  \Lambda B_c^*      & 0             & -{1\over\sqrt{3}}\lambda_3           & -{1\over\sqrt{6}}\lambda_1       & {1\over\sqrt{2}}\lambda_1 \\
  \Lambda_c B_s^*   & -{1\over\sqrt{3}}\lambda_3      & 0          & \sqrt{2}         & 0 \\
  \Xi_c B^*          & -{1\over\sqrt{6}}\lambda_1     & \sqrt{2}   & -1     & 0 \\
  \Xi_c^\prime B^*   & {1\over\sqrt{2}}\lambda_1      & 0          & 0      & -1
  \end{array}
  \right),
\\
\\
  C^{PB_{3/2}}_{1}&=&\left(
  \renewcommand{\arraystretch}{2}
  \renewcommand{\arraycolsep}{1.3pt}
  \begin{array}{c|c}
  J={3\over2}            & \Xi_c^* B
  \\ \hline
  \Xi_c^* B             & -1
  \end{array}
  \right),
\\
  C^{VB_{3/2}}_{1}&=&\left(
  \renewcommand{\arraystretch}{2}
  \begin{array}{c|c}
  J={1\over2}/{3\over2}/{5\over2}            & \Xi_c^* B^*
  \\ \hline
  \Xi_c^* B^*             & -1
  \end{array}
  \right).
\end{eqnarray}
For the latter $b\bar{c}sud$ system we consider
\begin{enumerate}[a)]
  \item $PB_{1/2}$ channels: $\Lambda \bar{B}_c$, $\Lambda_b \bar{D}_s$, $\Xi_b \bar{D}$, $\Xi_b^\prime \bar{D}$;
  \item $VB_{1/2}$ channels: $\Lambda \bar{B}_c^*$, $\Lambda_b \bar{D}_s^*$, $\Xi_b \bar{D}^*$, $\Xi_b^\prime \bar{D}^*$;
  \item $PB_{3/2}$ channels: $\Xi_b^* \bar{D}$;
  \item $VB_{3/2}$ channels: $\Xi_b^* \bar{D}^*$.
\end{enumerate}
Their corresponding $C^{\prime \mathcal{M}\mathcal{B}}_{1}$ matrices are
\begin{eqnarray}
  \nonumber
  C^{\prime PB_{1/2}}_{1}&=&\left(
  \renewcommand{\arraystretch}{2}
  \renewcommand{\arraycolsep}{1.3pt}
  \begin{array}{c|cccc}
  J={1\over2}      & \Lambda \bar{B}_c      & \Lambda_b \bar{D}_s       & \Xi_b \bar{D}       & \Xi_b^\prime \bar{D}
  \\ \hline
  \Lambda \bar{B}_c      & 0             & -{1\over\sqrt{3}}\lambda_4           & -{1\over\sqrt{6}}\lambda_2       & {1\over\sqrt{2}}\lambda_2 \\
  \Lambda_b \bar{D}_s   & -{1\over\sqrt{3}}\lambda_4      & 0          & \sqrt{2}         & 0 \\
  \Xi_b \bar{D}          & -{1\over\sqrt{6}}\lambda_2     & \sqrt{2}   & -1     & 0 \\
  \Xi_b^\prime \bar{D}   & {1\over\sqrt{2}}\lambda_2      & 0          & 0      & -1
  \end{array}
  \right),
\\
\\
  \nonumber
  C^{\prime VB_{1/2}}_{1}&=&\left(
  \renewcommand{\arraystretch}{2}
  \renewcommand{\arraycolsep}{1.3pt}
  \begin{array}{c|cccc}
  J={1\over2}/{3\over2}      & \Lambda \bar{B}_c^*      & \Lambda_b \bar{D}_s^*       & \Xi_b \bar{D}^*       & \Xi_b^\prime \bar{D}^*
  \\ \hline
  \Lambda \bar{B}_c^*      & 0             & -{1\over\sqrt{3}}\lambda_4           & -{1\over\sqrt{6}}\lambda_2       & {1\over\sqrt{2}}\lambda_2 \\
  \Lambda_b \bar{D}_s^*   & -{1\over\sqrt{3}}\lambda_4      & 0          & \sqrt{2}         & 0 \\
  \Xi_b \bar{D}^*          & -{1\over\sqrt{6}}\lambda_2     & \sqrt{2}   & -1     & 0 \\
  \Xi_b^\prime \bar{D}^*   & {1\over\sqrt{2}}\lambda_2      & 0          & 0      & -1
  \end{array}
  \right),
\\
\\
  C^{\prime PB_{3/2}}_{1}&=&\left(
  \renewcommand{\arraystretch}{2}
  \renewcommand{\arraycolsep}{1.3pt}
  \begin{array}{c|c}
  J={3\over2}            & \Xi_b^* \bar{D}
  \\ \hline
  \Xi_b^* \bar{D}             & -1
  \end{array}
  \right),
\\
  C^{\prime VB_{3/2}}_{1}&=&\left(
  \renewcommand{\arraystretch}{2}
  \renewcommand{\arraycolsep}{1.3pt}
  \begin{array}{c|c}
  J={1\over2}/{3\over2}/{5\over2}            & \Xi_b^* \bar{D}^*
  \\ \hline
  \Xi_b^* \bar{D}^*             & -1
  \end{array}
  \right).
\end{eqnarray}
The two reduction factors contained in the above expressions are
\begin{equation}
  \lambda_3 = \frac{-M_V^2}{(m_{B_s}-m_{B_c})^2 - m_{D_s^*}^2} \simeq 0.18,
\end{equation}
for the exchange of the $D_s^*$ meson, and
\begin{equation}
  \lambda_4 = \frac{-M_V^2}{(m_{\bar{D}_s}-m_{\bar{B}_c})^2 - m_{\bar{B}_s^*}^2} \simeq 0.06,
\end{equation}
for the exchange of the $\bar{B}_s^*$ meson.

\begin{table*}[htbp]
  \centering
  \renewcommand{\arraystretch}{1.4}
  \caption{The poles extracted from the $c\bar{b}sud$ and $b\bar{c}sud$ systems with $I=0$, obtained using $a(\mu=1 \, \rm GeV)=-3.2$. Pole positions, binding energies ($E_B$), widths, and threshold masses of various coupled channels are in units of MeV. The couplings $g_i$ and $P_i$ have no dimension.}
  \setlength{\tabcolsep}{1mm}{
  \begin{tabular}{c|c|c|c|c|c|c|c|c|c}
  \hline\hline
  \multicolumn{2}{c|}{\,\,Content:$c\bar{b}sud$\,\,} & \,$(I)J^P$\, &Pole   &~~~$E_B$~~~& \,Width\, & \,Channel \, & Threshold &~~~$|g_i|$~~~ & $P_i$
  \\ \hline\hline
  \multirow{4}{*}{$PB_{1/2}$} & \multirow{4}{*}{$|\Xi_c B\rangle$} & \multirow{4}{*}{$(0)\frac{1}{2}^-$} & \multirow{4}{*}{7685.9 + i3.0}  & \multirow{4}{*}{63} & \multirow{4}{*}{6.0} & $\Lambda B_c$    & 7390    & 0.22 & $-$ 0.00 $-$ i0.00
  \\ \cline{7-10}
                                                                                                                                               &&&&     &  & $\Lambda_c B_s$        & 7653 & 0.57 & 0.03 $-$ i0.02 	
  \\ \cline{7-10}
                                                                                                                                               &&&&
  &  & $\Xi_c B$        & 7749 & 3.48 & 0.97 + i0.02 	
  \\ \cline{7-10}
                                                                                                                                               &&&&   &  & $\Xi_c^\prime B$        & 7858 & 0.03 & $-$ 0.00 + i0.00
  \\ \hline
  \multirow{4}{*}{$PB_{1/2}$} & \multirow{4}{*}{$|\Xi_c^\prime B\rangle$} & \multirow{4}{*}{$(0)\frac{1}{2}^-$} & \multirow{4}{*}{7828.8 + i2.3}  & \multirow{4}{*}{29} & \multirow{4}{*}{4.6 } & $\Lambda B_c$    & 7390    & 0.45 & $-$ 0.00 $-$ i0.00
  \\ \cline{7-10}
                                                                                                                                               &&&&     &  & $\Lambda_c B_s$        & 7653 & 0.03 & $-$ 0.00 $-$ i0.00 	
  \\ \cline{7-10}
                                                                                                                                               &&&&
  &  & $\Xi_c B$        & 7749 & 0.02 & $-$ 0.00 + i0.00 	
  \\ \cline{7-10}
                                                                                                                                               &&&&   &  & $\Xi_c^\prime B$        & 7858 & 2.79 & 1.00 + i0.00
  \\ \hline
  \multirow{4}{*}{$VB_{1/2}$} & \multirow{4}{*}{$|\Xi_c B^*\rangle$} & \multirow{4}{*}{$(0)\frac{1}{2}^-, \frac{3}{2}^-$} & \multirow{4}{*}{7736.1 + i3.0}  & \multirow{4}{*}{58} & \multirow{4}{*}{6.0} & $\Lambda B_c^*$   & 7447    & 0.22 & $-$ 0.00 $-$ i0.00
  \\ \cline{7-10}
                                                                                                                                               &&&&   &  & $\Lambda_c B_s^*$        & 7702 & 0.56 & 0.03 $-$ i0.01	
  \\ \cline{7-10}
                                                                                                                                               &&&&
  &  & $\Xi_c B^*$        & 7794 & 3.40 & 0.97 + i0.02 	
  \\ \cline{7-10}
                                                                                                                                               &&&&   &  & $\Xi_c^\prime B^*$        & 7903 & 0.03 & $-$ 0.00 + i0.00
                                                                                                                                                 \\ \hline                                                                     \multirow{4}{*}{$VB_{1/2}$} & \multirow{4}{*}{$|\Xi_c^\prime B^*\rangle$} & \multirow{4}{*}{$(0)\frac{1}{2}^-, \frac{3}{2}^-$} & \multirow{4}{*}{7877.1 + i2.1}  & \multirow{4}{*}{26} & \multirow{4.2}{*}{4} & $\Lambda B_c^*$    & 7447    & 0.44 & $-$ 0.00 $-$ i0.00
  \\ \cline{7-10}
                                                                                                                                               &&&&   &  & $\Lambda_c B_s^*$        & 7702 & 0.03 & $-$ 0.00 $-$ i0.00 	
  \\ \cline{7-10}
                                                                                                                                               &&&&
  &  & $\Xi_c B^*$        & 7794 & 0.02 &	$-$ 0.00 + i0.00
  \\ \cline{7-10}
                                                                                                                                               &&&&   &  & $\Xi_c^\prime B^*$        & 7903 & 2.71 & 1.00 + i0.00
  \\\hline
  $PB_{3/2}$ & $|\Xi_c^* B\rangle$ & $(0)\frac{3}{2}^-$ & 7897.4 + i0.0  & 28 & 0 & $\Xi_c^* B$    & 7925    & 2.71 & 1.00 + i0.00
  \\\hline
  $VB_{3/2}$ & $|\Xi_c^* B^*\rangle$ & $(0)\frac{1}{2}^-, \frac{3}{2}^-, \frac{5}{2}^-$ & 7945.6 + i0.0  & 24 & 0 & $\Xi_c^* B^*$    & 7970    &2.63 & 1.00 + i0.00
  \\\hline\hline
  \multicolumn{2}{c|}{\,\,Content:$b\bar{c}sud$\,\,} & \,$(I)J^P$\, &Pole   &~~~$E_B$~~~& \,Width\, & \,Channel \, & Threshold  &~~~$|g_i|$~~~ & $P_i$
  \\ \hline\hline
  \multirow{4}{*}{$PB_{1/2}$} & \multirow{4}{*}{$|\Xi_b \bar{D}\rangle$} & \multirow{4}{*}{$(0)\frac{1}{2}^-$} & \multirow{4}{*}{7632.5 + i3.1}  & \multirow{4}{*}{29} & \multirow{4}{*}{6.2 } & $\Lambda \bar{B}_c$     & 7390   & 0.08 & $-$ 0.00 $-$ i0.00
  \\ \cline{7-10}
                                                                                                                                               &&&&      &  & $\Lambda_b \bar{D}_s$        & 7588 & 0.37 & 0.03 $-$ i0.01 	
  \\ \cline{7-10}
                                                                                                                                               &&&&
  &  & $\Xi_b \bar{D}$        & 7662 & 1.89 & 0.97 + i0.01 	
  \\ \cline{7-10}
                                                                                                                                               &&&&   &  & $\Xi_b^\prime \bar{D}$        & 7802 & 0 & $-$ 0.00 + i0.00
  \\ \hline
  \multirow{4}{*}{$PB_{1/2}$} & \multirow{4}{*}{$|\Xi_b^\prime \bar{D}\rangle$} & \multirow{4}{*}{$(0)\frac{1}{2}^-$} & \multirow{4}{*}{7799.7 + i0.1}  & \multirow{4}{*}{2} & \multirow{4}{*}{0.2 } & $\Lambda \bar{B}_c$    & 7390    & 0.11 & $-$ 0.00 $-$ i0.00
  \\ \cline{7-10}
                                                                                                                                               &&&&      &  & $\Lambda_b \bar{D}_s$        & 7588 & 0 & 0.00 + i0.00 	
  \\ \cline{7-10}
                                                                                                                                               &&&&
  &  & $\Xi_b \bar{D}$        & 7662 & 0 & 0.00 + i0.00 	
  \\ \cline{7-10}
                                                                                                                                               &&&&   &  & $\Xi_b^\prime \bar{D}$        & 7802 & 0.99 & 0.95 $-$ i0.02
  \\ \hline
  \multirow{4}{*}{$VB_{1/2}$} & \multirow{4}{*}{$|\Xi_b \bar{D}^*\rangle$} & \multirow{4}{*}{$(0)\frac{1}{2}^-, \frac{3}{2}^-$} & \multirow{4}{*}{7775.8 + i2.7}  & \multirow{4}{*}{27} & \multirow{4}{*}{5.4} & $\Lambda \bar{B}_c^*$    & 7447    & 0.08 & $-$ 0.00 $-$ i0.00
  \\ \cline{7-10}
                                                                                                                                               &&&&   &  & $\Lambda_b \bar{D}_s^*$        & 7732 & 0.35 & 0.03 $-$ i0.01 	
  \\ \cline{7-10}
                                                                                                                                               &&&&
  &  & $\Xi_b \bar{D}^*$        & 7803 & 1.86 & 0.97 + i0.01 	
  \\ \cline{7-10}
                                                                                                                                               &&&&   &  & $\Xi_b^\prime \bar{D}^*$        & 7944 & 0 & $-$ 0.00 + i0.00
                                                                                                                                                 \\ \hline                                                             \multirow{4}{*}{$VB_{1/2}$} & \multirow{4}{*}{$|\Xi_b^\prime \bar{D}^*\rangle$} & \multirow{4}{*}{$(0)\frac{1}{2}^-, \frac{3}{2}^-$} & \multirow{4}{*}{7941.0 + i0.1} & \multirow{4}{*}{3} & \multirow{4}{*}{0.2} & $\Lambda \bar{B}_c^*$    & 7447     & 0.10 & $-$ 0.00 $-$ i0.00
  \\ \cline{7-10}
                                                                                                                                               &&&&   &  & $\Lambda_b \bar{D}_s^*$        & 7732 & 0 & 0.00 + i0.00 	
  \\ \cline{7-10}
                                                                                                                                               &&&&
  &  & $\Xi_b \bar{D}^*$        & 7803 & 0 & 0.00 + i0.00 	
  \\ \cline{7-10}
                                                                                                                                               &&&&   &  & $\Xi_b^\prime \bar{D}^*$        & 7944 & 1.00 & 0.95 $-$ i0.02
  \\\hline
  $PB_{3/2}$ & $|\Xi_b^* \bar{D}\rangle$ & $(0)\frac{3}{2}^-$ & 7818.9 + i0.0  & 2 & 0 & $\Xi_b^* \bar{D}$    & 7821    & 0.97 & 1.00 + i0.00
  \\\hline
  $VB_{3/2}$ & $|\Xi_b^* \bar{D}^*\rangle$ & $(0)\frac{1}{2}^-, \frac{3}{2}^-, \frac{5}{2}^-$ & 7960.2 + i0.0 & 3 & 0 & $\Xi_b^* \bar{D}^*$     & 7963    & 0.98 & 1.00 + i0.00
  \\ \hline\hline
  \end{tabular}}
  \label{tab:resultsud}
\end{table*}

We use the subtraction constant $a(\mu) = -3.2$ to perform numerical analysis. For the former $c\bar{b}sud$ system, we find ten bound states with the binding energies about $24\sim63 \mev$, which strongly couple to the $\Xi_c^{(\prime,*)} B^{(*)}$ channels. For the latter $b\bar{c}sud$ system, we find ten bound states with the binding energies about $2\sim29 \mev$, which strongly couple to the $\Xi_b^{(\prime,*)} \bar D^{(*)}$ channels. Detailed results are summarized in Table~\ref{tab:resultsud}. As we can see, the $|\Xi_c B^{(*)} \rangle$ and $|\Xi_b \bar{D}^{(*)} \rangle$ states have larger binding energies due to the additional attraction induced by the effect of the $| \Lambda_c B_s^{(*)} \rangle$ and $| \Lambda_b \bar{D}_s^{(*)} \rangle$ coupled channels, but the $|\Xi_c^\prime B^{(*)} \rangle$ and $|\Xi_b^\prime \bar{D}^{(*)} \rangle$ states do not benefit from this since they couple weakly to the $\Lambda B_c^{(*)}$ and $\Lambda \bar{B}_c^{(*)}$ channels. We refer to Ref.~\cite{Feijoo:2022rxf} for detailed discussions. Note that the $PB_{1/2}$ and $VB_{1/2}$ sectors of the $c \bar b sud$ system with $I=0$ have been studied in Ref.~\cite{Shen:2022rpn} with the subtraction constant $a(\mu)=-3.1$, and the results obtained there are similar to ours.

\subsection{Poles in the $c\bar{b}suu$ and $b\bar{c}suu$ system}
\label{sec:suu}

\begin{table*}[htbp]
  \centering
  \renewcommand{\arraystretch}{1.4}
  \caption{The poles extracted from the $c\bar{b}suu$ and $b\bar{c}suu$ systems with $I=1$, obtained using $a(\mu=1 \, \rm GeV)=-3.2$. Pole positions, binding energies ($E_B$), widths, and threshold masses of various coupled channels are in units of MeV. The couplings $g_i$ and $P_i$ have no dimension.}
  \setlength{\tabcolsep}{1mm}{
  \begin{tabular}{c|c|c|c|c|c|c|c|c|c}
  \hline\hline
  \multicolumn{2}{c|}{\,\,Content:$c\bar{b}suu$\,\,} & \,$(I)J^P$\, & Pole &~~~$E_B$~~~& \,Width\, & \,Channel \, & Threshold &~~~$|g_i|$~~~ & $P_i$
  \\ \hline\hline
  \multirow{4}{*}{$PB_{1/2}$} & \multirow{4}{*}{$|\Xi_c^\prime B\rangle$} & \multirow{4}{*}{$(1)\frac{1}{2}^-$} & \multirow{4}{*}{7810.5 + i1.4} & \multirow{4}{*}{48} & \multirow{4}{*}{2.8} & $\Sigma B_c$  & 7468 & 0.33 & $-$ 0.00 $-$ i0.00
  \\ \cline{7-10}
                                                                                                                                               &&&&   &  & $\Xi_c B$        & 7749 & 0.03 & 0.00 + i0.00 	
  \\ \cline{7-10}
                                                                                                                                               &&&&   &  & $\Sigma_c B_s$        & 7820 & 1.25 & 0.34 + i0.02
\\ \cline{7-10}
                                                                                                                                               &&&&   &  & $\Xi_c^\prime B$        & 7858 & 2.29 & 0.51 $-$ i0.01
  \\ \hline
  \multirow{4}{*}{$VB_{1/2}$} & \multirow{4}{*}{$|\Xi_c^\prime B^*\rangle$} & \multirow{4}{*}{$(1)\frac{1}{2}^-,\frac{3}{2}^-$} & \multirow{4}{*}{7860.3 + i1.3} & \multirow{4}{*}{43} & \multirow{4}{*}{2.6} & $\Sigma B_c^*$  & 7524 & 0.32 & $-$ 0.00 $-$ i0.00
  \\ \cline{7-10}
                                                                                                                                               &&&&   &  & $\Xi_c B^*$        & 7794 & 0.03 & 0.00 + i0.00 	
  \\ \cline{7-10}
                                                                                                                                               &&&&   &  & $\Sigma_c B_s^*$        & 7869 & 1.23 & 0.35 + i0.02
\\ \cline{7-10}
                                                                                                                                               &&&&   &  & $\Xi_c^\prime B^*$        & 7903 & 2.19 & 0.49 $-$ i0.01
  \\\hline
  \multirow{3}{*}{$PB_{3/2}$} & \multirow{3}{*}{$|\Xi_c^* B\rangle$} & \multirow{3}{*}{$(1)\frac{3}{2}^-$} & \multirow{3}{*}{7878.4 + i0.1} & \multirow{3}{*}{47} & \multirow{3}{*}{0.2} & $\Sigma^* B_c$  & 7659 & 0.11 & $-$ 0.00 $-$ i0.00
  \\ \cline{7-10}
                                                                                                                                               &&&&   &  & $\Sigma_c^* B_s$        & 7885 & 1.13 & 0.35 $-$ i0.02  	
  \\ \cline{7-10}
                                                                                                                                               &&&&   &  & $\Xi_c^* B$        & 7925 & 2.44 & 0.61 + i0.02
  \\\hline
  \multirow{3}{*}{$VB_{3/2}$} & \multirow{3}{*}{$|\Xi_c^* B^*\rangle$} & \multirow{3}{*}{$(1)\frac{1}{2}^-, \frac{3}{2}^-, \frac{5}{2}^-$} & \multirow{3}{*}{7928.0 + i0.1} & \multirow{3}{*}{42} & \multirow{3}{*}{0.2} & $\Sigma^* B_c^*$  & 7716 & 0.10 & $-$ 0.00 $-$ i0.00
  \\ \cline{7-10}
                                                                                                                                               &&&&   &  & $\Sigma_c^* B_s^*$        & 7934 & 1.11 & 0.37 $-$ i0.02 	
  \\ \cline{7-10}
                                                                                                                                               &&&&   &  & $\Xi_c^* B^*$        & 7970 & 2.35 & 0.59 + i0.02
  \\ \hline\hline
  \multicolumn{2}{c|}{Content:$b\bar{c}suu$} & \,$(I)J^P$\, &Pole  &~~~$E_B$~~~& \,Width\, & \,Channel \,  & Threshold &~~~$|g_i|$~~~ & $P_i$
  \\ \hline\hline
  \multirow{4}{*}{$PB_{1/2}$} & \multirow{4}{*}{$|\Xi_b^\prime \bar{D}\rangle$} & \multirow{4}{*}{$(1)\frac{1}{2}^-$} & \multirow{4}{*}{7787.0 + i3.3} & \multirow{4}{*}{15} & \multirow{4}{*}{6.6} & $\Sigma \bar{B}_c$  & 7468 & 0.12 & 0.00 $-$ i0.00
  \\ \cline{7-10}
                                                                                                                                               &&&&   &  & $\Xi_b \bar{D}$        & 7662 & 0  & $-$ 0.00 $-$ i0.00	
  \\ \cline{7-10}
                                                                                                                                               &&&&   &  & $\Sigma_b \bar{D}_s$        & 7781 & 0.58 & 0.15 $-$ i0.15
\\ \cline{7-10}
                                                                                                                                               &&&&   &  & $\Xi_b^\prime \bar{D}$        & 7802 & 1.50 & 0.84 + i0.15
  \\ \hline
  \multirow{4}{*}{$VB_{1/2}$} & \multirow{4}{*}{$|\Xi_b^\prime \bar{D}^*\rangle$} & \multirow{4}{*}{$(1)\frac{1}{2}^-,\frac{3}{2}^-$} & \multirow{4}{*}{7929.5 + i2.7} & \multirow{4}{*}{15} & \multirow{4}{*}{5.4} & $\Sigma \bar{B}_c^*$  & 7524 & 0.11 & 0.00 $-$ i0.00
  \\ \cline{7-10}
                                                                                                                                               &&&&   &  & $\Xi_b \bar{D}^*$        & 7803 & 0 & $-$0.00 $-$ i0.00 	
  \\ \cline{7-10}
                                                                                                                                               &&&&   &  & $\Sigma_b \bar{D}_s^*$        & 7925 & 0.56 & 0.15 $-$ i0.17
\\ \cline{7-10}
                                                                                                                                               &&&&   &  & $\Xi_b^\prime \bar{D}^*$        & 7944 & 1.47 & 0.84 + i0.17
  \\\hline
  \multirow{3}{*}{$PB_{3/2}$} & \multirow{3}{*}{$|\Xi_b^* \bar{D}\rangle$} & \multirow{3}{*}{$(1)\frac{3}{2}^-$} & \multirow{3}{*}{7807.3 + i3.0} & \multirow{3}{*}{14} & \multirow{3}{*}{6.0} & $\Sigma^* \bar{B}_c$  & 7659 & 0.05 & $-$ 0.00 $-$ i0.00
  \\ \cline{7-10}
                                                                                                                                               &&&&   &  & $\Sigma_b^* \bar{D}_s$        & 7803 & 0.59 & 0.16 $-$ i0.18 	
  \\ \cline{7-10}
                                                                                                                                               &&&&   &  & $\Xi_b^* \bar{D}$        & 7821 & 1.45 & 0.83 + i0.18
  \\\hline
  \multirow{3}{*}{$VB_{3/2}$} & \multirow{3}{*}{$|\Xi_b^* \bar{D}^*\rangle$} & \multirow{3}{*}{$(1)\frac{1}{2}^-, \frac{3}{2}^-, \frac{5}{2}^-$} & \multirow{3}{*}{7949.7 + i2.3} & \multirow{3}{*}{13} & \multirow{3}{*}{4.6} & $\Sigma^* \bar{B}_c^*$  & 7716 & 0.05 & $-$ 0.00 $-$ i0.00
  \\ \cline{7-10}
                                                                                                                                               &&&&   &  & $\Sigma_b^* \bar{D}_s^*$        & 7947 & 0.57 & 0.16 $-$ i0.21
  \\ \cline{7-10}
                                                                                                                                               &&&&   &  & $\Xi_b^* \bar{D}^*$        & 7963 & 1.43 & 0.83 + i0.21
\\ \hline\hline
  \end{tabular}}
  \label{tab:resultsud1}
\end{table*}

In this subsection we investigate the $c\bar{b}suu$ and $b\bar{c}suu$ systems with $I=1$. For the former $c\bar{b}suu$ system we consider
\begin{enumerate}[a)]
  \item $PB_{1/2}$ channels: $\Sigma B_c$, $\Xi_c B$, $\Sigma_c B_s$, $\Xi_c^\prime B$;
  \item $VB_{1/2}$ channels: $\Sigma B_c^*$, $\Xi_c B^*$, $\Sigma_c B_s^*$, $\Xi_c^\prime B^*$;
  \item $PB_{3/2}$ channels: $\Sigma^* B_c$, $\Sigma_c^* B_s$, $\Xi_c^* B$;
  \item $VB_{3/2}$ channels: $\Sigma^* B_c^*$, $\Sigma_c^* B_s^*$, $\Xi_c^* B^*$;
\end{enumerate}
Their corresponding $C^{\mathcal{M}\mathcal{B}}_{2}$ matrices are
\begin{eqnarray}
  \label{eq:CsuuPB}
  C^{PB_{1/2}}_{2}&=&\left(
  \renewcommand{\arraystretch}{2}
  \renewcommand{\arraycolsep}{1.3pt}
  \begin{array}{c|cccc}
  J={1\over2}      & \Sigma B_c      & \Xi_c B       & \Sigma_c B_s         & \Xi_c^\prime B
  \\ \hline
  \Sigma B_c       & 0               & {1\over\sqrt{2}}\lambda_1           & -{1\over\sqrt{3}}\lambda_3       & {1\over\sqrt{6}}\lambda_1 \\
  \Xi_c B          & {1\over\sqrt{2}}\lambda_1      & 1                    & 0                               & 0 \\
  \Sigma_c B_s     & -{1\over\sqrt{3}}\lambda_3       & 0                    & 0                               & \sqrt{2} \\
  \Xi_c^\prime B   & {1\over\sqrt{6}}\lambda_1      & 0                    & \sqrt{2}                        & 1
  \end{array}
  \right),
\\
  \nonumber
  C^{VB_{1/2}}_{2}&=&\left(
  \renewcommand{\arraystretch}{2}
  \renewcommand{\arraycolsep}{1.3pt}
  \begin{array}{c|cccc}
  J={1\over2}/{3\over2}      & \Sigma B_c^*      & \Xi_c B^*       & \Sigma_c B_s^*       & \Xi_c^\prime B^*
  \\ \hline
  \Sigma B_c^*               & 0                 & {1\over\sqrt{2}}\lambda_1             & -{1\over\sqrt{3}}\lambda_3                  & {1\over\sqrt{6}}\lambda_1 \\
  \Xi_c B^*                  & {1\over\sqrt{2}}\lambda_1          & 1                    & 0               & 0 \\
  \Sigma_c B_s^*             & -{1\over\sqrt{3}}\lambda_3           & 0                    & 0               & \sqrt{2} \\
  \Xi_c^\prime B^*           & {1\over\sqrt{6}}\lambda_1          & 0                    & \sqrt{2}        & 1
  \end{array}
  \right),
\\
\\
  C^{PB_{3/2}}_{2}&=&\left(
  \renewcommand{\arraystretch}{2}
  \renewcommand{\arraycolsep}{1.3pt}
  \begin{array}{c|ccc}
  J={3\over2}            & \Sigma^* B_c     & \Sigma_c^* B_s      & \Xi_c^* B
  \\ \hline
  \Sigma^* B_c           & 0                & {1\over\sqrt{3}}\lambda_3                 & \sqrt{2\over3}\lambda_1 \\
  \Sigma_c^* B_s         & {1\over\sqrt{3}}\lambda_3              & 0                   & \sqrt{2} \\
  \Xi_c^* B              & \sqrt{2\over3}\lambda_1                & \sqrt{2}            & 1
  \end{array}
  \right),
\\
  C^{VB_{3/2}}_{2}&=&\left(
  \renewcommand{\arraystretch}{2}
  \renewcommand{\arraycolsep}{1.3pt}
  \begin{array}{c|ccc}
    J={1\over2}/{3\over2}/{5\over2}            & \Sigma^* B_c^*     & \Sigma_c^* B_s^*      & \Xi_c^* B^*
    \\ \hline
    \Sigma^* B_c^*           & 0                & {1\over\sqrt{3}}\lambda_3                 & \sqrt{2\over3}\lambda_1 \\
    \Sigma_c^* B_s^*         & {1\over\sqrt{3}}\lambda_3              & 0                   & \sqrt{2} \\
    \Xi_c^* B^*              & \sqrt{2\over3}\lambda_1                & \sqrt{2}            & 1
  \end{array}
  \right).
\end{eqnarray}
For the latter $b\bar{c}suu$ system we consider
\begin{enumerate}[a)]
  \item $PB_{1/2}$ channels: $\Sigma \bar{B}_c$, $\Xi_b \bar{D}$, $\Sigma_b \bar{D}_s$, $\Xi_b^\prime \bar{D}$;
  \item $VB_{1/2}$ channels: $\Sigma \bar{B}_c^*$, $\Xi_b \bar{D}^*$, $\Sigma_b \bar{D}_s^*$, $\Xi_b^\prime \bar{D}^*$;
  \item $PB_{3/2}$ channels: $\Sigma^* \bar{B}_c$, $\Sigma_b^* \bar{D}_s$, $\Xi_b^* \bar{D}$;
  \item $VB_{3/2}$ channels: $\Sigma^* \bar{B}_c^*$, $\Sigma_b^* \bar{D}_s^*$, $\Xi_b^* \bar{D}^*$.
\end{enumerate}
Their corresponding $C^{\prime \mathcal{M}\mathcal{B}}_{2}$ matrices are
\begin{eqnarray}
  C^{\prime PB_{1/2}}_{2}&=&\left(
  \renewcommand{\arraystretch}{2}
  \renewcommand{\arraycolsep}{1.3pt}
  \begin{array}{c|cccc}
  J={1\over2}      & \Sigma \bar{B}_c      & \Xi_b \bar{D}       & \Sigma_b \bar{D}_s         & \Xi_b^\prime \bar{D}
  \\ \hline
  \Sigma \bar{B}_c       & 0               & {1\over\sqrt{2}}\lambda_2           & -{1\over\sqrt{3}}\lambda_4       & {1\over\sqrt{6}}\lambda_2 \\
  \Xi_b \bar{D}          & {1\over\sqrt{2}}\lambda_2      & 1                    & 0                               & 0 \\
  \Sigma_b \bar{D}_s     & -{1\over\sqrt{3}}\lambda_4       & 0                    & 0                               & \sqrt{2} \\
  \Xi_b^\prime \bar{D}   & {1\over\sqrt{6}}\lambda_2      & 0                    & \sqrt{2}                        & 1
  \end{array}
  \right),
\\
  \nonumber
  C^{\prime VB_{1/2}}_{2}&=&\left(
  \renewcommand{\arraystretch}{2}
  \renewcommand{\arraycolsep}{1.3pt}
  \begin{array}{c|cccc}
    J={1\over2}/{3\over2}      & \Sigma \bar{B}_c^*      & \Xi_b \bar{D}^*       & \Sigma_b \bar{D}_s^*         & \Xi_b^\prime \bar{D}^*
    \\ \hline
    \Sigma \bar{B}_c^*       & 0               & {1\over\sqrt{2}}\lambda_2           & -{1\over\sqrt{3}}\lambda_4       & {1\over\sqrt{6}}\lambda_2 \\
    \Xi_b \bar{D}^*          & {1\over\sqrt{2}}\lambda_2      & 1                    & 0                               & 0 \\
    \Sigma_b \bar{D}_s^*     & -{1\over\sqrt{3}}\lambda_4       & 0                    & 0                               & \sqrt{2} \\
    \Xi_b^\prime \bar{D}^*   & {1\over\sqrt{6}}\lambda_2      & 0                    & \sqrt{2}                        & 1
  \end{array}
  \right),
\\
\\
  C^{\prime PB_{3/2}}_{2}&=&\left(
  \renewcommand{\arraystretch}{2}
  \renewcommand{\arraycolsep}{1.3pt}
  \begin{array}{c|ccc}
  J={3\over2}            & \Sigma^* \bar{B}_c     & \Sigma_b^* \bar{D}_s      & \Xi_b^* \bar{D}
  \\ \hline
  \Sigma^* \bar{B}_c           & 0                & {1\over\sqrt{3}}\lambda_4                 & \sqrt{2\over3}\lambda_2 \\
  \Sigma_b^* \bar{D}_s         & {1\over\sqrt{3}}\lambda_4              & 0                   & \sqrt{2} \\
  \Xi_b^* \bar{D}              & \sqrt{2\over3}\lambda_2                & \sqrt{2}            & 1
  \end{array}
  \right),
\\
\label{eq:Cbcsuu}
  C^{\prime VB_{3/2}}_{2}&=&\left(
  \renewcommand{\arraystretch}{2}
  \renewcommand{\arraycolsep}{1.3pt}
  \begin{array}{c|ccc}
    J={1\over2}/{3\over2}/{5\over2}            & \Sigma^* \bar{B}_c^*     & \Sigma_b^* \bar{D}_s^*      & \Xi_b^* \bar{D}^*
    \\ \hline
    \Sigma^* \bar{B}_c^*           & 0                & {1\over\sqrt{3}}\lambda_4                 & \sqrt{2\over3}\lambda_2 \\
    \Sigma_b^* \bar{D}_s^*         & {1\over\sqrt{3}}\lambda_4              & 0                   & \sqrt{2} \\
    \Xi_b^* \bar{D}^*              & \sqrt{2\over3}\lambda_2                & \sqrt{2}            & 1
  \end{array}
  \right).
\end{eqnarray}

We use the subtraction constant $a(\mu)=-3.2$ to perform numerical analysis. For the former $c\bar b suu$ system, we find seven bound states with the binding energies about $42 \sim 48 \mev$, which strongly couple to the $\Xi_c^{\prime(*)} B^{(*)}$ channels. For the latter $b\bar c suu$ system, we find seven bound states with the binding energies about $13 \sim 15 \mev$, which strongly couple to the $\Xi_b^{\prime(*)} \bar D^{(*)}$ channels. In fact, the diagonal terms of Eqs.~\eqref{eq:CsuuPB}-\eqref{eq:Cbcsuu} are null or repulsive, which would not dynamically generate the $|\Xi_c^{\prime, *} B_{(*)} \rangle$ and the $|\Xi_b^{\prime, *} \bar{D}^{(*)} \rangle$ states. However, their non-diagonal terms have sizable coefficients, $C_{\Sigma_c^{(*)} B_s^{(*)} \to \Xi_c^{\prime, *} B^{(*)}} = C_{\Sigma_b^{(*)} \bar{D}_s^{(*)} \to \Xi_b^{\prime, *} \bar{D}^{(*)}} = \sqrt{2}$, which via coupled channel mechanism providing enough attraction to produce a molecular state, this coupled channel mechanism was already discussed in Ref.~\cite{Marse-Valera:2022khy}. Detailed results are summarized in Table~\ref{tab:resultsud1}.

It is interesting to notice that the isoscalar states $| \Xi_c B^{(*)} \rangle$ and $| \Xi_b \bar{D}^{(*)} \rangle$ exist, while the isovector states $| \Xi_c B^{(*)} \rangle$ and $| \Xi_b \bar{D}^{(*)} \rangle$ do not. This is because the isoscalar $\Xi_c B^{(*)}$ and $\Xi_b \bar{D}^{(*)}$ channels are attractive, while the isovector $\Xi_c B^{(*)}$ and $\Xi_b \bar{D}^{(*)}$ channels are repulsive. Oppositely, the isovector $\Xi_c^{\prime} B^{(*)}$ and $\Xi_b^{\prime} \bar{D}^{(*)}$ channels are also repulsive by themselves, but they become attractive after taking into account the coupled-channel effects, {\it i.e.}, the $\Sigma_c B_s^{(*)}$ and $\Sigma_b \bar{D}_s^{(*)}$ channels with the off-diagonal coefficients $C^{PB_{1/2}}_{2;\Sigma_c B_s \rightarrow \Xi_c^{\prime} B} = C^{\prime PB_{1/2}}_{2; \Sigma_b \bar{D}_s \rightarrow \Xi_b^{\prime} \bar{D}} = \sqrt2$ and $C^{VB_{1/2}}_{2;\Sigma_c B_s^* \rightarrow \Xi_c^{\prime} B^*} = C^{\prime VB_{1/2}}_{2; \Sigma_b \bar{D}_s^* \rightarrow \Xi_b^{\prime} \bar{D}^*} = \sqrt2$. This coupled-channel effect was already discussed in Ref.~\cite{Marse-Valera:2022khy}. Accordingly, the isoscalar states $| \Xi_c^{\prime} B^{(*)} \rangle$ and $| \Xi_b^{\prime} \bar{D}^{(*)} \rangle$ as well as the isovector states $| \Xi_c^{\prime} B^{(*)} \rangle$ and $| \Xi_b^{\prime} \bar{D}^{(*)} \rangle$ all exist.

\subsection{Poles in the $c\bar{b}ssu$ and $b\bar{c}ssu$ system}
\label{sec:ssu}

In this subsection we investigate the $c\bar{b}ssu$ and $b\bar{c}ssu$ systems with $I=1/2$. For the former $c\bar{b}ssu$ system we consider
\begin{enumerate}[a)]
  \item $PB_{1/2}$ channels: $\Xi B_c$, $\Xi_c B_s$, $\Xi_c^\prime B_s$, $\Omega_c B$;
  \item $VB_{1/2}$ channels: $\Xi B_c^*$, $\Xi_c B_s^*$, $\Xi_c^\prime B_s^*$, $\Omega_c B^*$;
  \item $PB_{3/2}$ channels: $\Xi^* B_c$, $\Xi_c^* B_s$, $\Omega_c^* B$;
  \item $VB_{3/2}$ channels: $\Xi^* B_c^*$, $\Xi_c^* B_s^*$, $\Omega_c^* B^*$.
\end{enumerate}
Their corresponding $C^{\mathcal{M}\mathcal{B}}_{3}$ matrices are
\begin{eqnarray}
  \nonumber
  C^{PB_{1/2}}_{3}&=&\left(
  \renewcommand{\arraystretch}{2}
  \renewcommand{\arraycolsep}{1.3pt}
  \begin{array}{c|cccc}
  J={1\over2}      & \Xi B_c     & \Xi_c B_s       & \Xi_c^\prime B_s         & \Omega_c B
  \\ \hline
  \Xi B_c       & 0               & -{1\over\sqrt{2}}\lambda_3           & {1\over\sqrt{6}}\lambda_3       & -{1\over\sqrt{3}}\lambda_1 \\
  \Xi_c B_s          & -{1\over\sqrt{2}}\lambda_3      & 1                    & 0                               & 0 \\
  \Xi_c^\prime B_s     & {1\over\sqrt{6}}\lambda_3       & 0                    & 1                               & \sqrt{2} \\
  \Omega_c B   & -{1\over\sqrt{3}}\lambda_1      & 0                    & \sqrt{2}                        & 0
  \end{array}
  \right),
\\
\\
  \nonumber
  C^{VB_{1/2}}_{3}&=&\left(
  \renewcommand{\arraystretch}{2}
  \renewcommand{\arraycolsep}{1.3pt}
  \begin{array}{c|cccc}
  J={1\over2}/{3\over2}      & \Xi B_c^*     & \Xi_c B_s^*       & \Xi_c^\prime B_s^*         & \Omega_c B^*
  \\ \hline
  \Xi B_c^*       & 0               & -{1\over\sqrt{2}}\lambda_3           & {1\over\sqrt{6}}\lambda_3       & -{1\over\sqrt{3}}\lambda_1 \\
  \Xi_c B_s^*          & -{1\over\sqrt{2}}\lambda_3      & 1                    & 0                               & 0 \\
  \Xi_c^\prime B_s^*     & {1\over\sqrt{6}}\lambda_3       & 0                    & 1                               & \sqrt{2} \\
  \Omega_c B^*   & -{1\over\sqrt{3}}\lambda_1      & 0                    & \sqrt{2}                        & 0
  \end{array}
  \right),
\\
\\
  C^{PB_{3/2}}_{3}&=&\left(
  \renewcommand{\arraystretch}{2}
  \renewcommand{\arraycolsep}{1.3pt}
  \begin{array}{c|ccc}
  J={3\over2}            & \Xi^* B_c     & \Xi_c^* B_s      & \Omega_c^* B
  \\ \hline
  \Xi^* B_c          & 0                & \sqrt{2\over3}\lambda_3                 & {1\over\sqrt{3}}\lambda_1 \\
  \Xi_c^* B_s         & \sqrt{2\over3}\lambda_3              & 1                   & \sqrt{2} \\
  \Omega_c^* B              & {1\over\sqrt{3}}\lambda_1                & \sqrt{2}            & 0
  \end{array}
  \right),
\\
  C^{VB_{3/2}}_{3}&=&\left(
  \renewcommand{\arraystretch}{2}
  \renewcommand{\arraycolsep}{1.3pt}
  \begin{array}{c|ccc}
  J={1\over2}/{3\over2}/{5\over2}            & \Xi^* B_c^*     & \Xi_c^* B_s^*      & \Omega_c^* B^*
  \\ \hline
  \Xi^* B_c^*          & 0                & \sqrt{2\over3}\lambda_3                 & {1\over\sqrt{3}}\lambda_1 \\
  \Xi_c^* B_s^*         & \sqrt{2\over3}\lambda_3              & 1                   & \sqrt{2} \\
  \Omega_c^* B^*              & {1\over\sqrt{3}}\lambda_1                & \sqrt{2}            & 0
  \end{array}
  \right).
\end{eqnarray}
For the latter $b\bar{c}ssu$ system we consider
\begin{enumerate}[a)]
  \item $PB_{1/2}$ channels: $\Xi \bar{B}_c$, $\Xi_b \bar{D}_s$, $\Xi_b^\prime \bar{D}_s$, $\Omega_b \bar{D}$;
  \item $VB_{1/2}$ channels: $\Xi \bar{B}_c^*$, $\Xi_b \bar{D}_s^*$, $\Xi_b^\prime \bar{D}_s^*$, $\Omega_b \bar{D}^*$;
  \item $PB_{3/2}$ channels: $\Xi^* \bar{B}_c$, $\Xi_b^* \bar{D}_s$, $\Omega_b^* \bar{D}$;
  \item $VB_{3/2}$ channels: $\Xi^* \bar{B}_c^*$, $\Xi_b^* \bar{D}_s^*$, $\Omega_b^* \bar{D}^*$.
\end{enumerate}
Their corresponding $C^{\prime \mathcal{M}\mathcal{B}}_{3}$ matrices are
\begin{eqnarray}
  \nonumber
  C^{\prime PB_{1/2}}_{3}&=&\left(
  \renewcommand{\arraystretch}{2}
  \renewcommand{\arraycolsep}{1.3pt}
  \begin{array}{c|cccc}
  J={1\over2}      & \Xi \bar{B}_c     & \Xi_b \bar{D}_s       & \Xi_b^\prime \bar{D}_s         & \Omega_b \bar{D}
  \\ \hline
  \Xi \bar{B}_c       & 0               & -{1\over\sqrt{2}}\lambda_4           & {1\over\sqrt{6}}\lambda_4       & -{1\over\sqrt{3}}\lambda_2 \\
  \Xi_b \bar{D}_s          & -{1\over\sqrt{2}}\lambda_4      & 1                    & 0                               & 0 \\
  \Xi_b^\prime \bar{D}_s     & {1\over\sqrt{6}}\lambda_4       & 0                    & 1                               & \sqrt{2} \\
  \Omega_b \bar{D}   & -{1\over\sqrt{3}}\lambda_2      & 0                    & \sqrt{2}                        & 0
  \end{array}
  \right),
\\
\\
  \nonumber
  C^{\prime VB_{1/2}}_{3}&=&\left(
  \renewcommand{\arraystretch}{2}
  \renewcommand{\arraycolsep}{1.3pt}
  \begin{array}{c|cccc}
  J={1\over2}/{3\over2}      & \Xi \bar{B}_c^*     & \Xi_b \bar{D}_s^*       & \Xi_b^\prime \bar{D}_s^*         & \Omega_b \bar{D}^*
  \\ \hline
  \Xi \bar{B}_c^*       & 0               & -{1\over\sqrt{2}}\lambda_4           & {1\over\sqrt{6}}\lambda_4       & -{1\over\sqrt{3}}\lambda_2 \\
  \Xi_b \bar{D}_s^*          & -{1\over\sqrt{2}}\lambda_4      & 1                    & 0                               & 0 \\
  \Xi_b^\prime \bar{D}_s^*     & {1\over\sqrt{6}}\lambda_4       & 0                    & 1                               & \sqrt{2} \\
  \Omega_b \bar{D}^*   & -{1\over\sqrt{3}}\lambda_2      & 0                    & \sqrt{2}                        & 0
  \end{array}
  \right),
\\
\\
  C^{\prime PB_{3/2}}_{3}&=&\left(
  \renewcommand{\arraystretch}{2}
  \renewcommand{\arraycolsep}{1.3pt}
  \begin{array}{c|ccc}
  J={3\over2}            & \Xi^* \bar{B}_c     & \Xi_b^* \bar{D}_s      & \Omega_b^* \bar{D}
  \\ \hline
  \Xi^* \bar{B}_c          & 0                & \sqrt{2\over3}\lambda_4                 & {1\over\sqrt{3}}\lambda_2 \\
  \Xi_b^* \bar{D}_s         & \sqrt{2\over3}\lambda_4              & 1                   & \sqrt{2} \\
  \Omega_b^* \bar{D}              & {1\over\sqrt{3}}\lambda_2                & \sqrt{2}            & 0
  \end{array}
  \right),
\\
  C^{\prime VB_{3/2}}_{3}&=&\left(
  \renewcommand{\arraystretch}{2}
  \renewcommand{\arraycolsep}{1.3pt}
  \begin{array}{c|ccc}
  J={1\over2}/{3\over2}/{5\over2}            & \Xi^* \bar{B}_c^*     & \Xi_b^* \bar{D}_s^*      & \Omega_b^* \bar{D}^*
  \\ \hline
  \Xi^* \bar{B}_c^*          & 0                & \sqrt{2\over3}\lambda_4                 & {1\over\sqrt{3}}\lambda_2 \\
  \Xi_b^* \bar{D}_s^*         & \sqrt{2\over3}\lambda_4              & 1                   & \sqrt{2} \\
  \Omega_b^* \bar{D}^*              & {1\over\sqrt{3}}\lambda_2                & \sqrt{2}            & 0
  \end{array}
  \right).
\end{eqnarray}

We use the subtraction constant $a(\mu)=-3.2$ to perform numerical analysis. For the former $c \bar b ssu$ system, we find seven bound states with the binding energies about $30 \sim 34 \mev$, which strongly couple to the $\Omega_c^{(*)} B^{(*)}$ channels. For the latter $b \bar c ssu$ system, we find seven bound states with the binding energies about $2 \sim 3 \mev$, which strongly couple to $\Omega_b^{(*)} \bar D^{(*)}$ channels. Detailed results are summarized in Table~\ref{tab:resultssu}. 
Note that the $PB_{1/2}$ and $VB_{1/2}$ sectors of the $c \bar c ssq$ system have been studied in Ref.~\cite{Marse-Valera:2022khy}, where the $\Omega_c \bar D^{(*)}$ bound states are suggested to exist.

\begin{table*}[htbp]
  \centering
  \renewcommand{\arraystretch}{1.4}
  \caption{The poles extracted from the $c\bar{b}ssu$ and $b\bar{c}ssu$ systems with $I=1/2$, obtained using $a(\mu=1 \, \rm GeV)=-3.2$. Pole positions, binding energies ($E_B$), widths, and threshold masses of various coupled channels are in units of MeV. The couplings $g_i$ and $P_i$ have no dimension.}
  \setlength{\tabcolsep}{1mm}{
  \begin{tabular}{c|c|c|c|c|c|c|c|c|c}
  \hline\hline
  \multicolumn{2}{c|}{\,\,Content:$c\bar{b}ssu$\,\,} & \,$(I)J^P$\, & Pole &~~~$E_B$~~~& \,Width\, & \,Channel \, & Threshold &~~~$|g_i|$~~~ & $P_i$
  \\ \hline\hline
  \multirow{4}{*}{$PB_{1/2}$} & \multirow{4}{*}{$|\Omega_c B\rangle$} & \multirow{4}{*}{$(\frac{1}{2})\frac{1}{2}^-$} & \multirow{4}{*}{7941.4 + i1.7} & \multirow{4}{*}{34} & \multirow{4}{*}{3.4} & $\Xi B_c$  & 7593 & 0.36 & $-$ 0.00 $-$ i0.00
  \\ \cline{7-10}
                                                                                                                                               &&&&   &  & $\Xi_c B_s$        & 7836 & 0.03 & 0.00 + i0.00 	
  \\ \cline{7-10}
                                                                                                                                               &&&&   &  & $\Xi_c^\prime B_s$        & 7945 & 0.75 & 0.20 + i0.03
\\ \cline{7-10}
                                                                                                                                               &&&&   &  & $\Omega_c B$        & 7975 & 2.37 & 0.70 $-$ i0.02
  \\ \hline
  \multirow{4}{*}{$VB_{1/2}$} & \multirow{4}{*}{$|\Omega_c B^*\rangle$} & \multirow{4}{*}{$(\frac{1}{2})\frac{1}{2}^-, \frac{3}{2}^-$} & \multirow{4}{*}{7990.2 + i1.6} & \multirow{4}{*}{30} & \multirow{4}{*}{3.2} & $\Xi B_c^*$  & 7649 & 0.35 & $-$ 0.00 $-$ i0.00
  \\ \cline{7-10}
                                                                                                                                               &&&&   &  & $\Xi_c B_s^*$        & 7884 & 0.03 & 0.00 + i0.00 	
  \\ \cline{7-10}
                                                                                                                                               &&&&   &  & $\Xi_c^\prime B_s^*$        & 7994 & 0.75 & 0.20 + i0.02
\\ \cline{7-10}
                                                                                                                                               &&&&   &  & $\Omega_c B^*$        & 8020 & 2.25 & 0.67 $-$ i0.04
  \\\hline
  \multirow{3}{*}{$PB_{3/2}$} & \multirow{3}{*}{$|\Omega_c^* B\rangle$} & \multirow{3}{*}{$(\frac{1}{2})\frac{3}{2}^-$} & \multirow{3}{*}{8012.6 + i0.3} & \multirow{3}{*}{32} & \multirow{3}{*}{0.6} & $\Xi^* B_c$  & 7808 & 0.06 & $-$ 0.00 $-$ i0.00
  \\ \cline{7-10}
                                                                                                                                               &&&&   &  & $\Xi_c^* B_s$        & 8013 & 0.52 & $-$0.14$-$ i0.40 	
  \\ \cline{7-10}
                                                                                                                                               &&&&   &  & $\Omega_c^* B$        & 8045 & 2.29 & 0.69 + i0.14
  \\\hline
  \multirow{3}{*}{$VB_{3/2}$} & \multirow{3}{*}{$|\Omega_c^* B^*\rangle$} & \multirow{3}{*}{$(\frac{1}{2})\frac{1}{2}^-, \frac{3}{2}^-, \frac{5}{2}^-$} & \multirow{3}{*}{8061.0 + i0.4} & \multirow{3}{*}{30} & \multirow{3}{*}{0.8} & $\Xi^* B_c^*$  & 7864 & 0.06 & $-$ 0.00 $-$ i0.00
  \\ \cline{7-10}
                                                                                                                                               &&&&   &  & $\Xi_c^* B_s^*$        & 8061 & 0.53 & $-$0.00 $-$ i0.32 	
  \\ \cline{7-10}
                                                                                                                                               &&&&   &  & $\Omega_c^* B^*$        & 8091 & 2.26 & 0.69 + i0.13
  \\ \hline\hline
  \multicolumn{2}{c|}{Content:$b\bar{c}ssu$} & \,$(I)J^P$\, &Pole  &~~~$E_B$~~~& \,Width\, & \,Channel \,  & Threshold &~~~$|g_i|$~~~ & $P_i$
  \\ \hline\hline
  \multirow{4}{*}{$PB_{1/2}$} & \multirow{4}{*}{$|\Omega_b \bar{D}\rangle$} & \multirow{4}{*}{$(\frac{1}{2})\frac{1}{2}^-$} & \multirow{4}{*}{7909.9 + i1.3} & \multirow{4}{*}{2} & \multirow{4}{*}{2.6} & $\Xi \bar{B}_c$  & 7593 & 0.10 &$-$ 0.00 $-$ i0.00
  \\ \cline{7-10}
                                                                                                                                               &&&&   &  & $\Xi_b \bar{D}_s$        & 7763 & 0 & 0.00 + i0.00 	
  \\ \cline{7-10}
                                                                                                                                               &&&&   &  & $\Xi_b^\prime \bar{D}_s$        & 7903 & 0.33 & 0.06 $-$ i0.05
\\ \cline{7-10}
                                                                                                                                               &&&&   &  & $\Omega_c B$        & 7912 & 0.97 & 0.88 $-$ i0.06
  \\ \hline
  \multirow{4}{*}{$VB_{1/2}$} & \multirow{4}{*}{$|\Omega_b \bar{D}^*\rangle$} & \multirow{4}{*}{$(\frac{1}{2})\frac{1}{2}^-, \frac{3}{2}^-$} & \multirow{4}{*}{8051.4 + i1.1} & \multirow{4}{*}{3} & \multirow{4}{*}{2.2} & $\Xi \bar{B}_c^*$  & 7649 & 0.09 & $-$ 0.00 $-$ i0.00
  \\ \cline{7-10}
                                                                                                                                               &&&&   &  & $\Xi_b \bar{D}_s^*$        & 7907 & 0 & 0.00 + i0.00	
  \\ \cline{7-10}
                                                                                                                                               &&&&   &  & $\Xi_b^\prime \bar{D}_s^*$        & 8047 & 0.32 & 0.06 $-$ i0.07
\\ \cline{7-10}
                                                                                                                                               &&&&   &  & $\Omega_b \bar{D}^*$        & 8054 & 0.96 & 0.89 $-$ i0.06
  \\\hline
  \multirow{3}{*}{$PB_{3/2}$} & \multirow{3}{*}{$|\Omega_b^* \bar{D}\rangle$} & \multirow{3}{*}{$(\frac{1}{2})\frac{3}{2}^-$} & \multirow{3}{*}{7936.0+ i1.4} & \multirow{3}{*}{2} & \multirow{3}{*}{2.8} & $\Xi^* \bar{B}_c$  & 7808 & 0.02 & $-$ 0.00 + i0.00
  \\ \cline{7-10}
                                                                                                                                               &&&&   &  & $\Xi_b^* \bar{D}_s$        & 7922 & 0.30 & 0.04 $-$ i0.02 	
  \\ \cline{7-10}
                                                                                                                                               &&&&   &  & $\Omega_b^* \bar{D}$        & 7938 & 0.99 & 0.94 $-$ i0.13
  \\\hline
  \multirow{3}{*}{$VB_{3/2}$} & \multirow{3}{*}{$|\Omega_b^* \bar{D}^*\rangle$} & \multirow{3}{*}{$(\frac{1}{2})\frac{1}{2}^-, \frac{3}{2}^-, \frac{5}{2}^-$} & \multirow{3}{*}{8077.4+ i1.3} & \multirow{3}{*}{3} & \multirow{3}{*}{2.6} & $\Xi^* \bar{B}_c^*$  & 7864 & 0.02 & $-$ 0.00 + i0.00
  \\ \cline{7-10}
                                                                                                                                               &&&&   &  & $\Xi_b^* \bar{D}_s^*$        & 8066 & 0.30 & 0.04 $-$ i0.02	
  \\ \cline{7-10}
                                                                                                                                               &&&&   &  & $\Omega_b^* \bar{D}^*$        & 8080 & 0.99 & 0.94 $-$ i0.14
\\ \hline\hline
  \end{tabular}}
  \label{tab:resultssu}
\end{table*}

\subsection{Poles in the $c\bar{b}uuu$ and $b\bar{c}uuu$ system}
\label{sec:uuu}

In this subsection we investigate the $c\bar{b}uuu$ and $b\bar{c}uuu$ systems with $I=3/2$. For the former $c\bar{b}uuu$ system we consider
\begin{enumerate}[a)]
  \item $PB_{1/2}$ channel: $\Sigma_c B$;
  \item $VB_{1/2}$ channel: $\Sigma_c B^*$;
  \item $PB_{3/2}$ channels: $\Delta B_c$, $\Sigma_c^* B$;
  \item $VB_{3/2}$ channels: $\Delta B_c^*$, $\Sigma_c^* B^*$.
\end{enumerate}
Their corresponding $C^{\mathcal{M}\mathcal{B}}_{4}$ matrices are
\begin{eqnarray}
  C^{PB_{1/2}}_{4}&=&\left(
  \renewcommand{\arraystretch}{2}
  \renewcommand{\arraycolsep}{1.3pt}
  \begin{array}{c|c}
  J={1\over2}            & \Sigma_c B
  \\ \hline
  \Sigma_c B                & 2
  \end{array}
  \right),
\\
  C^{VB_{1/2}}_{4}&=&\left(
  \renewcommand{\arraystretch}{2}
  \renewcommand{\arraycolsep}{1.3pt}
  \begin{array}{c|c}
  J={1\over2}/{3\over2}       & \Sigma_c B^*
  \\ \hline
  \Sigma_c B^*                & 2
  \end{array}
  \right),
\\
  C^{PB_{3/2}}_{4}&=&\left(
  \renewcommand{\arraystretch}{2}
  \renewcommand{\arraycolsep}{1.3pt}
  \begin{array}{c|cc}
  J={3\over2}            & \Delta B_c         & \Sigma_c^* B
  \\ \hline
  \Delta B_c             & 0                  & \lambda_1
  \\
  \Sigma_c^* B           & \lambda_1          & 2
  \end{array}
  \right),
\\
  C^{VB_{3/2}}_{4}&=&\left(
  \renewcommand{\arraystretch}{2}
  \renewcommand{\arraycolsep}{1.3pt}
  \begin{array}{c|cc}
   J={1\over2}/{3\over2}/{5\over2}            & \Delta B_c^*         & \Sigma_c^* B^*
  \\ \hline
  \Delta B_c^*             & 0                  & \lambda_1
  \\
  \Sigma_c^* B^*           & \lambda_1          & 2
  \end{array}
  \right).
\end{eqnarray}
For the latter $b\bar{c}uuu$ system we consider
\begin{enumerate}[a)]
  \item $PB_{1/2}$ channel: $\Sigma_b \bar{D}$;
  \item $VB_{1/2}$ channel: $\Sigma_b \bar{D}^*$;
  \item $PB_{3/2}$ channels: $\Delta \bar{B}_c$, $\Sigma_b^* \bar{D}$;
  \item $VB_{3/2}$ channels: $\Delta \bar{B}_c^*$, $\Sigma_b^* \bar{D}^*$.
\end{enumerate}
Their corresponding $C^{\prime \mathcal{M}\mathcal{B}}_{4}$ matrices are
\begin{eqnarray}
  C^{\prime PB_{1/2}}_{4}&=&\left(
  \renewcommand{\arraystretch}{2}
  \renewcommand{\arraycolsep}{1.3pt}
  \begin{array}{c|c}
  J={1\over2}            & \Sigma_b \bar{D}
  \\ \hline
  \Sigma_b \bar{D}                & 2
  \end{array}
  \right),
\\
  C^{\prime VB_{1/2}}_{4}&=&\left(
  \renewcommand{\arraystretch}{2}
  \renewcommand{\arraycolsep}{1.3pt}
  \begin{array}{c|c}
  J={1\over2}/{3\over2}       & \Sigma_b \bar{D}^*
  \\ \hline
  \Sigma_b \bar{D}^*                & 2
  \end{array}
  \right),
\\
  C^{\prime PB_{3/2}}_{4}&=&\left(
  \renewcommand{\arraystretch}{2}
  \renewcommand{\arraycolsep}{1.3pt}
  \begin{array}{c|cc}
  J={3\over2}            & \Delta \bar{B}_c         & \Sigma_b^* \bar{D}
  \\ \hline
  \Delta \bar{B}_c             & 0                  & \lambda_2
  \\
  \Sigma_b^* \bar{D}           & \lambda_2          & 2
  \end{array}
  \right),
\\
  C^{\prime VB_{3/2}}_{4}&=&\left(
  \renewcommand{\arraystretch}{2}
  \renewcommand{\arraycolsep}{1.3pt}
  \begin{array}{c|cc}
  J={1\over2}/{3\over2}/{5\over2}            & \Delta \bar{B}_c^*         & \Sigma_b^* \bar{D}^*
  \\ \hline
  \Delta \bar{B}_c^*             & 0                  & \lambda_2
  \\
  \Sigma_b^* \bar{D}^*           & \lambda_2          & 2
  \end{array}
  \right).
\end{eqnarray}

We find the interactions of the $c\bar{b}uuu$ and $b\bar{c}uuu$ systems to be both repulsive, so there are no poles extracted in these two systems.

\subsection{Poles in the $c\bar{b}sss$ and $b\bar{c}sss$ system}
\label{sec:sss}

In this subsection we investigate the $c\bar{b}sss$ and $b\bar{c}sss$ systems. For the former $c\bar{b}sss$ system we consider
\begin{enumerate}[a)]
  \item $PB_{1/2}$ channel: $\Omega_c B_s$;
  \item $VB_{1/2}$ channel: $\Omega_c B_s^*$;
  \item $PB_{3/2}$ channels: $\Omega_c^* B_s$, $\Omega B_c$;
  \item $VB_{3/2}$ channels: $\Omega_c^* B_s^*$, $\Omega B_c^*$.
\end{enumerate}
Their corresponding $C^{\mathcal{M}\mathcal{B}}_{5}$ matrices are
\begin{eqnarray}
  C^{PB_{1/2}}_{5}&=&\left(
  \renewcommand{\arraystretch}{2}
  \renewcommand{\arraycolsep}{1.3pt}
  \begin{array}{c|c}
  J={1\over2}            & \Omega_c B_s
  \\ \hline
  \Omega_c B_s                & 2
  \end{array}
  \right),
\\
  C^{VB_{1/2}}_{5}&=&\left(
  \renewcommand{\arraystretch}{2}
  \renewcommand{\arraycolsep}{1.3pt}
  \begin{array}{c|c}
  J={1\over2}/{3\over2}       & \Omega_c B_s^*
  \\ \hline
  \Omega_c B_s^*                & 2
  \end{array}
  \right),
\\
  C^{PB_{3/2}}_{5}&=&\left(
  \renewcommand{\arraystretch}{2}
  \renewcommand{\arraycolsep}{1.3pt}
  \begin{array}{c|cc}
  J={3\over2}            & \Omega B_c         & \Omega_c^* B_s
  \\ \hline
  \Omega B_c             & 0                  & \lambda_3
  \\
  \Omega_c^* B_s           & \lambda_3          & 2
  \end{array}
  \right),
\\
  C^{VB_{3/2}}_{5}&=&\left(
  \renewcommand{\arraystretch}{2}
  \renewcommand{\arraycolsep}{1.3pt}
  \begin{array}{c|cc}
  J={1\over2}/{3\over2}/{5\over2}            & \Omega B_c^*         & \Omega_c^* B_s^*
  \\ \hline
  \Omega B_c^*             & 0                  & \lambda_3
  \\
  \Omega_c^* B_s^*           & \lambda_3          & 2
  \end{array}
  \right).
\end{eqnarray}
For the latter $b\bar{c}sss$ system we consider
\begin{enumerate}[a)]
  \item $PB_{1/2}$ channel: $\Omega_b \bar{D}_s$;
  \item $VB_{1/2}$ channel: $\Omega_b \bar{D}_s^*$;
  \item $PB_{3/2}$ channels: $\Omega \bar{B}_c$, $\Omega_b^* \bar{D}_s$;
  \item $VB_{3/2}$ channels: $\Omega \bar{B}_c^*$, $\Omega_b^* \bar{D}_s^*$.
\end{enumerate}
Their corresponding $C^{\prime \mathcal{M}\mathcal{B}}_{5}$ matrices are
\begin{eqnarray}
  C^{\prime PB_{1/2}}_{5}&=&\left(
  \renewcommand{\arraystretch}{2}
  \renewcommand{\arraycolsep}{1.3pt}
  \begin{array}{c|c}
  J={1\over2}            & \Omega_b \bar{D}_s
  \\ \hline
  \Omega_b \bar{D}_s                & 2
  \end{array}
  \right),
\\
  C^{\prime VB_{1/2}}_{5}&=&\left(
  \renewcommand{\arraystretch}{2}
  \renewcommand{\arraycolsep}{1.3pt}
  \begin{array}{c|c}
  J={1\over2}/{3\over2}       & \Omega_b \bar{D}_s^*
  \\ \hline
  \Omega_b \bar{D}_s^*                & 2
  \end{array}
  \right),
\\
  C^{\prime PB_{3/2}}_{5}&=&\left(
  \renewcommand{\arraystretch}{2}
  \renewcommand{\arraycolsep}{1.3pt}
  \begin{array}{c|cc}
  J={3\over2}            & \Omega \bar{B}_c         & \Omega_b^* \bar{D}_s
  \\ \hline
  \Omega \bar{B}_c             & 0                  & \lambda_4
  \\
  \Omega_b^* \bar{D}_s           & \lambda_4          & 2
  \end{array}
  \right),
\\
  C^{\prime VB_{3/2}}_{5}&=&\left(
  \renewcommand{\arraystretch}{2}
  \renewcommand{\arraycolsep}{1.3pt}
  \begin{array}{c|cc}
  J={1\over2}/{3\over2}/{5\over2}            & \Omega \bar{B}_c^*         & \Omega_b^* \bar{D}_s^*
  \\ \hline
  \Omega \bar{B}_c^*             & 0                  & \lambda_4
  \\
  \Omega_b^* \bar{D}_s^*           & \lambda_4          & 2
  \end{array}
  \right).
\end{eqnarray}

We find the interactions of the $c\bar{b}sss$ and $b\bar{c}sss$ systems to be both repulsive, so there are no poles extracted in these two systems.

%
\section{Conclusion}
\label{sec:Con}
%

In this paper we systematically study the possibly-existing molecular pentaquark states with open charm and bottom flavors, {\it i.e.}, the states with the quark contents $c\bar{b}qqq$ and $b\bar{c}qqq$ ($q=u,d,s$). We apply the coupled-channel unitary approach within the local hidden-gauge formalism to derive the transition potentials from the meson-baryon interactions. Then we solve the Bethe-Salpeter equation in coupled channels to extract the poles in the complex plane. These poles qualify as dynamically generated molecular pentaquark states. We calculate their masses and widths as well as their couplings to various coupled channels. Especially, we use the subtraction constant $a(\mu=1 \, \rm GeV)=-3.2$ to obtain:
\begin{itemize}

\item[$\bullet$] In the $c\bar{b}uud$ system with $I=1/2$, we consider the $N B_c^{(*)}$, $\Lambda_c B^{(*)}$, and $\Sigma_c^{(*)} B^{(*)}$ channels. We extract seven poles, which strongly couple to the $\Sigma_c^{(*)} B^{(*)}$ channels, suggesting that they can be associated to $\Sigma_c^{(*)} B^{(*)}$ molecular pentaquark states:
    \begin{eqnarray}
    && |\Sigma_c B; (I)J^P=({1\over2})\, {1\over2}^- \rangle \, ,
    \\ && |\Sigma_c B^*; (I)J^P=({1\over2})\, {1\over2}^-\Big/{3\over2}^-\rangle \, ,
    \\ && |\Sigma_c^* B; (I)J^P=({1\over2})\, {3\over2}^-\rangle \, ,
    \\ && |\Sigma_c^* B^*; (I)J^P=({1\over2})\, {1\over2}^-\Big/{3\over2}^-\Big/{5\over2}^-\rangle \, .
    \end{eqnarray}
    Their binding energies are estimated to be about $28\sim32$~MeV.

\item[$\bullet$] In the $b\bar{c}uud$ system with $I=1/2$, we consider the $N \bar{B}_c^{(*)}$, $\Lambda_b \bar{D}^{(*)}$, and $\Sigma_b^{(*)}\bar{D}^{(*)}$ channels. We also extract seven poles, which strongly couple to the $\Sigma_b^{(*)} \bar{D}^{(*)}$ channels, suggesting that they can be associated to $\Sigma_b^{(*)} \bar{D}^{(*)}$ molecular pentaquark states:
\begin{eqnarray}
&& |\Sigma_b \bar{D}; (I)J^P=({1\over2})\, {1\over2}^- \rangle \, ,
\\ && |\Sigma_b \bar{D}^*; (I)J^P=({1\over2})\, {1\over2}^-\Big/{3\over2}^-\rangle \, ,
\\ && |\Sigma_b^* \bar{D}; (I)J^P=({1\over2})\, {3\over2}^-\rangle \, ,
\\ && |\Sigma_b^* \bar{D}^*; (I)J^P=({1\over2})\, {1\over2}^-\Big/{3\over2}^-\Big/{5\over2}^-\rangle \, .
\end{eqnarray}
Their binding energies are estimated to be about $6\sim7$~MeV.

\item[$\bullet$] In the $c\bar{b}sud$ system with $I=0$, we consider the $\Lambda B_c^{(*)}$, $\Lambda_c B_s^{(*)}$, $\Xi_c^\prime B^{(*)}$, and $\Xi_c^{(*)} B^{(*)}$ channels. We extract ten poles, which strongly couple to the $\Xi_c^{(\prime,*)} B^{(*)}$ channels, suggesting that they can be associated to $\Xi_c^{(\prime,*)} B^{(*)}$ molecular pentaquark states:
\begin{eqnarray}
  && |\Xi_c B; (I)J^P=(0)\, {1\over2}^- \rangle \, ,
  \\ && |\Xi_c^\prime B; (I)J^P=(0)\, {1\over2}^- \rangle \, ,
  \\ && |\Xi_c B^*; (I)J^P=(0)\, {1\over2}^-\Big/{3\over2}^-\rangle \, ,
  \\ && |\Xi_c^\prime B^*; (I)J^P=(0)\, {1\over2}^-\Big/{3\over2}^-\rangle \, ,
  \\ && |\Xi_c^* B; (I)J^P=(0)\, {3\over2}^-\rangle \, ,
  \\ && |\Xi_c^* B^*; (I)J^P=(0)\, {1\over2}^-\Big/{3\over2}^-\Big/{5\over2}^-\rangle \, .
  \end{eqnarray}
  Their binding energies are estimated to be about $24\sim63$~MeV.

\item[$\bullet$] In the $b\bar{c}sud$ system with $I=0$, we consider the $\Lambda \bar{B}_c^{(*)}$, $\Lambda_b \bar{D}_s^{(*)}$, $\Xi_b^\prime \bar{D}^{(*)}$, and $\Xi_b^{(*)} \bar{D}^{(*)}$ channels. We also extract ten poles, which strongly couple to the $\Xi_b^{(\prime,*)} \bar{D}^{(*)}$ channels, suggesting that they can be associated to $\Xi_b^{(\prime,*)} \bar{D}^{(*)}$ molecular pentaquark states:
\begin{eqnarray}
  && |\Xi_b \bar{D}; (I)J^P=(0)\, {1\over2}^- \rangle \, ,
  \\ && |\Xi_b^\prime \bar{D}; (I)J^P=(0)\, {1\over2}^- \rangle \, ,
  \\ && |\Xi_b \bar{D}^*; (I)J^P=(0)\, {1\over2}^-\Big/{3\over2}^-\rangle \, ,
  \\ && |\Xi_b^\prime \bar{D}^*; (I)J^P=(0)\, {1\over2}^-\Big/{3\over2}^-\rangle \, ,
  \\ && |\Xi_b^* \bar{D}; (I)J^P=(0)\, {3\over2}^-\rangle \, ,
  \\ && |\Xi_b^* \bar{D}^*; (I)J^P=(0)\, {1\over2}^-\Big/{3\over2}^-\Big/{5\over2}^-\rangle \, .
  \end{eqnarray}
  Their binding energies are estimated to be about $2\sim29$~MeV.

\item[$\bullet$] In the $c\bar{b}sud$ system with $I=1$, we consider the $\Sigma^{(*)} B_c^{(*)}$, $\Xi_c^{(*)} B^{(*)}$, $\Sigma_c^{(*)} B_s^{(*)}$, and $\Xi_c^{\prime} B^{(*)}$ channels. We extract seven poles, which strongly couple to the $\Xi_c^{\prime, *} B^{(*)}$ channels, suggesting that they can be associated to $\Xi_c^{\prime, *} B^{(*)}$ molecular pentaquark states:
\begin{eqnarray}
  && |\Xi_c^\prime B; (I)J^P=(1)\, {1\over2}^- \rangle \, ,
  \\ && |\Xi_c^\prime B^*; (I)J^P=(1)\, {1\over2}^-\Big/{3\over2}^-\rangle \, ,
  \\ && |\Xi_c^* B; (I)J^P=(1)\, {3\over2}^-\rangle \, ,
  \\ && |\Xi_c^* B^*; (I)J^P=(1)\, {1\over2}^-\Big/{3\over2}^-\Big/{5\over2}^-\rangle \, .
  \end{eqnarray}
  Their binding energies are estimated to be about $42\sim48$~MeV.

\item[$\bullet$] In the $b\bar{c}sud$ system with $I=1$, we consider the $\Sigma^{(*)} \bar B_c^{(*)}$, $\Xi_b^{(*)} \bar D^{(*)}$, $\Sigma_b^{(*)} \bar D_s^{(*)}$, and $\Xi_b^{\prime} \bar D^{(*)}$ channels. We also extract seven poles, which strongly couple to the $\Xi_b^{\prime, *} \bar D^{(*)}$ channels, suggesting that they can be associated to $\Xi_b^{\prime, *} \bar D^{(*)}$ molecular pentaquark states:
\begin{eqnarray}
  && |\Xi_b^\prime \bar{D}; (I)J^P=(1)\, {1\over2}^- \rangle \, ,
  \\ && |\Xi_b^\prime \bar{D}^*; (I)J^P=(1)\, {1\over2}^-\Big/{3\over2}^-\rangle \, ,
  \\ && |\Xi_b^* \bar{D}; (I)J^P=(1)\, {3\over2}^-\rangle \, ,
  \\ && |\Xi_b^* \bar{D}^*; (I)J^P=(1)\, {1\over2}^-\Big/{3\over2}^-\Big/{5\over2}^-\rangle \, .
  \end{eqnarray}
  Their binding energies are estimated to be about $13\sim15$~MeV.

\item[$\bullet$] In the $c\bar{b}ssu$ system with $I=1/2$, we consider the $\Xi^{(*)} B_c^{(*)}$, $\Xi_c^\prime B_s^{(*)}$, $\Xi_c^{(*)} B_s^{(*)}$, and $\Omega_c^{(*)} B^{(*)}$ channels. We extract seven poles, which strongly couple to the $\Omega_c^{(*)} B^{(*)}$ channels, suggesting that they can be associated to $\Omega_c^{(*)} B^{(*)}$ molecular pentaquark states:
\begin{eqnarray}
  && |\Omega_c B; (I)J^P=({1\over2})\, {1\over2}^- \rangle \, ,
  \\ && |\Omega_c B^*; (I)J^P=({1\over2})\, {1\over2}^-\Big/{3\over2}^-\rangle \, ,
  \\ && |\Omega_c^* B; (I)J^P=({1\over2})\, {3\over2}^-\rangle \, ,
  \\ && |\Omega_c^* B^*; (I)J^P=({1\over2})\, {1\over2}^-\Big/{3\over2}^-\Big/{5\over2}^-\rangle \, .
  \end{eqnarray}
  Their binding energies are estimated to be about $30\sim34$~MeV.

\item[$\bullet$] In the $b\bar{c}ssu$ system with $I=1/2$, we consider the $\Xi^{(*)} \bar B_c^{(*)}$, $\Xi_b^\prime \bar D_s^{(*)}$, $\Xi_b^{(*)} \bar D_s^{(*)}$, and $\Omega_b^{(*)} \bar D^{(*)}$ channels. We extract seven poles, which strongly couple to the $\Omega_b^{(*)} \bar D^{(*)}$ channels, suggesting that they can be associated to $\Omega_b^{(*)} \bar D^{(*)}$ molecular pentaquark states:
\begin{eqnarray}
  && |\Omega_b \bar D; (I)J^P=({1\over2})\, {1\over2}^- \rangle \, ,
  \\ && |\Omega_b \bar D^*; (I)J^P=({1\over2})\, {1\over2}^-\Big/{3\over2}^-\rangle \, ,
  \\ && |\Omega_b^* \bar D; (I)J^P=({1\over2})\, {3\over2}^-\rangle \, ,
  \\ && |\Omega_b^* \bar D^*; (I)J^P=({1\over2})\, {1\over2}^-\Big/{3\over2}^-\Big/{5\over2}^-\rangle \, .
  \end{eqnarray}
  Their binding energies are estimated to be about $2\sim3$~MeV.

\end{itemize}
Most of the above poles are tied to the attractive interaction indicated by a negative diagonal coefficient in the corresponding matrix of coefficients. Besides, some of the above poles are tied to the coupled channel mechanism indicated by a non-diagonal coefficient, even when the diagonal terms are repulsive. The above results are summarized in Tables~\ref{tab:resultuud} - \ref{tab:resultssu}. Note that some of the binding energies are not so large, {\it e.g.}, the binding energy of $|\Sigma_b \bar{D}; (I)J^P=({1\over2})\, {1\over2}^- \rangle$ is just 6~MeV, the binding energy of $|\Xi_b^\prime \bar{D}; (I)J^P=(0)\, {1\over2}^- \rangle$ is just 2~MeV, and the binding energy of $|\Omega_b \bar{D}; (I)J^P=({1\over2})\, {1\over2}^- \rangle$ is just 2~MeV. These states may not be bound, and they become the near-threshold virtual states when taking $a(\mu)=-3.0$. However, the binding energies of the $\Sigma_c^{(*)} B^{(*)}$ molecular states with $I=1/2$, the $\Xi_c^{(\prime,*)} B^{(*)}$ molecular states with $I=0$ and $I=1$, and the $\Omega_c^{(*)} B^{(*)}$ molecular states with $I=1/2$ are all quite large, so these bound states are more likely to exist.

In total, sixty-two molecular pentaquark states with open charm and bottom flavors are predicted in this work, including seven $\Sigma_c^{(*)} B^{(*)}$ molecular states with $I=1/2$, seven $\Sigma_b^{(*)} \bar{D}^{(*)}$ molecular states with $I=1/2$, ten $\Xi_c^{(\prime,*)} B^{(*)}$ molecular states with $I=0$, ten $\Xi_b^{(\prime,*)} \bar{D}^{(*)}$ molecular states with $I=0$, seven $\Xi_c^{\prime ,*} B^{(*)}$ molecular states with $I=1$, seven $\Xi_b^{\prime ,*} \bar{D}^{(*)}$ molecular states with $I=1$, seven $\Omega_c^{(*)} B^{(*)}$ molecular states with $I=1/2$, and seven $\Omega_b^{(*)} \bar{D}^{(*)}$ molecular states with $I=1/2$. We propose to search for these possibly-existing molecular pentaquark states in the $\Upsilon$ decays in the future LHCb experiments. Besides, we propose to search for the $\Sigma_c B^{(*)}$, $\Sigma_b \bar{D}^{(*)}$, $\Xi_c B^{(*)}$, $\Xi_c^{\prime} B^{(*)}$, $\Xi_b \bar{D}^{(*)}$, $\Xi_b^{\prime} \bar{D}^{(*)}$, $\Omega_c B^{(*)}$, and $\Omega_b \bar D^{(*)}$ molecular states in their $S/D$-wave two-body decay channels $N B_c$, $N \bar{B}_c$, $\Lambda_c B_s$, $\Lambda B_c/\Sigma B_c$, $\Lambda_b \bar{D}_s$, $\Lambda \bar{B}_c/\Sigma \bar B_c$, $\Xi B_c$, and $\Xi \bar B_c$, respectively; we propose to search for the $\Sigma_c^{*} B^{(*)}$, $\Sigma_b^{*} \bar{D}^{(*)}$, $\Xi_c^{*} B^{(*)}$, and $\Xi_b^{*} \bar{D}^{(*)}$ molecular states through their $P$-wave three-body decay channels $\Lambda_c B^{(*)} \pi$, $\Lambda_b \bar{D}^{(*)} \pi$, $\Xi_c B^{(*)} \pi$, and $\Xi_b \bar{D}^{(*)} \pi$, respectively; we also propose to search for the $\Omega_c^* B^{(*)}$ and $\Omega_b^* \bar D^{(*)}$ molecular states in their $S$-wave two-body decay channels $\Xi^* B_c^{(*)}$ and $\Xi^* \bar B_c^{(*)}$, respectively.

%
\section*{Acknowledgments}
%

We are grateful to Eulogio Oset for the very helpful discussion.
This project is supported by
the National Natural Science Foundation of China under Grants No.~12075019 and No.~11975083,
the Jiangsu Provincial Double-Innovation Program under Grant No.~JSSCRC2021488,
and
the Fundamental Research Funds for the Central Universities.
This project is also supported by the Central Government Guidance Funds for Local Scientific and Technological Development, China (No. Guike ZY22096024).

%
\appendix
\section{Baryon wave functions}
\label{Sec:baryon}

We summarize all the relevant baryon wave functions in this appendix. The wave functions for the $J^P=1/2^+$ baryons are:
\begin{eqnarray}
  \big| \Xi_c^+ \big\rangle &=& \Big| {1\over\sqrt{2}} c(us-su) \Big\rangle \big| \chi_{MA} \big\rangle, \nonumber\\ [2mm]
  \big| \Xi_c^0 \big\rangle &=& \Big| {1\over\sqrt{2}} c(ds-sd) \Big\rangle \big| \chi_{MA} \big\rangle, \nonumber\\ [2mm]
  \big| \Xi_c^{\prime +} \big\rangle &=& \Big| {1\over\sqrt{2}} c(us+su) \Big\rangle \big| \chi_{MS} \big\rangle, \nonumber\\ [2mm]
  \big| \Xi_c^{\prime 0} \big\rangle &=& \Big| {1\over\sqrt{2}} c(ds+sd) \Big\rangle \big| \chi_{MS} \big\rangle, \nonumber\\ [2mm]
  \big| \Omega_c^{0} \big\rangle &=& \Big| css \Big\rangle \big| \chi_{MS} \big\rangle, \nonumber\\  [2mm]
  \big| \Xi_b^{0} \big\rangle &=& \Big| {1\over\sqrt{2}} b(us-su) \Big\rangle \big| \chi_{MA} \big\rangle, \nonumber\\ [2mm]
  \big| \Xi_b^{-} \big\rangle &=& \Big| {1\over\sqrt{2}} b(ds-sd) \Big\rangle \big| \chi_{MA} \big\rangle, \nonumber\\ [2mm]
  \big| \Xi_b^{\prime 0} \big\rangle &=& \Big| {1\over\sqrt{2}} b(us+su) \Big\rangle \big| \chi_{MS} \big\rangle, \nonumber\\ [2mm]
  \big| \Xi_b^{\prime -} \big\rangle &=& \Big| {1\over\sqrt{2}} b(ds+sd) \Big\rangle \big| \chi_{MS} \big\rangle, \nonumber\\ [2mm]
  \big| \Omega_b^{-} \big\rangle &=& \Big| bss \Big\rangle \big| \chi_{MS} \big\rangle, \nonumber\\ [2mm]
  \big| \Lambda^{0} \big\rangle &=& {1\over\sqrt{2}} \left(\big| \phi_{MS} \rangle \big| \chi_{MS} \big\rangle + \big| \phi_{MA} \big\rangle \big| \chi_{MA} \big\rangle\right), \nonumber\\ [2mm]
  &&\big| \phi_{MS} \rangle = {1\over2} \left(\big| dus \big\rangle + \big| dsu \big\rangle - \big| uds \big\rangle - \big| usd \big\rangle\right), \nonumber\\  [2mm]
  &&\big| \phi_{MA} \rangle = {1\over2\sqrt{3}} \left(\big| u(ds-sd) \big\rangle + \big| d(su-us) \big\rangle\right. \nonumber\\  [2mm]
  && \left.- 2\big| s(ud-du) \big\rangle\right), \nonumber\\  [2mm]
  \big| \Sigma^{+} \big\rangle &=& {1\over\sqrt{2}} \left(\big| \phi_{MS} \rangle \big| \chi_{MS} \big\rangle + \big| \phi_{MA} \big\rangle \big| \chi_{MA} \big\rangle\right), \nonumber\\ [2mm]
  &&\big| \phi_{MS} \rangle = -{1\over\sqrt{6}} \left(\big| u(us+su) \big\rangle - 2\big| suu \big\rangle\right), \nonumber\\  [2mm]
  &&\big| \phi_{MA} \rangle = {1\over\sqrt{2}} \big| u(su-us) \big\rangle, \nonumber\\  [2mm]
  \big| \Sigma^{0} \big\rangle &=& {1\over\sqrt{2}} \left(\big| \phi_{MS} \rangle \big| \chi_{MS} \big\rangle + \big| \phi_{MA} \big\rangle \big| \chi_{MA} \big\rangle\right), \nonumber\\ [2mm]
  &&\big| \phi_{MS} \rangle = {1\over2\sqrt{3}} \left(\big| u(ds+sd) \big\rangle + \big| d(su+us) \big\rangle \right. \nonumber\\ [2mm]
  &&\left.- 2\big| s(du+ud) \big\rangle\right), \nonumber\\  [2mm]
  &&\big| \phi_{MA} \rangle = {1\over2} \left(\big| u(ds-sd) \big\rangle - \big| d(su-us) \big\rangle\right), \nonumber\\  [2mm]
  \big| \Xi^{0} \big\rangle &=& {1\over\sqrt{2}} \left(\big| \phi_{MS} \rangle \big| \chi_{MS} \big\rangle + \big| \phi_{MA} \big\rangle \big| \chi_{MA} \big\rangle\right), \nonumber\\ [2mm]
  &&\big| \phi_{MS} \rangle = {1\over\sqrt{6}} \left(\big| s(us+su) \big\rangle - 2\big| uss \big\rangle\right), \nonumber\\  [2mm]
  &&\big| \phi_{MA} \rangle = -{1\over\sqrt{2}} \big| s(us-su) \big\rangle.
\end{eqnarray}
The wave functions for the $J^P=3/2^+$ baryons:
\begin{eqnarray}
  \big| \Xi_c^{*+} \big\rangle &=& \Big| {1\over\sqrt{2}} c(us+su) \Big\rangle \big| \chi_S \big\rangle, \nonumber\\ [2mm]
  \big| \Xi_c^{*0} \big\rangle &=& \Big| {1\over\sqrt{2}} c(ds+sd) \Big\rangle \big| \chi_S \big\rangle, \nonumber\\ [2mm]
  \big| \Omega_c^{*0} \big\rangle &=& \big| css \big\rangle \big| \chi_S \big\rangle, \nonumber\\ [2mm]
  \big| \Xi_b^{*0} \big\rangle &=& \Big| {1\over\sqrt{2}} b(us+su) \Big\rangle \big| \chi_S \big\rangle, \nonumber\\ [2mm]
  \big| \Xi_b^{*-} \big\rangle &=& \Big| {1\over\sqrt{2}} b(ds+sd) \Big\rangle \big| \chi_S \big\rangle, \nonumber\\ [2mm]
  \big| \Omega_b^{*-} \big\rangle &=& \big| bss \big\rangle \big| \chi_S \big\rangle, \nonumber\\ [2mm]
  \big| \Sigma^{*+} \big\rangle &=& {1\over\sqrt{3}}\big| u(su+us)+suu \big\rangle \big| \chi_S \big\rangle, \nonumber\\ [2mm]
  \big| \Sigma^{*0} \big\rangle &=& {1\over\sqrt{6}} \big( \big| s(du+ud)+d(su+us) \big\rangle, \nonumber\\ [2mm]
  && + \big| u(sd+ds) \big\rangle \big)  \big| \chi_S \big\rangle \nonumber\\ [2mm]
  \big| \Xi^{*0} \big\rangle &=& {1\over\sqrt{3}}\big| s(us+su)+uss \big\rangle \big| \chi_S \big\rangle, \nonumber\\ [2mm]
  \big| \Delta^{++} \big\rangle &=& \big| uuu \big\rangle \big| \chi_S \big\rangle, \nonumber\\ [2mm]
  \big| \Omega^{-} \big\rangle &=& \big| sss \big\rangle \big| \chi_S \big\rangle.
\end{eqnarray}

\section{Discussion on the dimensional and cut-off regularizations}
\label{Sec:discussion}

\begin{table*}[htb!]
  \centering
  \renewcommand{\arraystretch}{1.4}
  \caption{The poles extracted from the $PB_{1/2}$ sector of the $b\bar{c}uud$ system with $I={1/2}$, obtained using the dimensional regularization with $a(\mu=1 \, \rm GeV)=-3.2$ and the cut-off regularization with $q_{\max}=450\mev$. Pole positions, binding energies ($E_B$), widths, and threshold masses of various coupled channels are in units of MeV. The couplings $g_i$ have no dimension.}
  \setlength{\tabcolsep}{1mm}{
  \begin{tabular}{c|c|c|c|c|c|c|c|c}
  \hline\hline
  \multicolumn{2}{c|}{\,Content:$b\bar{c}uud$\,} & \,$(I)J^P$\, &Pole    &~~~$E_B$~~~& \,Width\, & \,Channel \, & Threshold &~~~$|g_i|$~~~
  \\ \hline\hline
  \multirow{6}{*}{Dimensional} & \multirow{3}{*}{$|\Lambda_b \bar{D}\rangle$} & \multirow{3}{*}{$({1\over2})\frac{1}{2}^-$} & \multirow{3}{*}{7372.6 $-$ i0.6}  & \multirow{3}{*}{114} & \multirow{3}{*}{1.2} & $N \bar{B}_c$    & 7213    & 0.30
  \\ \cline{7-9}
                                                                                                                                               &&&&     &  & $\Lambda_b \bar{D}$        & 7487 & 2.69  	
  \\ \cline{7-9}
                                                                                                                                               &&&&
  &  & $\Sigma_b \bar{D}$        & 7680 & 0.01  	
  \\\cline{2-9}
 & \multirow{3}{*}{$|\Sigma_b \bar{D}\rangle$} & \multirow{3}{*}{$({1\over2})\frac{1}{2}^-$} & \multirow{3}{*}{7674.2 $-$ i0.2}  & \multirow{3}{*}{6} & \multirow{3}{*}{0.4} & $N \bar{B}_c$    & 7213    & 0.13
  \\ \cline{7-9}
                                                                                                                                               &&&&     &  & $\Lambda_b \bar{D}$        & 7487 & 0  	
  \\ \cline{7-9}
                                                                                                                                               &&&&
  &  & $\Sigma_b \bar{D}$        & 7680 & 1.23  	
  \\\hline
  \multirow{3}{*}{Cut-off} & \multirow{3}{*}{$|\Sigma_b \bar{D}\rangle$} & \multirow{3}{*}{$({1\over2})\frac{1}{2}^-$} & \multirow{3}{*}{7670.2 $-$ i0.4}  & \multirow{3}{*}{10} & \multirow{3}{*}{0.8} & $N \bar{B}_c$    & 7213    & 0.21
  \\ \cline{7-9}
                                                                                                                                               &&&&     &  & $\Lambda_b \bar{D}$        & 7487 & 0  	
  \\ \cline{7-9}
                                                                                                                                               &&&&
  &  & $\Sigma_b \bar{D}$        & 7680 & 1.88
  \\ \hline\hline
  \end{tabular}}
  \label{tab:resultcduud}
\end{table*}

\begin{figure*}[htb!]
  \centering
  \subfigure[Re{[$G^{\mathcal{MB}}$]} for $b\bar{c}uud$]{\includegraphics[width=0.24\hsize, height=0.24\hsize]{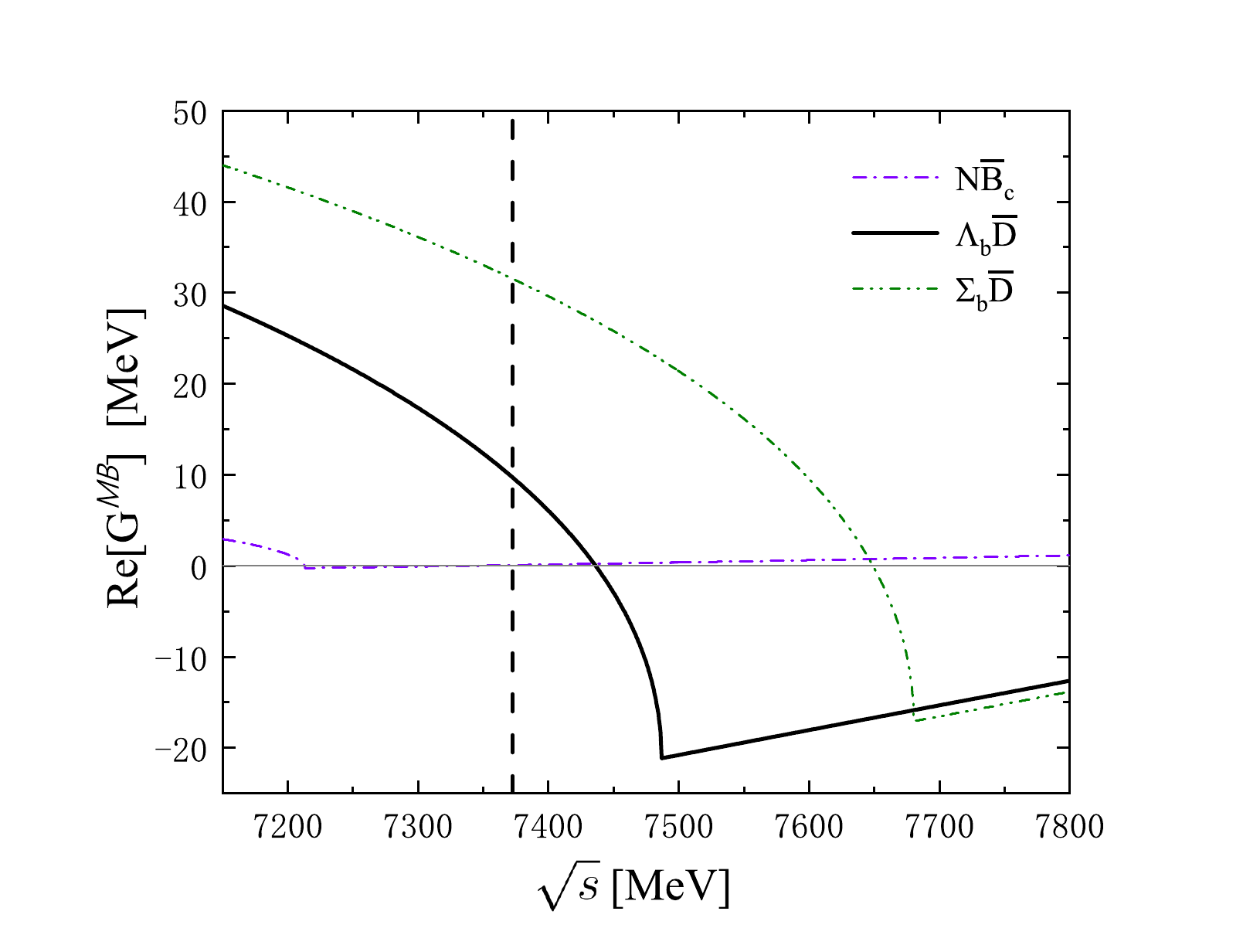}\label{fig:Reduud}}
  \subfigure[Re{[$G^{\prime \mathcal{MB}}$]} for $b\bar{c}uud$]{\includegraphics[width=0.24\hsize, height=0.24\hsize]{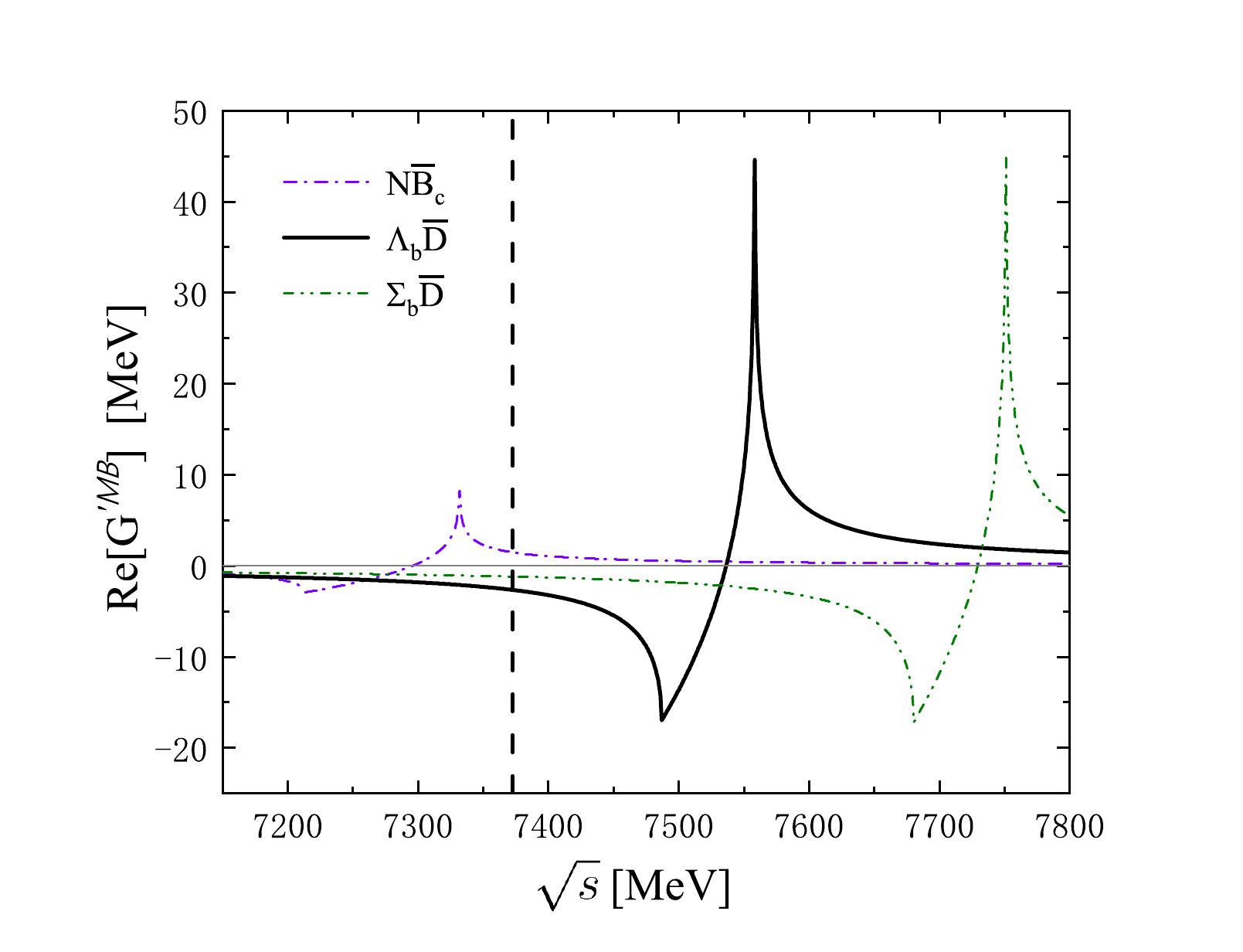}\label{fig:Recuud}}
  \subfigure[Im{[$G^{\mathcal{MB}}$]} for $b\bar{c}uud$]{\includegraphics[width=0.24\hsize, height=0.24\hsize]{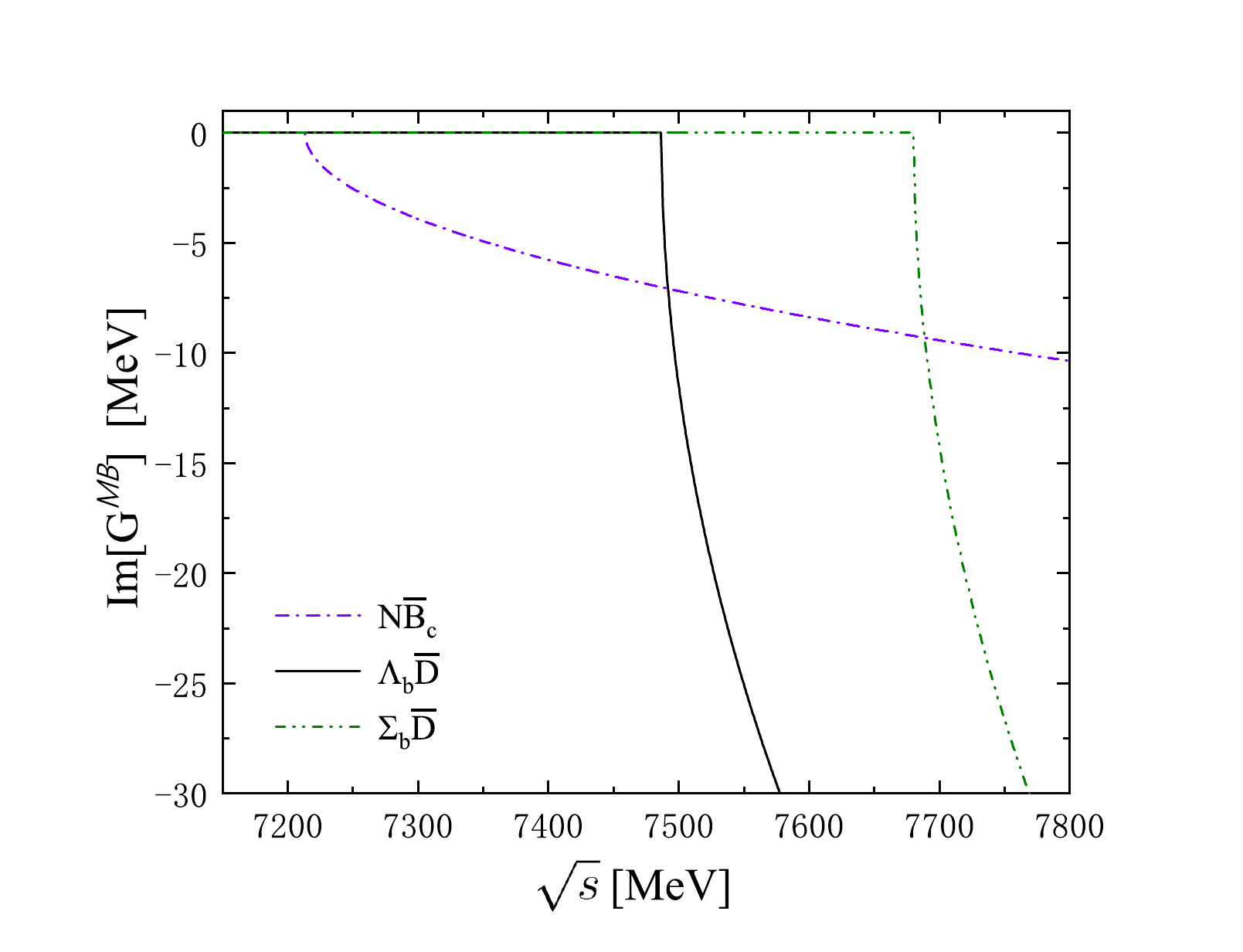}\label{fig:Imduud}}
  \subfigure[Im{[$G^{\prime \mathcal{MB}}$]} for $b\bar{c}uud$]{\includegraphics[width=0.24\hsize, height=0.24\hsize]{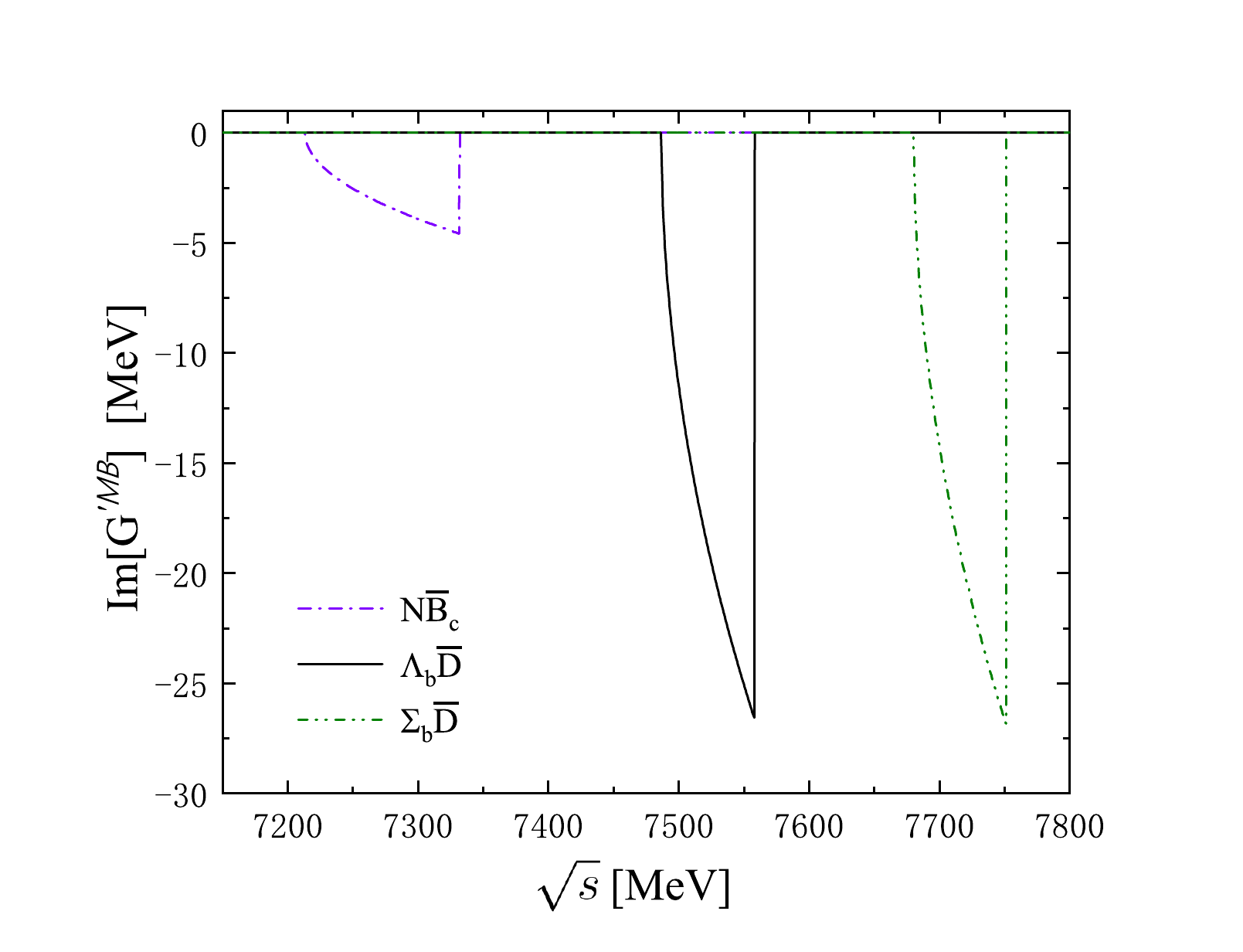}\label{fig:Imcuud}}
\\
  \subfigure[Re{[$G^{\mathcal{MB}}$]} for $b\bar{c}sud$]{\includegraphics[width=0.24\hsize, height=0.24\hsize]{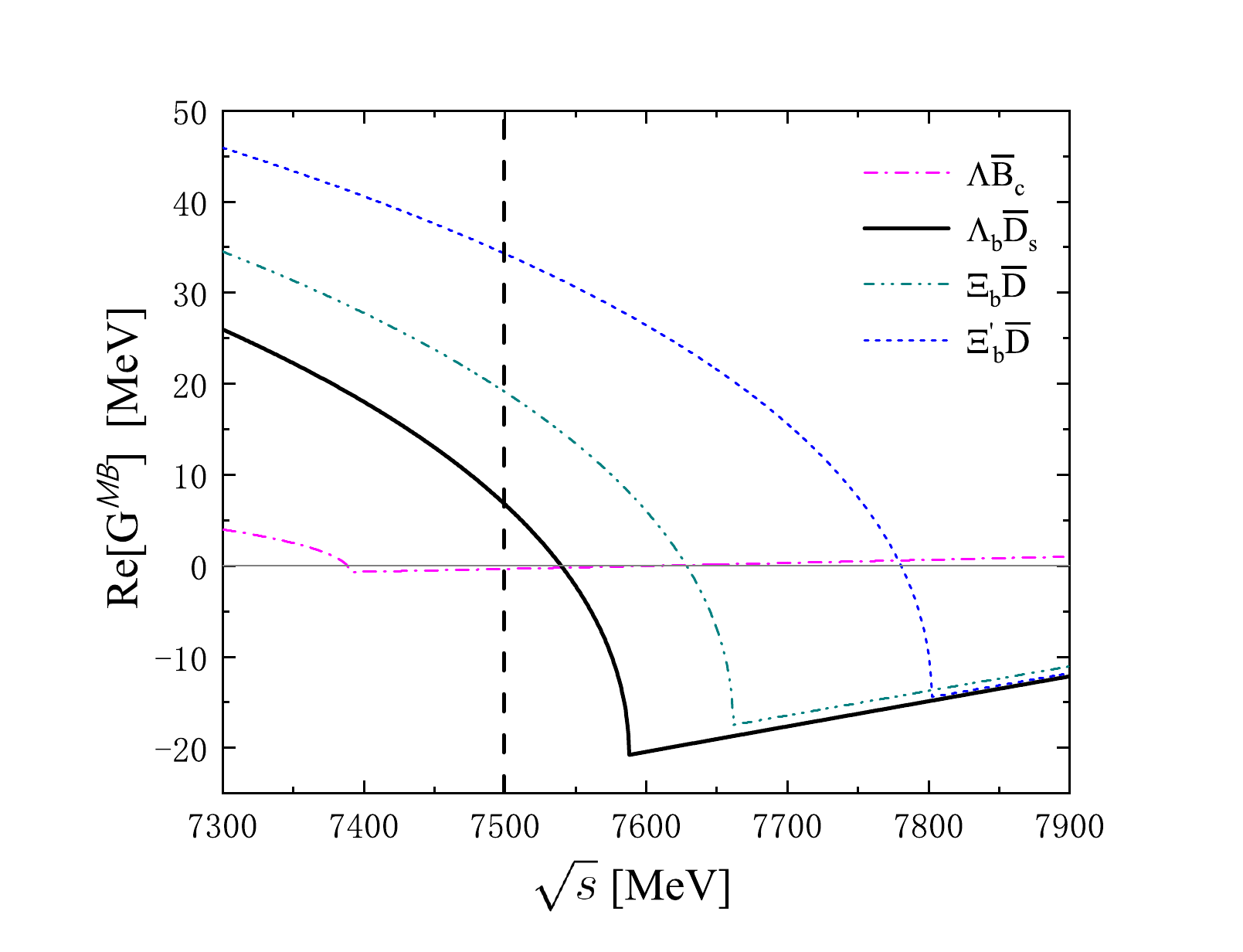}\label{fig:Redsud}}
  \subfigure[Re{[$G^{\prime \mathcal{MB}}$]} for $b\bar{c}sud$]{\includegraphics[width=0.24\hsize, height=0.24\hsize]{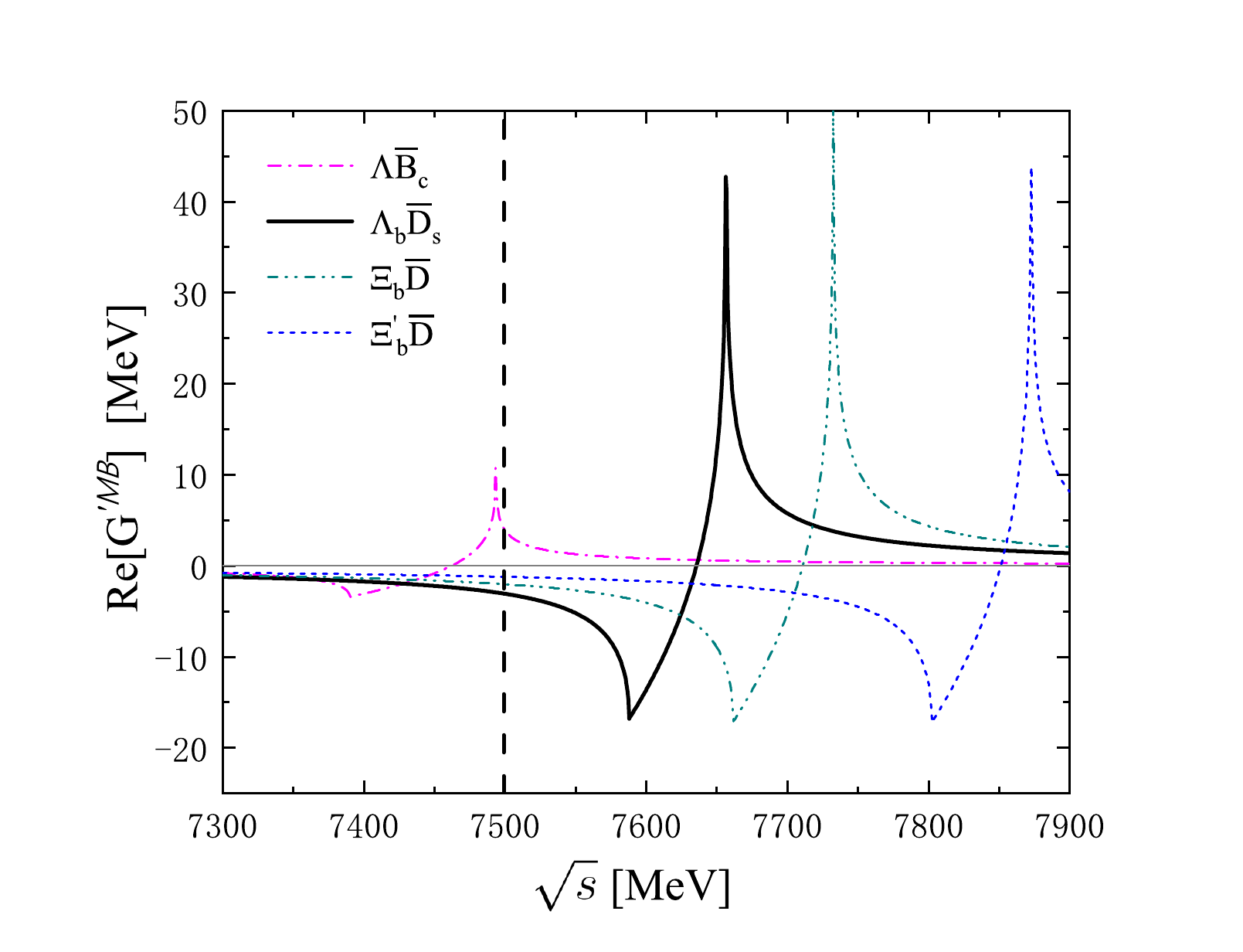}\label{fig:Recsud}}
  \subfigure[Im{[$G^{\mathcal{MB}}$]} for $b\bar{c}sud$]{\includegraphics[width=0.24\hsize, height=0.24\hsize]{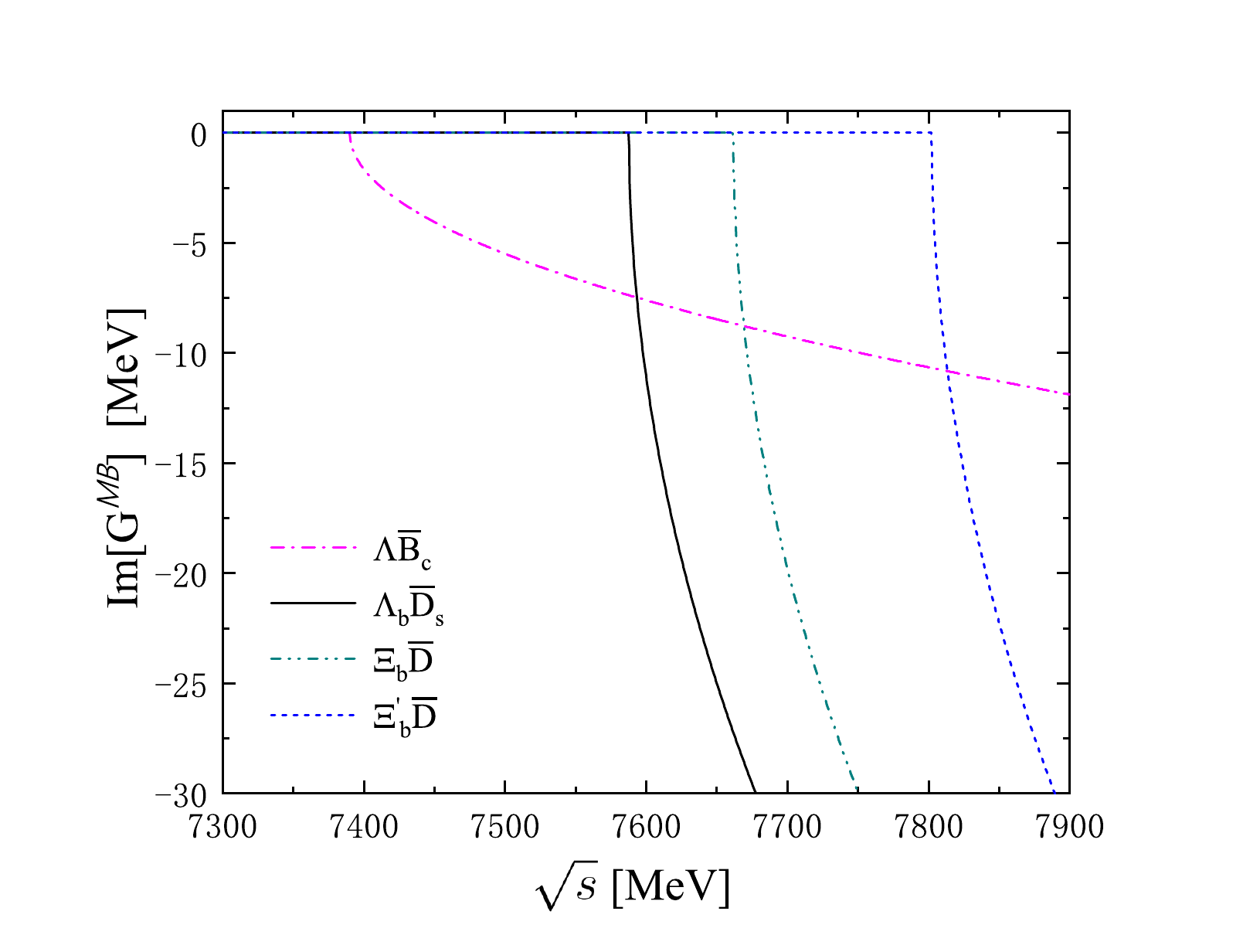}\label{fig:Imdsud}}
  \subfigure[Im{[$G^{\prime \mathcal{MB}}$]} for $b\bar{c}sud$]{\includegraphics[width=0.24\hsize, height=0.24\hsize]{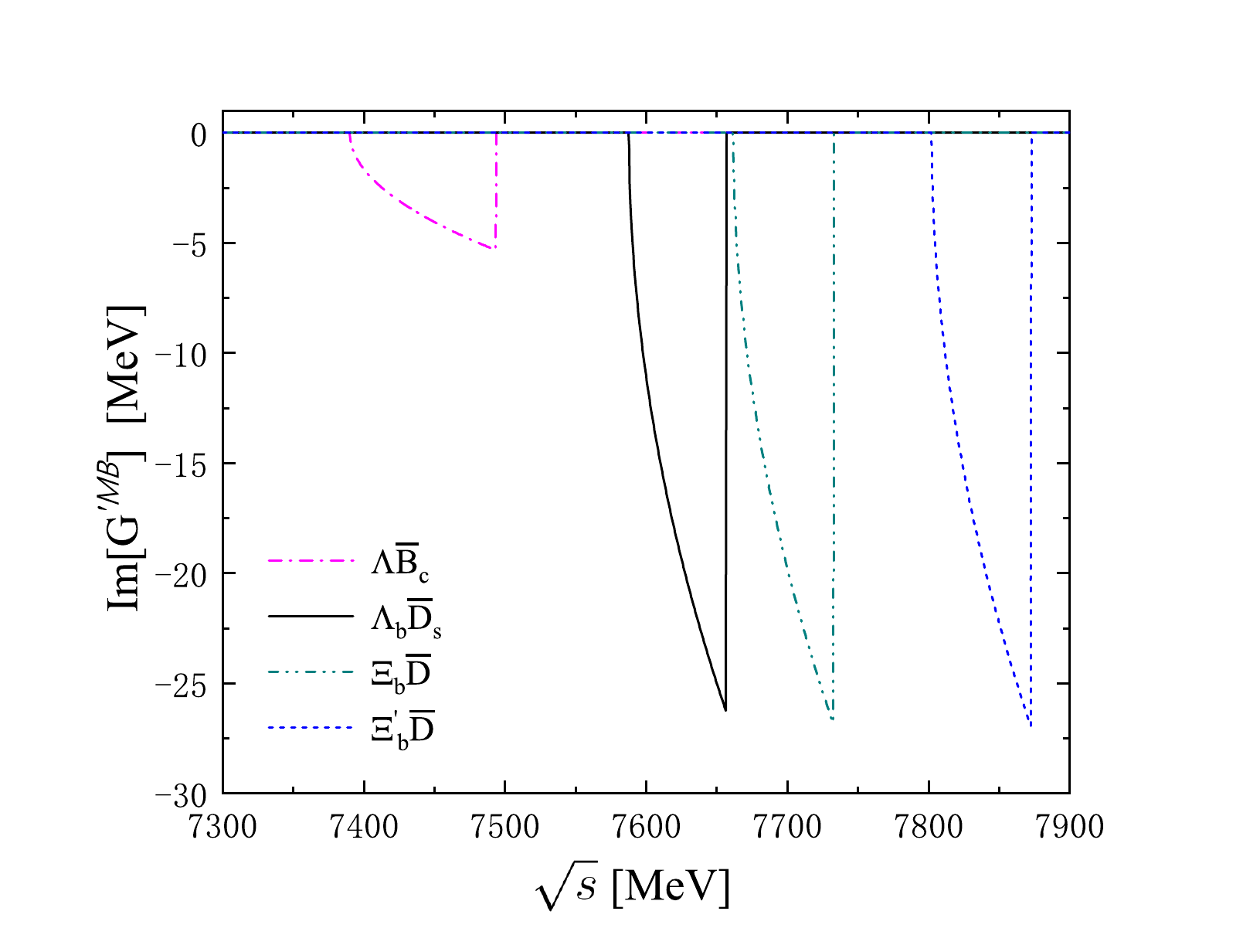}\label{fig:Imcsud}}
  \caption{Re[$G^{\mathcal{MB}}$] (Im[$G^{\mathcal{MB}}$]) and Re[$G^{\prime \mathcal{MB}}$] (Im[$G^{\prime \mathcal{MB}}$]) stand for the real (imaginary) parts of the loop function for various coupled channels, calculated through the dimensional and cut-off regularizations, respectively. The dash black lines indicate the poles positions.}
  \label{fig:ReIm}
\end{figure*}

\begin{table*}[htb!]
  \centering
  \renewcommand{\arraystretch}{1.4}
  \caption{The poles extracted from the $PB_{1/2}$ sector of the $b\bar{c}sud$ system with $I=0$, obtained using the dimensional regularization with $a(\mu=1 \, \rm GeV)=-3.2$ and the cut-off regularization with $q_{\max}=450\mev$. Pole positions, binding energies ($E_B$), widths, and threshold masses of various coupled channels are in units of MeV. The couplings $g_i$ have no dimension.}
  \setlength{\tabcolsep}{1mm}{
  \begin{tabular}{c|c|c|c|c|c|c|c|c}
  \hline\hline
  \multicolumn{2}{c|}{\,Content:$c\bar{b}sud$\,} & \,$(I)J^P$\, &Pole    &~~~$E_B$~~~& \,Width\, & \,Channel \, & Threshold &~~~$|g_i|$~~~
  \\ \hline\hline
  \multirow{12}{*}{Dimensional} & \multirow{4}{*}{$|\Lambda_b \bar{D}_s\rangle$} & \multirow{4}{*}{$(0)\frac{1}{2}^-$} & \multirow{4}{*}{7499.1 $-$ i0.4}  & \multirow{4}{*}{89} & \multirow{4}{*}{0.8} & $\Lambda \bar{B}_c$    & 7390    & 0.58
  \\ \cline{7-9}
                                                                                                                                               &&&&     &  & $\Lambda_b \bar{D}_s$        & 7588 & 5.71  	
  \\ \cline{7-9}
                                                                                                                                               &&&&
  &  & $\Xi_b \bar{D}$        & 7662 & 2.00  	
  \\ \cline{7-9}
                                                                                                                                               &&&&   &  & $\Xi_b^\prime \bar{D}$        & 7802 & 0
  \\ \cline{2-9}
  & \multirow{4}{*}{$|\Xi_b \bar{D}\rangle$} & \multirow{4}{*}{$(0)\frac{1}{2}^-$} & \multirow{4}{*}{7632.5 $-$ i3.1}  & \multirow{4}{*}{29} & \multirow{4}{*}{6.2} & $\Lambda \bar{B}_c$    & 7390    & 0.08
  \\ \cline{7-9}
                                                                                                                                               &&&&     &  & $\Lambda_b \bar{D}_s$        & 7588 & 0.37  	
  \\ \cline{7-9}
                                                                                                                                               &&&&
  &  & $\Xi_b \bar{D}$        & 7662 & 1.89  	
  \\ \cline{7-9}
                                                                                                                                               &&&&   &  & $\Xi_b^\prime \bar{D}$        & 7802 & 0
  \\ \cline{2-9}
  & \multirow{4}{*}{$|\Xi_b^\prime \bar{D}\rangle$} & \multirow{4}{*}{$(0)\frac{1}{2}^-$} & \multirow{4}{*}{7799.7 $-$ i0.1}  & \multirow{4}{*}{2} & \multirow{4}{*}{0.2} & $\Lambda \bar{B}_c$    & 7390    & 0.12
  \\ \cline{7-9}
                                                                                                                                               &&&&     &  & $\Lambda_b \bar{D}_s$        & 7588 & 0  	
  \\ \cline{7-9}
                                                                                                                                               &&&&
  &  & $\Xi_b \bar{D}$        & 7662 & 0  	
  \\ \cline{7-9}
                                                                                                                                               &&&&   &  & $\Xi_b^\prime \bar{D}$        & 7802 & 0.99
  \\\hline
  \multirow{8}{*}{Cut-off} & \multirow{4}{*}{$|\Xi_b \bar{D}\rangle$} & \multirow{4}{*}{$(0)\frac{1}{2}^-$} & \multirow{4}{*}{7576.1 $-$ i0}  & \multirow{4}{*}{86} & \multirow{4}{*}{0} & $\Lambda \bar{B}_c$    & 7390    & 0.05
  \\ \cline{7-9}
                                                                                                                                               &&&&     &  & $\Lambda_b \bar{D}_s$        & 7588 & 1.68 	
  \\ \cline{7-9}
                                                                                                                                               &&&&
  &  & $\Xi_b \bar{D}$        & 7662 & 3.35 	
  \\ \cline{7-9}
                                                                                                                                               &&&&   &  & $\Xi_b^\prime \bar{D}$        & 7802 & 0
  \\ \cline{2-9}
  & \multirow{4}{*}{$|\Xi_b^\prime \bar{D}\rangle$} & \multirow{4}{*}{$(0)\frac{1}{2}^-$} & \multirow{4}{*}{7792.0 $-$ i0.4}  & \multirow{4}{*}{10} & \multirow{4}{*}{0.8} & $\Lambda \bar{B}_c$    & 7390    & 0.21
  \\ \cline{7-9}
                                                                                                                                               &&&&     &  & $\Lambda_b \bar{D}_s$        & 7588 & 0  	
  \\ \cline{7-9}
                                                                                                                                               &&&&
  &  & $\Xi_b \bar{D}$        & 7662 & 0  	
  \\ \cline{7-9}
                                                                                                                                               &&&&   &  & $\Xi_b^\prime \bar{D}$        & 7802 & 1.89
  \\ \hline\hline
  \end{tabular}}
  \label{tab:resultcdsud}
\end{table*}

During the present study, we find that the results obtained using the dimensional and cut-off regularizations are not always the same. In this appendix we use the $PB_{1/2}$ sector of the $b\bar{c}uud$ system with $I={1/2}$ and the $PB_{1/2}$ sector of the $b\bar{c}sud$ system with $I=0$ as two examples to discuss this difference.

The dimensional regularization for the meson-baryon loop function $G^{\mathcal{MB}}(s)$ has been given in Eq.~(\ref{eq:G}), while the cut-off regularization for this loop function is
\begin{eqnarray}
  G_{ll}^{\prime\mathcal{M}\mathcal{B}}(s) &=& i \int \frac{\mathrm{d}^4 q}{(2\pi)^4}\frac{M_l}{E_l(\mathbf{q})} \frac{1}{k^0_i + p^0_i - q^0-E_l(\mathbf{q})+i\epsilon} \nonumber\\[2mm]
  && \times \frac{1}{q^2-m_l^2+i\epsilon} \nonumber\\[2mm]
  &=& \int_{|\mathbf{q}|<q_{\rm max}} \frac{\mathrm{d}^3 q}{(2\pi)^3} \frac{1}{2\omega_l(\mathbf{q})} \frac{M_l}{E_l(\mathbf{q})} \nonumber\\[2mm]
  && \times \frac{1}{k^0_i + p^0_i - \omega_l(\mathbf{q})-E_l(\mathbf{q})+i\epsilon},
  \label{eq:Gcut}
\end{eqnarray}
where $l$ denotes the intermediate channel; $p^0_i$ and $k^0_i$ are the energies of the initial meson and baryon, respectively; $m_l$ and $M_l$ are the masses of the intermediate meson and baryon, respectively; $\omega_l=\sqrt{m_l^2 + \mathbf{q}^2}$ and $E_l=\sqrt{M_l^2 + \mathbf{q}^2}$ are the energies of the intermediate meson and baryon, respectively.

For the $PB_{1/2}$ sector of the $b\bar{c}uud$ system with $I={1/2}$, we consider the $N \bar{B}_c$, $\Lambda_b \bar{D}$, and $\Sigma_b \bar{D}$ coupled channels. The results are summarized in Table~\ref{tab:resultcduud}, where we choose the subtraction constant for the dimensional regularization to be $a(\mu=1 \gev)=-3.2$, and we choose the three-momentum cutoff for the cut-off regularization to be $q_{\max}=450 \mev$. Using these two parameters, the loop functions $G_{\Sigma_b \bar{D}}^{\mathcal{MB}}(s)$ and $G_{\Sigma_b \bar{D}}^{\prime \mathcal{MB}}(s)$ have the same value at the $\Sigma_b \bar{D}$ threshold. Compared to the cut-off regularization, there exists an extra bound state below the $\Lambda_b \bar{D}$ threshold when using the dimensional regularization. This pole should be discarded, as already discussed in Ref.~\cite{Wu:2010rv}. The reason can be clearly seen in Fig.~\ref{fig:ReIm}, where we show the real and imaginary parts of the $G^{\mathcal{MB}}(s)$ and $G^{\prime \mathcal{MB}}(s)$ loop functions. As shown in Fig.~\ref{fig:Imduud} and Fig.~\ref{fig:Imcuud}, their imaginary parts are exactly the same within the effective range. In the absence of coupled channel effects, the scattering matrix is given by $T=(V^{-1} - G)^{-1}$, and a pole is expected when the real part of the loop function equals the inverse of the interaction. As shown in Fig.~\ref{fig:Reduud}, we use the dash line to indicate the extra bound state located at $7372.6 \mev$, where the real part of $G_{\Lambda_b \bar{D}}^{\mathcal{MB}}(s)$ is positive and far below the $\Lambda_b \bar{D}$ threshold. Moreover, the potential $V_{\Lambda_b \bar{D} \to \Lambda_b \bar{D}}^{P\mathcal{B}}(s)$ is also positive around this energy region, indicating the interaction to be repulsive. Therefore, this pole can not be a bound state and should be discarded. Oppositely, as shown in Fig.~\ref{fig:Recuud}, the real part of $G_{\Lambda_b \bar{D}}^{\prime \mathcal{MB}}(s)$ is negative below the $\Lambda_b \bar{D}$ threshold, so that the above bound state does not appear at all.

For the $PB_{1/2}$ sector of the $b\bar{c}sud$ system with $I=0$, we consider the $\Lambda \bar{B}_c$, $\Lambda_b \bar{D}_s$, $\Xi_b \bar{D}$, and $\Xi_b^\prime \bar{D}$ coupled channels. The results are summarized in Table~\ref{tab:resultcdsud}, where we choose the subtraction constant for the dimensional regularization to be $a(\mu=1 \gev)=-3.2$, and we choose the three-momentum cutoff for the cut-off regularization to be $q_{\max}=450 \mev$. Similarly, compared to the cut-off regularization, there exists an extra bound state below the $\Lambda_b \bar{D}_s$ threshold when using the dimensional regularization, and this pole should also be discarded. Moreover, in this case the results obtained using the dimensional regularization are significantly different from those obtained using the cut-off regularization, which needs further investigations in the future studies.

\bibliographystyle{elsarticle-num}
\bibliography{ref}

\end{sloppypar}
\end{document}